\documentclass[acmsmall]{acmart}
\AtBeginDocument{%
  }

\usepackage{booktabs} 
\usepackage{color} 
\usepackage{balance} 
\usepackage{url} 
\usepackage[all]{nowidow}
\usepackage{tabularx}
\newcolumntype{Y}{>{\centering\arraybackslash}X}
\usepackage{siunitx} 
\usepackage{tabularx,subfigure,multirow, graphicx}
\usepackage{epsfig}
\usepackage{epstopdf} 
\usepackage{tikz}
\usepackage{pgfplots}
\usepackage{enumitem}
\usepackage{caption}
\usepackage{colortbl}
\usepackage{tcolorbox}

\usepackage[a-2b]{pdfx}

\graphicspath{{./figures/}}

\usepackage{hyperref} 

\usepackage{makecell}
\usepackage{colortbl}

\usepackage{amsthm,amsmath,amsfonts}

\newcolumntype{L}[1]{>{\raggedright\arraybackslash}p{#1}}
\newcolumntype{C}[1]{>{\centering\arraybackslash}p{#1}}
\newcolumntype{R}[1]{>{\raggedleft\arraybackslash}p{#1}}

\usepackage[linesnumbered,ruled,vlined]{algorithm2e}
\SetKwRepeat{Do}{do}{while}

\long\def\comment#1{}

\newcommand{\nop}[1]{}

\newcommand{\figureTopMargin}{\vspace{-3ex}}
\newcommand{\figureCaptionMargin}{\vspace{-1ex}}
\newcommand{\figureBelowMargin}{\vspace{-2ex}}

\theoremstyle{remark}

\theoremstyle{definition}
\newtheorem{definition}{\bf Definition}

\begin{document}

\title{FGIM: a Fast Graph-based Indexes Merging Framework for Approximate Nearest Neighbor Search}

\author{Zekai Wu}
\affiliation{
	\institution{East China Normal University}
	\city{Shanghai}
	\country{China}
}
\email{zekaiwu@stu.ecnu.edu.cn}

\author{Jiabao Jin}
\affiliation{
	\institution{Ant Group}
	\city{Shanghai}
	\country{China}
}
\email{jinjiabao.jjb@antgroup.com}

\author{Peng Cheng}
\authornote{Corresponding Author.}
\affiliation{
	\institution{Tongji University}
	\city{Shanghai}
	\country{China}
}
\email{cspcheng@tongji.edu.cn}

\author{Xiaoyao Zhong}
\affiliation{
	\institution{Ant Group}
	\city{Shanghai}
	\country{China}
}
\email{zhongxiaoyao.zxy@antgroup.com}

\author{Lei Chen}
\affiliation{
	\institution{HKUST (GZ)}
	\city{Guangzhou}
	\country{China}
}
\affiliation{
	\institution{HKUST}
	\city{Hong Kong SAR}
	\country{China}
}
\email{leichen@cse.ust.hk}

\author{Yongxin Tong}
\affiliation{
	\institution{Beihang University}
	\city{Beijing}
	\country{China}
}
\email{yxtong@buaa.edu.cn}

\author{Zhitao Shen}
\affiliation{
	\institution{Ant Group}
	\city{Shanghai}
	\country{China}
}
\email{zhitao.szt@antgroup.com}

\author{Jingkuan Song}
\affiliation{
	\institution{Tongji University}
	\city{Shanghai}
	\country{China}
}
\email{jingkuan.song@gmail.com}

\author{Heng Tao Shen}
\affiliation{
	\institution{Tongji University}
	\city{Shanghai}
	\country{China}
}
\email{shenhengtao@hotmail.com}

\author{Xuemin Lin}
\affiliation{
	\institution{Shanghai Jiaotong University}
	\city{Shanghai}
	\country{China}
}
\email{xuemin.lin@gmail.com}

\renewcommand{\shortauthors}{Zekai Wu et al.}

\begin{abstract}
	As the state-of-the-art methods for high-dimensional data retrieval, Approximate Nearest Neighbor Search (ANNS) approaches with graph-based indexes have attracted increasing attention and play a crucial role in many real-world applications, e.g., \textit{retrieval-augmented generation} (RAG) and recommendation systems. Unlike the extensive works focused on designing efficient graph-based ANNS methods, this paper delves into merging multiple existing graph-based indexes into a single one, which is also crucial in many real-world scenarios (e.g., cluster consolidation in distributed systems and read-write contention in real-time vector databases). We propose a Fast Graph-based Indexes Merging (FGIM) framework with three core techniques: (1) \textit{Proximity Graphs (PGs) to $k$ Nearest Neighbor Graph ($k$-NNG) transformation} used to extract potential candidate neighbors from input graph-based indexes through \textit{cross-querying}, (2) \textit{$k$-NNG refinement} designed to identify overlooked high-quality neighbors and maintain graph connectivity, and (3) \textit{$k$-NNG to PG transformation} aimed at improving graph navigability and enhancing search performance. Then, we integrate our FGIM framework with the state-of-the-art ANNS method, HNSW, and other existing mainstream graph-based methods to demonstrate its generality and merging efficiency. Extensive experiments on six real-world datasets show that our FGIM framework is applicable to various mainstream graph-based ANNS methods, achieves up to 3.5$\times$ speedup over HNSW's incremental construction and an average of 7.9$\times$ speedup for methods without incremental support, while maintaining comparable or superior search performance.
\end{abstract}



\setcopyright{cc}
\setcctype{by-nc-nd}
\acmJournal{PACMMOD}
\acmYear{2026} \acmVolume{4} \acmNumber{1 (SIGMOD)} \acmArticle{37} \acmMonth{2} \acmPrice{}\acmDOI{10.1145/3786651}

\maketitle

\section{Introduction}
\label{sec:introduction}

\textit{Approximate Nearest Neighbor Search} (ANNS) \cite{arya1993approximate,arya1998optimal} is a fundamental problem across various fields such as machine learning \cite{cost1993weighted, wang2020survey}, information retrieval \cite{wang2012scalable, zhu2019accelerating}, recommendation systems \cite{das2007google,meng2020pmd}, vector databases \cite{wang2021milvus}, and Large Language Models (LLMs) \cite{dobson2023scaling, asai2023retrieval}. Recent deep-learning methods have revolutionized data representation by embedding unstructured data (e.g., images and texts) into high-dimensional vectors that preserve semantic relationships, which are crucial for similarity-based retrieval and various downstream machine-learning tasks. ANNS can efficiently retrieve semantically related data points from large-scale vectorized datasets, and is indispensable for modern AI applications \cite{asai2023retrieval, lewis2020retrieval, liu2024retrievalattention}. Given a vector dataset $\mathcal{X}$ and a query vector $\vec{x}_q \in \mathbb{R}^d$, ANNS aims to efficiently and effectively retrieve a vector $\vec{x}_r$ from $\mathcal{X}$ with the minimum distance to $\vec{x}_q$. According to recent studies \cite{naidan2015permutation,aumuller2020ann,li2019approximate,wang2021comprehensive}, graph-based indexes have emerged as one of the most effective techniques for ANNS in view of their superior search accuracy.

Despite the good performance of graph-based approaches, their practical deployment in industrial settings remains challenging. Industrial applications often involve massive datasets (e.g., 1B\textasciitilde10B vectors), where \textit{distance computation} across candidate vectors dominates query latency. For small indexes constructed from data shards, merging them into a larger index can improve computational efficiency (e.g., Hierarchical Navigable Small World (HNSW) \cite{malkov2018efficient} consumes $O(\log{n})$ time for search~\cite{wang2021comprehensive}, and since $\sum \log{n_i} < \log{\sum n_i}$, a larger index enables faster search than partitioned ones). Moreover, real-world systems often operate in dynamic environments, where both data and device conditions change frequently. As a result, merging multiple indexes into a single one is required in various scenarios, which poses significant challenges.

\begin{figure}[t]
	\centering
	\begin{minipage}[t]{0.48\linewidth}
		\centering
		\includegraphics[width=\linewidth]{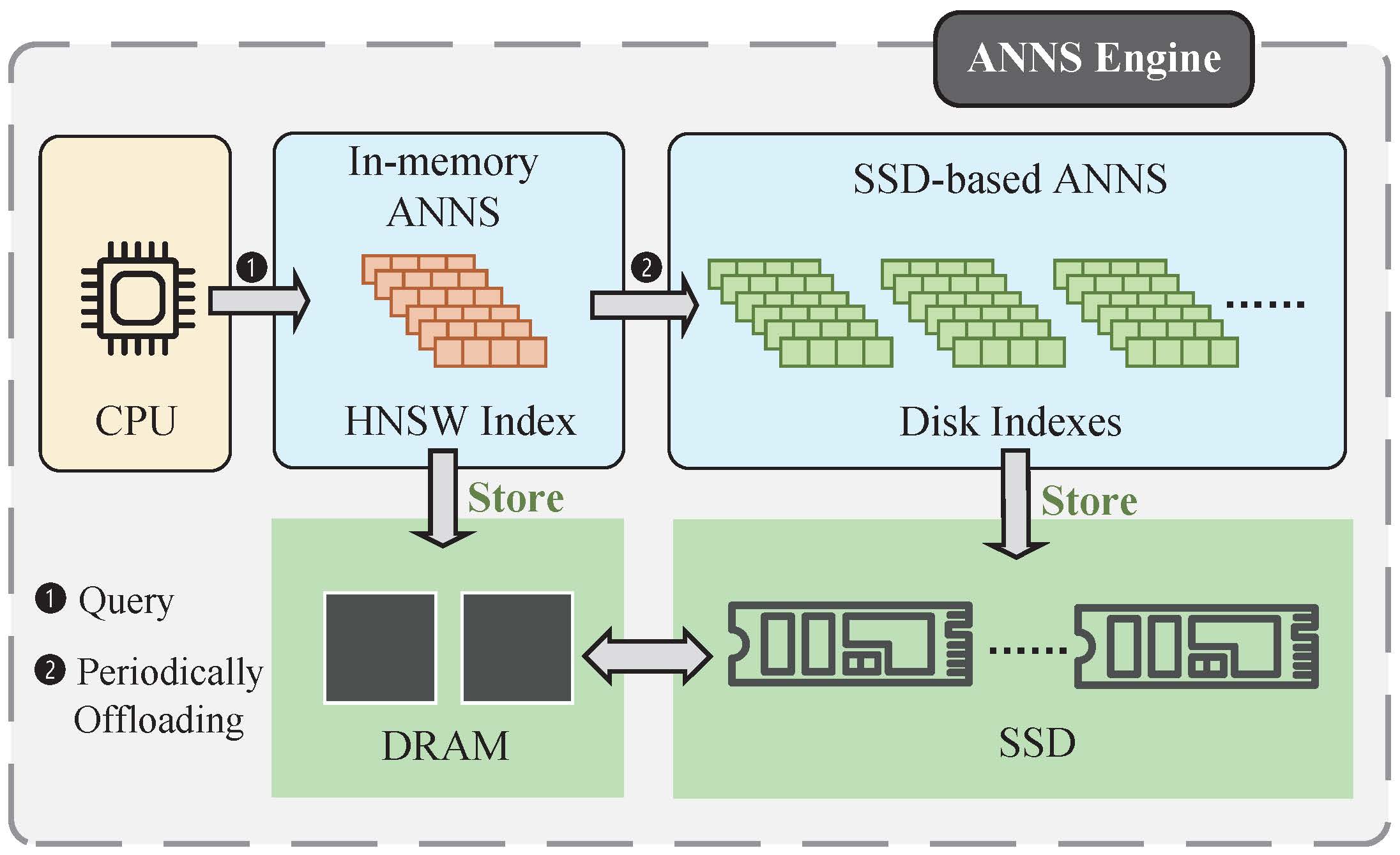}
		\vspace{-2ex}
		\captionof{figure}{A hybrid engine for ANNS.}
		\label{fig:engine}
	\end{minipage}
	\hfill
	\begin{minipage}[t]{0.48\linewidth}
		\centering
		\includegraphics[width=\linewidth]{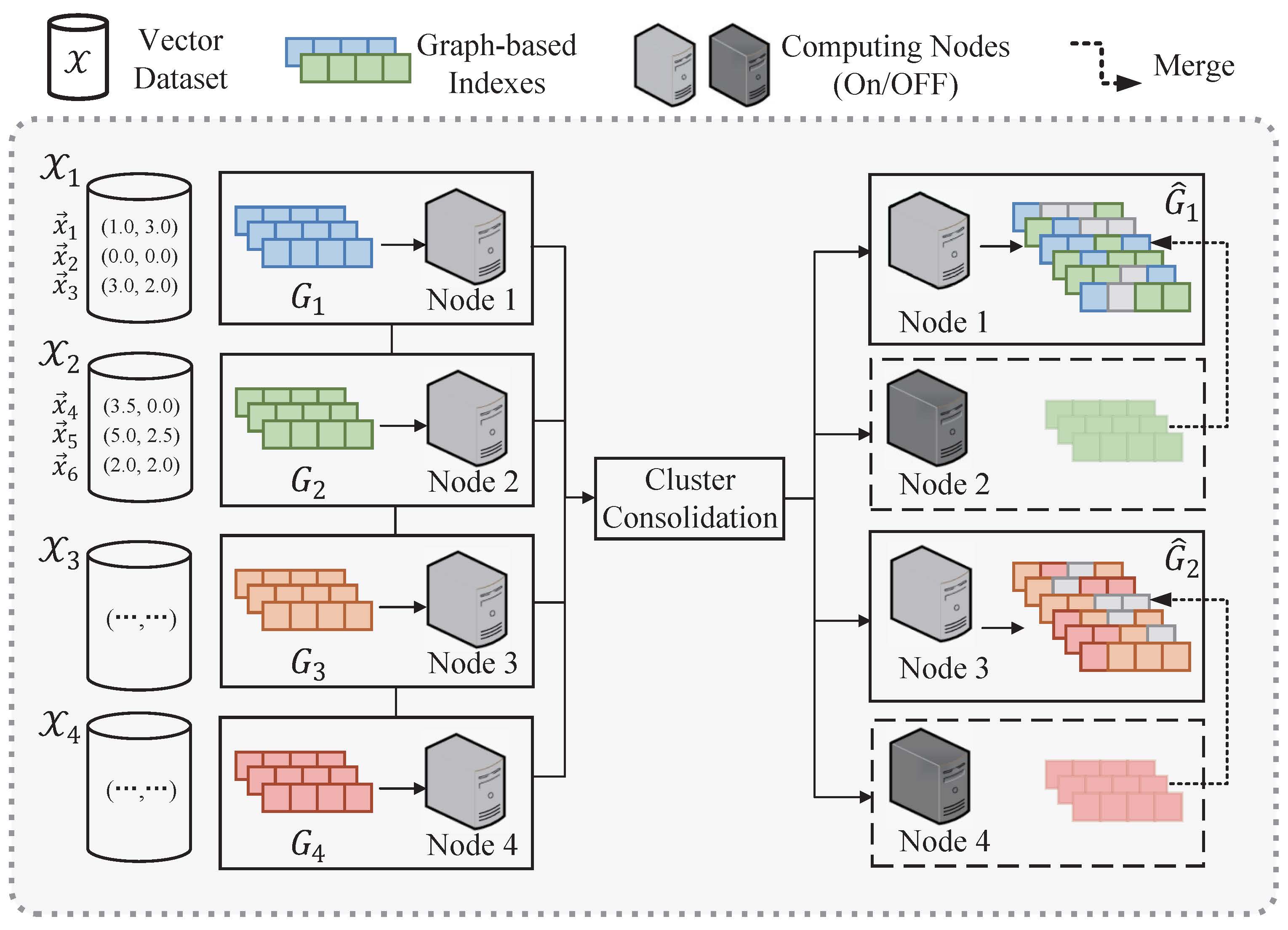}
		\vspace{-2ex}
		\captionof{figure}{An illustration of cluster consolidation.}
		\label{fig:cluster}
	\end{minipage}
	\figureBelowMargin
\end{figure}

We first consider a cost-effective and high-performance scenario of real-time industrial applications (e.g., \textit{Retrieval-Augmented Generation} (RAG) \cite{lewis2020retrieval, asai2023retrieval}), where indexing must be performed online to support continuous data writes. While HNSW \cite{malkov2018efficient} is widely adopted as a real-time indexing solution for its incremental insertion capability, its memory-intensive multi-layer structure motivates complementary SSD-based solutions (e.g., DiskANN \cite{jayaram2019diskann}), which trade query speed for cost efficiency. An industrial hybrid approach, adopted by Milvus \cite{wang2021milvus} and VSAG \cite{zhong2025vsag} illustrated in Figure~\ref{fig:engine}, follows a Log-Structured Merge (LSM) architecture \cite{o1996log}: a size-bounded HNSW index (typically 1M–5M vectors) handles real-time writes, with periodic offloading to disk-based indexes. However, accumulating disk-based indexes requires background reconstruction to preserve search quality. \textit{This process is computationally expensive, thus difficult to be applied in online scenarios.}

In another distributed scenario illustrated in Figure~\ref{fig:cluster}, to handle large-scale data efficiently, distributed storage and indexing systems~\cite{li2023more, yu2025approximate} have become increasingly prevalent. In these systems, each node constructs its index based on locally available data. As the distributed system evolves, scaling requirements may arise due to \textit{cluster consolidation}~\cite{hermenier2009entropy,lin2013consolidated} or \textit{resource reallocation}~\cite{Lee2014Resource}. Specifically, computing nodes within a cluster are often merged to reduce costs and simplify management, which requires merging the existing indexes into a single one. The existing methods~\cite{fu2017fast,peng2023efficient,dong2011efficient} rely on dataset-oriented merging operations (i.e., rebuilding the index from scratch) due to their unique construction strategies, which incurs \textit{substantial computational overhead and limited scalability.}

\bgroup
\setlength{\parskip}{2pt}
\noindent\textbf{Challenges.} To the best of our knowledge, no existing studies \cite{malkov2018efficient,fu2017fast,jayaram2019diskann,peng2023efficient,dong2011efficient,fu2016efanna} can meet the aforementioned requirements of industrial deployment to efficiently merge the graph-based indexes. To bridge this gap, we develop a practical approach to significantly reduce the time overhead of the merging process 
assuming the involved graph-based indexes fit in main memory, which is realistic in modern systems. While vectors may reside on SSDs, merging is typically performed on bounded, memory-resident graph shards \cite{zhong2025vsag}, and growing DRAM capacity makes such in-memory consolidation increasingly feasible \cite{wang2021comprehensive, azizi2025graph}.
Our work focuses on the efficient merging of multiple existing graph indexes into a single one, which is non-trivial and has two key challenges.

\noindent\underline{Challenge I:} \textit{How can the neighbor relationships stored in the existing indexes be used to facilitate the construction of the merged graph-based index?} Intuitively, an effective merging approach should extract valuable information (e.g., neighborhood relationship) from the existing multiple indexes to avoid as many computational overheads as possible, which is particularly important for high-dimensional datasets (e.g., GloVe \cite{pennington2014glove}, Gist \cite{jegou2010product} and Crawl \cite{commoncrawl}). However, different ANNS indexing structures employ different graph construction strategies and pruning techniques \cite{wang2021comprehensive}. To ensure that, our approach is compatible with various indexing structures, it is necessary to map these heterogeneous structures into a unified merged structure in the preprocessing step, which facilitates high scalability and enables more efficient downstream procedures.

\noindent\underline{Challenge II:} \textit{How can the merged index achieve good search performance?} The merging process must preserve the structural properties that are crucial for efficient ANNS. The state-of-the-art (SOTA) graph-based methods (e.g., HNSW \cite{malkov2018efficient} and Vamana \cite{fu2017fast})  achieve high performance through careful edge selection and pruning strategies, which optimize the search path length and thus reduce distance computations. During index merging, we need to ensure the combined graph-based index maintains similar connectivity and navigability properties to those of the index directly constructed on the merged dataset to guarantee efficient search performance.

\noindent\textbf{Solution and Contributions.} We propose a \textit{Fast Graph-based Indexes Merging (FGIM)} framework to effectively merge graph-based indexes for ANNS. FGIM reformulates the index merging problem as a graph merging problem by mapping disconnected graph-based indexes onto a unified graph structure and optimizing it for ANNS tasks. Specifically, our framework consists of three core steps: \textit{PGs to $k$-NNG transformation}, \textit{$k$-NNG refinement}, and \textit{$k$-NNG to PG transformation}. Firstly, the \textit{PGs to $k$-NNG transformation} extracts enough potential candidate neighbors from the original graph-based indexes. It retains the existing neighborhood relationships and elaborately selects pivotal cross-graph candidate neighbors through \textit{cross-querying}, then maps multiple graphs into a unified $k$-NNG structure. In the next step, the obtained $k$-NNG is iteratively optimized through a streamlined \textit{$k$-NNG refinement} process to establish more accurate direct connections between cross-graph vertices, incorporating \textit{indegree-aware} mechanism to enhance graph connectivity. Then, the neighbor selection and graph optimization processes are applied in the \textit{$k$-NNG to PG Transformation} step, both designed to improve search performance and graph navigability. Besides, a merging strategy tailored for HNSW \cite{malkov2018efficient} is proposed to facilitate the integration of hierarchical structures. The FGIM framework produces a merged graph-based index optimized for ANNS tasks, ensuring both applicability and efficiency. To summarize, we make the following contributions:

\begin{itemize}[leftmargin=*,topsep=0pt, itemsep=0pt, parsep=0pt]
	\item We formulate the graph-based index merging problem in \textsection\ref{sec:problemDefinition}, then propose a universal framework FGIM for merging graph-based indexes in \textsection\ref{sec:framework}, which is designed to be compatible with a wide range of mainstream graph-based ANNS methods.
	\item We present the implementation details of our proposed framework in \textsection\ref{sec:implementation}, including the \textit{PGs to $k$-NNG transformation}, \textit{$k$-NNG refinement}, and \textit{$k$-NNG to PG transformation}. 
	\item Extensive experiments on real datasets show the efficiency and applicability of our proposed framework in \textsection\ref{sec:experimental}.
\end{itemize}

\egroup

\section{Preliminaries}
\label{sec:problemDefinition}

\begin{table}
	\vspace{-2ex}
	\centering
	{\small
		\caption{\small Symbols and Descriptions}
		\label{tab:symbols}
		\vspace{-2ex}
		\label{table0}
		\begin{tabular}{l|l}
			\toprule
			{\bf Symbol} & \multicolumn{1}{c}{\bf Description} \\
			\midrule
			${\mathcal{X}}$ & the base dataset \\
			$h$ & the number of indexes to be merged \\
			${G(V, E)}$ & the graph index with vertex set $V$ and edge set $E$ \\
			$\delta(\cdot, \cdot)$ & the distance function \\
			$\vec{x}, \vec{x}_b, \vec{x}_q$ & a normal, base, and query vector \\
			$\lvert \cdot \rvert$ & the cardinality of a set \\ 
			$L$ & the candidate pool size in search \\
			$k$ & the maximum outdegree of the graph \\
			$\phi(\cdot)$ & the id mapping function from old to new indexes \\
			$C_i(u)$ & the candidate neighbor set of vertex $u$ in $G_i$. \\
			$N_G(u)$ & the neighbors of vertex $u$ in graph $G$ \\
			$\mathcal{T}(u)$ & the indegree of vertex $u$ \\
			\bottomrule
		\end{tabular}
	}
	\vspace{-1ex}
\end{table}

We show problem definitions in \textsection\ref{subsec:problemDefinition} and discuss the current studies in \textsection\ref{subsec:currentStudies}. Table \ref{table0} lists the frequently used notations in this paper.

\subsection{Problem Definition}
\label{subsec:problemDefinition}

\subsubsection{ANNS Definition.} 

Let $d \in \mathbb{N}$ be the dimension of the vector, $n \in \mathbb{N}$ be the number of vectors in the dataset. The dataset is given by $\mathcal{X} = \{ \vec{x}_1, \vec{x}_2, \ldots, \vec{x}_n\} \subset \mathbb{R}^d$. Let $\delta(a, b) : \mathbb{R}^d \times \mathbb{R}^d \rightarrow \mathbb{R}$ denote the distance function that computes the distance between any two vectors $\vec{x}_a$ and $\vec{x}_b$ in $\mathcal{X}$ in a specific metric space (e.g., $\ell_2$ or cosine). The task of $k$-Nearest Neighbor Search ($k$-NNS) is defined as follows:

\begin{definition}[$k$-NNS]
	\textit{Given a finite dataset $\mathcal{X}$ and a query vector $\vec{x}_q \in \mathbb{R}^d$, and a parameter $k \le n$, $k$-NNS retrieve the set $\mathcal{R}$ consisting of the $k$ vectors from $\mathcal{X}$ that have the minimum distance to $\vec{x}_q$ based on $\mathcal{\delta}$. For $\forall \vec{x}_r \in \mathcal{R}$ and $\forall \vec{x}_s \in \mathcal{X} \setminus R$, we have $\delta(\vec{x}_q, \vec{x}_r) \le \delta(\vec{x}_q, \vec{x}_s)$. $\mathcal{R}$ can be formally described as follows:}
	
	\vspace{-2ex}
	\begin{equation}
		\mathcal{R} = \mathop{\arg\min}\limits_{\lvert \mathcal{R} \rvert = k, \mathcal{R} \subset \mathcal{X}} \sum_{x \in \mathcal{R}} \delta(q, x)
	\end{equation}
	\label{def:knns}
\end{definition}

In modern applications, the increasing size of datasets and the high dimensionality of vectors lead to significant amplification of the computational cost of exact $k$-NNS. To mitigate this, ANNS techniques construct optimized indexes that balance the search efficiency and result accuracy.

\begin{definition}[$k$-ANNS]
	\textit{Given a finite dataset $\mathcal{X}$ and a query vector $\vec{x}_q \in \mathbb{R}^d$, ANNS constructs an Index $I$ on $\mathcal{X}$. It then retrieves a subset $C$ of $\mathcal{X}$ by $I$, and evaluates $\delta(\vec{x}_i, \vec{x}_q)$ to obtain the approximate $k$ nearest neighbors $\mathcal{\tilde{R}}$ of $q$, where $\vec{x}_i \in C$.}
	
	\label{def:anns}
\end{definition}

The top-$k$ recall rate ($Recall@k$) is commonly used to evaluate ANNS performance. Given a query $\vec{x}_q$ and a value $k \le n$, $Recall@k$ is defined as:\vspace{-2ex}

\begin{equation}
	Recall@k = \frac{| \mathcal{R} \cap \tilde{ \mathcal{R} } |}{|k|}
\end{equation}

\begin{algorithm}[t]\small
	\DontPrintSemicolon
	\KwIn{query vector $\vec{x}_q$, graph index $G = (V, E)$, search pool size $L$, $k$ for top-$k$, optional enterpoint $ep$}
	\KwOut{$k$ nearest neighbors of $\vec{x}_q$}
	initialize $C \gets$ $\{(ep, \delta(ep, q))\}$ and $i$ $\gets$ 0\;
	\While{$i < L$}{
		$u$ $\gets$ $C[i]$\;
		mark $u$ as visited\;
		\ForEach{$v \in N(u)$ and $v$ is not visited}{
			insert($v, \delta(v, q)$) into $C$\;
		}
		sort $C$ by $\delta(q, x)$ and keep the top-$L$ results\;
		$i$ $\gets$ index of the first unexpanded vertex in $C$\;
	}
	\Return{the first $k$ results in $C$}\;
	\caption{KNNSearch($q$, $G$, $L$, $k$, $ep$)}
	\label{alg:anns}
\end{algorithm}

Algorithm \ref{alg:anns} shows a general search process of ANNS. Specifically, the algorithm maintains a candidate set $C$ with search pool size $L (\ge k)$ (also referred to as \textit{beam width}) to record the currently best $L$ nearest neighbors of the query $\vec{x}_q$. The algorithm adds the enterpoint $ep$ to the candidate set as the initialization of the candidate set (Line 1). The algorithm extracts the nearest neighbor from the candidate set at the beginning of each iteration (Lines 2-4), and then expands the selected vertex $u$ by inserting all unvisited vertices $v \in N(u)$ into the candidate set (Lines 4-5). If the size of the candidate set exceeds $L$, the algorithm keeps the top-$L$ nearest neighbors in the candidate set and removes the rest (Line 7). Afterward, the algorithm selects the next vertex to expand by discovering the first unvisited vertex in the candidate set at the end of each iteration (Line 8). The algorithm terminates when no unvisited vertex can be found in the $C$. Finally, the algorithm returns the top-$k$ nearest neighbors in the candidate set as the search results (Line 9).

\subsubsection{Graph-based Indexes.}
\label{subsubsec:graphBasedIndex}

Graph-based methods are reported to achieve superior search performance in ANNS tasks \cite{aumuller2020ann,li2019approximate,wang2021comprehensive}. They map base vectors into a graph space, and then construct a proximity graph (PG) to represent the similarity relationships between base vectors.

\begin{definition}[Proximity Graph]
	\textit{The PG of $\mathcal{X}$ is a graph $G = (V, E)$ with the vertex set $V$ and edge set $E$. Each vertex $v_i \in V$ corresponds to a vector $\vec{x}_i \in \mathcal{X}$. Each edge $e_{ij} \in E$ represents the proximity between vertices $v_i$ and $v_j$, which is determined by a specified distance metric (e.g., $\ell_2$). The neighbors of a vertex $v$ in $V$ are denoted as $N_G(v)$.}
	\label{def:proximity}
\end{definition}

\begin{definition}($k$-Nearest Neighbor Graph).
\textit{$k$-NNG $G = (V, E)$ connects each vertex $v_i \in V$ to its $k$ nearest neighbors in dataset $\mathcal{X}$, where $|N_G(v_i)| = k$.}
	\label{def:knn}
\end{definition}

The graph-based methods employ various strategies for PG construction, including search-based approaches (e.g., HNSW \cite{malkov2018efficient} and Vamana \cite{jayaram2019diskann}) and refinement-based techniques (e.g., NSG \cite{fu2017fast} and $\tau$-MNG \cite{peng2023efficient}). Specifically, the construction process typically involves obtaining a subset of vertices as candidate neighbor set $C$ for each vertex $v$ (referred to as \textit{Candidate Neighbor Acquisition} in the literature \cite{wang2021comprehensive, yang2024revisiting}) to construct an approximate $k$-NNG, which can be achieved through search or other heuristics, and then applying a pruning strategy to select the final neighbors from $C$ named \textit{Neighbor Selection}. 
For example, NSG \cite{fu2017fast} first constructs an approximate $k$-NNG and then prunes it to produce the PG. Similarly, while Vamana \cite{jayaram2019diskann} starts from a random graph, it obtains the $k$-NNG through search before performing two rounds of optimization. Previous studies~\cite{yang2024revisiting} show that increasing parameters such as $ef_{construction}$ in HNSW or $L$ in Vamana produces a more accurate $k$-NNG, leading to better PG quality and improved ANNS performance. Although higher-quality $k$-NNGs incur greater construction costs, it requires a balance between accuracy and efficiency.

A critical distinction among these methods lies in their pruning strategies, which play a crucial role in shaping the structural properties of the graph and directly influence search performance~\cite{wang2021comprehensive}. While some methods (e.g., HNSW) incorporate Relative Neighborhood Graph (RNG) \cite{toussaint1980relative}, others (e.g., NSG and Vamana) apply Monotonic Relative Neighborhood Graph (MRNG) \cite{fu2017fast}. RNG preserves edges based on relative proximity, whereas MRNG adds a monotonicity constraint to ensure greedy-search navigability.

PGs have become a focal point in recent ANNS studies \cite{wang2021comprehensive,li2019approximate,yang2024revisiting} due to their ability to effectively capture the local neighborhood structure of high-dimensional data, facilitating efficient and accurate search operations. By representing data points as vertices and establishing edges between nearby points based on a distance metric, PGs preserve essential proximity relationships. Their fundamental properties, such as sparsity, which reduces computational complexity, and connectivity, which ensures navigability, enable efficient graph traversal while minimizing distance computations. These characteristics make PGs effective for balancing speed and accuracy in large-scale retrieval tasks. 

Then, we have the following problem definition:
\begin{definition}[Merging Graph-based Indexes for ANNS]\label{def:merge_revised}
Given $h$ datasets $\{\mathcal{X}_i \subset \mathbb{R}^d\}_{i=1}^h$, we have their $h$ pre-built graph-based indexes $\{G_i=(V_i,E_i)\}_{i=1}^h$ using the same distance metric $\delta$. Let $B(\mathcal{X}_i)$ denote the index construction cost of dataset $\mathcal{X}_i$, and $M(\cdot)$ the cost of merging multiple indexes. The merge process is formulated as a multi-objective optimization problem:
$$
\max \Big\{\, B\!\left(\mathcal{X}\right) - M(G_1,\dots,G_h),\; Q(\hat{G}),\; R(\hat{G}) \,\Big\},
$$
where $\mathcal{X} = \bigcup_{i=1}^h \mathcal{X}_i$, the merged index is $\hat{G}=(V,E)$ with $V=\bigcup_{i=1}^h V_i$, $Q(\hat{G})$ is the query-per-second throughput of $\hat{G}$, and $R(\hat{G})$ is the recall rate of $\hat{G}$.
\end{definition}
The goal of merging graph-based indexes for ANNS is to build an index $\hat{G}$ for the union set $\mathcal{X}$ of the $h$ datasets with the maximum saving cost comparing with building $\hat{G}$ on $\mathcal{X}$ directly, while optimizing the query throughput $Q(\hat{G})$ and recall rate $R(\hat{G})$ of $\hat{G}$.

\subsection{Current Studies}
\label{subsec:currentStudies}

Apart from the brute-force reconstruction method, prior work in the literature can be summarized into the following three categories.

\noindent\underline{Incremental Indexing.} Incremental insertion provides a practical approach to index merging, where vectors from smaller datasets are sequentially added to a base index built on a larger dataset. Some search-based structures, such as HNSW \cite{malkov2018efficient}, naturally support this process through point-wise insertions. However, this approach has notable limitations. For instance, NSG \cite{fu2017fast} and Vamana~\cite{jayaram2019diskann} identify data centroids during preprocessing and use Algorithm~\ref{alg:anns} starting from these centroids to select candidate neighbors, ensuring a monotonic search path in the MRNG \cite{fu2017fast}. Incrementally inserting new points can shift centroids, potentially violating the MRNG property and requiring a full reconstruction of the index. Moreover, incremental insertion relies solely on the base index and ignores indexing information from other indexes, which limits its ability to fully leverage their potential contributions to the merged index.

\noindent\underline{$k$-NNG Merge.} Zhao et al. \cite{zhao2021merge} propose a method for merging two $k$-NNGs, which first introduces random edges between the two disjoint $k$-NNGs, followed by applying NNDescent \cite{dong2011efficient} as refinement. However, since this method is designed specifically for merging two $k$-NNGs, it is neither suitable for efficient ANNS tasks, according to recent studies \cite{li2019approximate,wang2021comprehensive}, nor compatible with mainstream graph-based indexing methods (e.g., HNSW \cite{malkov2018efficient}, Vamana \cite{jayaram2019diskann} and NSG \cite{fu2017fast}), which do not produce $k$-NNGs as their final index structures.

\noindent\underline{DiskANN.} DiskANN \cite{jayaram2019diskann} introduces a scalable index merging strategy to handle billion-scale datasets using multiple shards. The method partitions the base data into clusters via $k$-means and assigns each point to its $\ell$ nearest centers to form $k$ partitions, then constructs a Vamana index for each partition. Finally, merging is performed by taking the union of the subgraphs. However, this approach entails repeated indexing on the same data point across partitions, inflating the index size by a factor of $\ell$. Additionally, search performance becomes highly sensitive to the choice of clustering parameters. Since the method depends on spatial clustering of the original data and forms cross-partition connections through repeated indexing, it struggles to generalize to arbitrarily distributed data without dedicated preprocessing, resulting in substantial performance degradation.

\textit{Overall, the aforementioned methods exhibit limitations in terms of universality, efficiency, and effectiveness}. These limitations highlight the potential to design a novel merging method that can effectively leverage the strengths of existing graph-based indexes while addressing the challenges of efficiency and search performance.

\section{Fast Graph-based Indexes Merging Framework}
\label{sec:framework}

We propose a general framework for merging graph-based ANNS indexes. We first introduce the motivation behind our solution in \textsection\ref{subsec:motivation}. Next, in \textsection\ref{subsec:why_knng}, we explain why we convert the original indexes into a $k$-NNG. We then present the full pipeline in \textsection\ref{subsec:framework}.

\begin{figure}[t]
	\centering
	\figureTopMargin
	
	\centering
	\includegraphics[width=0.5\linewidth]{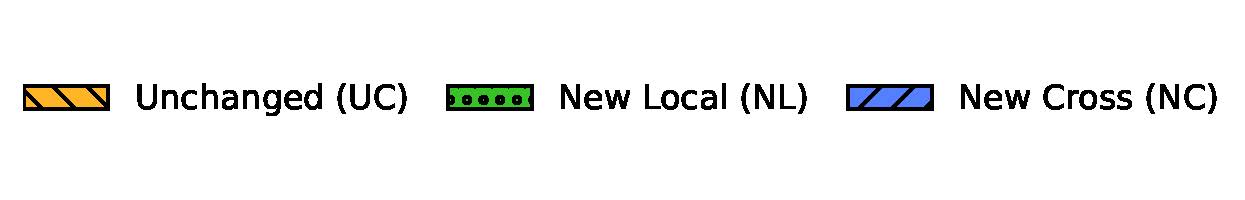}
	\vspace{-3ex}
	
	\begin{minipage}{0.497\linewidth}
		\centering
		\subfigure[][\scriptsize Number of Subsets Constructed by HNSW]{
			\hspace{-0.5em}
			\includegraphics[width=1.02\linewidth]{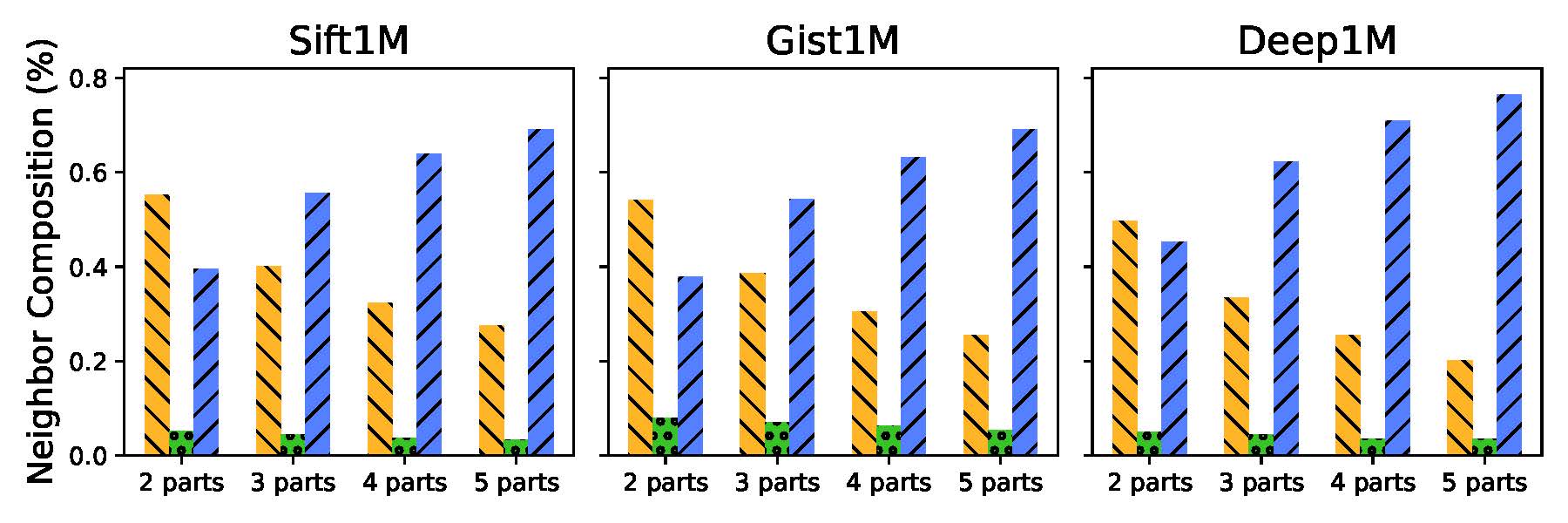}
			\hspace{-0.5em}
			\label{fig:hnsw_composition}
		}
	\end{minipage}
	\begin{minipage}{0.497\linewidth}
		\centering
		\subfigure[][\scriptsize Number of Subsets Constructed by Vamana]{
			\hspace{-0.5em}
			\includegraphics[width=1.02\linewidth]{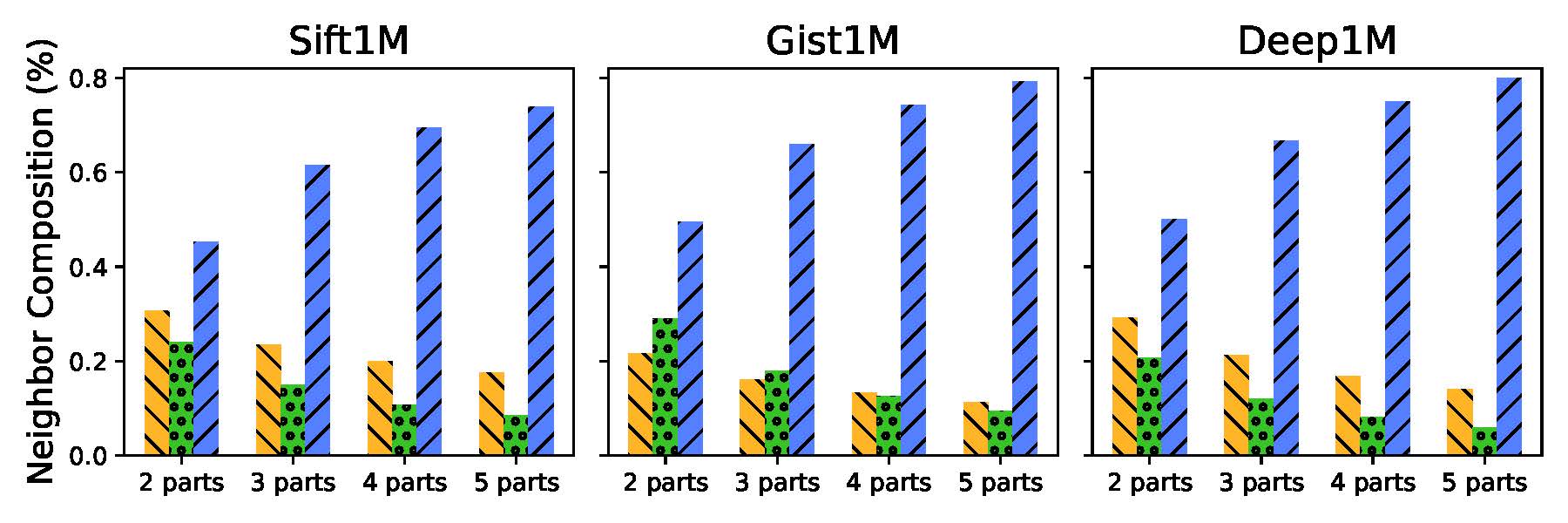}
			\hspace{-0.5em}
			\label{fig:vamana_composition}
		}
	\end{minipage}
	
	\figureCaptionMargin\vspace{-2ex}
	\caption{The neighbor composition of HNSW and Vamana.}
	\label{fig:composition}
	\figureBelowMargin
\end{figure}

\subsection{Motivation}
\label{subsec:motivation}

\textit{The primary reason why existing methods are incompetent in merging graph-based indexes is that they do not fully utilize the neighborhood information from the existing indexes.} Intuitively, connected vertices in the original indexes are likely to remain neighbors in the merged index. Even when certain neighbors cannot be retained as direct connections in the new graph due to outdegree constraints, where they are replaced by newly discovered closer vertices, they still provide valuable proximity information. This aligns with the fundamental principle in graph-based ANNS that \textit{a neighbor of a neighbor may also be a neighbor} \cite{dong2011efficient,li2019approximate,wang2021comprehensive}, which can be exploited to improve the merging process.

In Figure~\ref{fig:composition}, we compare the neighbor composition of small indexes to be merged with that of the large index constructed on the full dataset. Three real-world datasets (i.e., Sift \cite{jegou2010product}, Gist~\cite{jegou2010product} and Deep \cite{babenko2016efficient}) are randomly and equally partitioned into $h$ ($h = 2\ldots 5$) subsets, on each of which we build a graph index $G_1, G_2, \ldots, G_h$ using method $M$ (i.e., HNSW \cite{malkov2018efficient} and Vamana \cite{jayaram2019diskann}). Subsequently, we construct a graph index $G'$ on the full dataset using the same indexing method $M$ with identical parameters. For each vertex $v$ in index $G_i$, we identify the corresponding vertex $v'$ in $G'$ such that $\vec{x}_v = \vec{x}_{v'}$. We then compare the differences in neighbor sets $N_{G_i}(v)$ and $N_{G'}(v')$, where the proportion of shared neighbors is denoted as $\pi_{uc}$, newly discovered intra-subgraph neighbors as $\pi_{nl}$, and newly discovered cross-subgraph neighbors as $\pi_{nc}$. We aggregate these values across all vertices to obtain the final results. Our observations reveal that a consistent portion of neighbors remains unchanged, with HNSW demonstrating higher $\pi_{uc}$ (e.g., when using only two subgraphs, $\pi_{\text{uc}}$ reaches 55.3\%, 54.2\%, and 49.6\%, respectively). Moreover, among the newly added connections, cross-subgraph links consistently account for a larger share than intra-subgraph links, i.e., $\pi_{\text{nc}} > \pi_{\text{nl}}$. \textit{These findings suggest that our method should aim to preserve original neighbor relationships while also promoting the establishment of effective cross-subgraph connections.}

\subsection{Why Using $k$-NNG?}
\label{subsec:why_knng}

Another key challenge in merging multiple graph-based indexes arises from \textit{structural heterogeneity}, which stems from variations in the number of neighbors each vertex maintains across different indexes. These discrepancies originate from the distinct \textit{construction strategies} and \textit{pruning techniques} employed by various indexing methods \cite{wang2021comprehensive,li2019approximate}. For instance, HNSW \cite{malkov2018efficient} creates long edges to facilitate efficient routing, Vamana \cite{jayaram2019diskann} preserves medium-to-long edges during pruning, and $\tau$-MNG \cite{peng2023efficient} retains certain short-edge connections. These strategies result in intrinsic differences in graph structure and connectivity, making merging challenging: directly combining graphs can lead to imbalanced connectivity, disconnected components, or suboptimal routing. As discussed in \textsection\ref{subsubsec:graphBasedIndex}, mainstream graph construction methods typically follow a two-stage paradigm~\cite{wang2021comprehensive}: \textit{Candidate Neighbor Acquisition}, which builds a $k$-NNG, followed by \textit{Neighbor Selection}, which forms the final PG. This naturally raises the question: \textit{why not adopt a similar pattern for merging existing PG indexes?} Specifically, we convert each input PG into a unified $k$-NNG, ensuring uniform vertex degrees and consistent connectivity across graphs. This transformation facilitates efficient and effective graph merging by (i) providing a standardized intermediate representation that preserves the core local neighborhood structures of the original PGs, (ii) recovering informative edges that may have been pruned during individual PG constructions (i.e., from $k$-NNG to PG) \cite{yang2024revisiting}, as even edges omitted in all local indexes may still benefit the global index. By converting the PG back into a $k$-NNG, we no longer need to know the specific structural properties of the original graph-based index, thus mitigating heterogeneity across different indexes.

In summary, our approach preserves the existing neighborhood relationships from existing indexes and leverages this information to establish effective cross-graph connections, thereby pre-constructing a $k$-NNG that encodes the original graph structural information as the initial solution for merging.

\begin{figure*}[t!]
	\centering
	\includegraphics[width=1.0\linewidth]{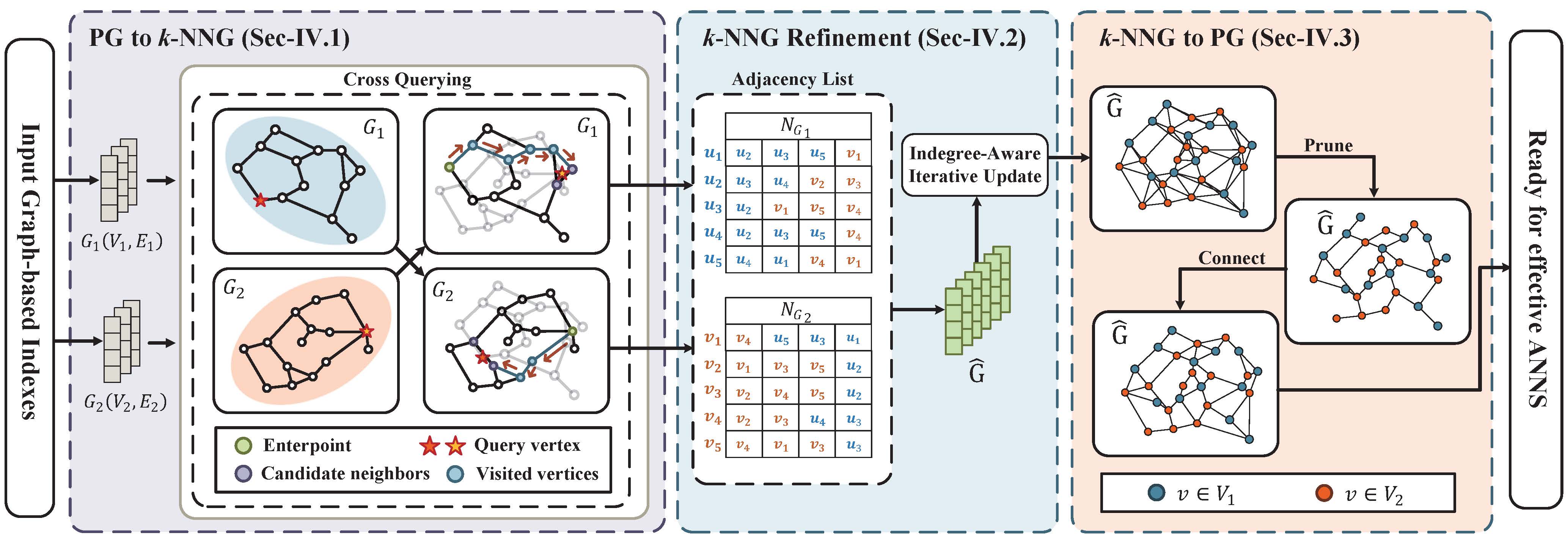}
	\figureBelowMargin\vspace{-2ex}
	\caption{The workflow of our FGIM framework with two graph-based indexes.}
	\label{fig:framework}
\end{figure*}

\subsection{Pipeline of the Framework}
\label{subsec:framework}

Figure~\ref{fig:framework} illustrates the FGIM pipeline, which consists of three steps: \textit{PGs to $k$-NNG}, \textit{$k$-NNG Refinement}, and \textit{$k$-NNG to PG}. The \textit{PGs to $k$-NNG} step plays a key role by generating \textit{Candidate Neighbor Acquisition} results from existing graph-based indexes, which are then used as input to the \textit{$k$-NNG Refinement} stage to obtain more accurate candidate neighbors. Finally, the \textit{$k$-NNG to PG} step constructs the final merged graph-based index, improving both search performance and connectivity.

\begin{itemize}[leftmargin=1em,itemsep=0pt]
	\item \textit{PGs to $k$-NNG transformation} (\textsection\ref{sec:pg2knng}): The transformation from PGs to $k$-NNG consists of two techniques: \textit{cross-querying} and \textit{minimum querying strategy}. Given a set of PG $\mathcal{G} = \{G_1, G_2, \ldots, G_h\}$, it first generates a candidate neighbor set $C_{i}(u)$ for each vertex $u$ in $G_i$ by efficiently extracting neighborhood relationships from the existing graph-based indexes. Then, the top-$k$ nearest neighbors from $C_{i}(u)$ are selected to form the \textit{Candidate Neighbor Acquisition} results, with each vertex obtaining $k$ candidate neighbors. From a macroscopic perspective (i.e., the entire graph), this process can be viewed as a transformation from the existing PGs to an interconnected $k$-NNG, which serves as a bridge between the original graph-based indexes and the final merged index.
	\item \textit{$k$-NNG Refinement} (\textsection\ref{sec:knngrefinement}): $k$-NNG refinement aims to refine the candidate neighbor set by identifying high-quality neighbors that may have been previously overlooked during the \textit{PGs to $k$-NNG transformation} while eliminating redundant connections to ensure that $k$-NNG maintains high accuracy. Specifically, we apply an improved refinement method to discover closer neighbors, which iteratively updates the neighbors of each vertex $u$ based on the current neighbor set $N_G(u)$. Moreover, to maintain graph connectivity, an indegree repair mechanism is integrated into the refinement process, enhancing the overall reachability.
	\item \textit{$k$-NNG to PG transformation} (\textsection\ref{sec:knng2pg}): Finally, we take above refined $k$-NNG as input to optimize the merged graph-based index with a \textit{Neighbor Selection} process. Specifically, candidate neighbors are carefully selected to further improve the search performance while minimizing the negative impact on graph navigability, thus facilitating the graph-based ANNS. It is worth mentioning that different pruning strategies (e.g., RNG \cite{toussaint1980relative}, MRNG \cite{fu2017fast}, and $\tau$-MNG \cite{peng2023efficient}) can be applied to our method, depending on the specific requirements for the merged graph-based index.
\end{itemize} 

\section{Components of the FGIM Framework}
\label{sec:implementation}

\subsection{PGs to $k$-NNG Transformation}
\label{sec:pg2knng}

In this subsection, we present the \textit{PGs to $k$-NNG transformation} process, which constructs a preliminary merged graph structure in the form of a $k$-NNG from the existing graph-based indexes.

\subsubsection{Candidate Neighbor Acquisition.} We first identify the candidate neighbors from the existing graph-based indexes. To be formal, we denote the candidate neighbor set of vertex $u$ in $G_i$ as $C_{i}(u)$. Since we have a set of graphs $\{G_1, G_2, \ldots, G_h\}$ with $n_1, n_2, \ldots, n_h$ vertices, two types of candidate neighbors can be obtained: the \textit{local candidate neighbors} $C_{i}^{+}(u)$ from the original graph $G_i$ and the \textit{cross candidate neighbors} $C_{i}^{-}(u)$ from the other graphs $G_j$ ($j \ne i$).

\begin{figure*}[t]
	\centering
	\includegraphics[width=1.0\linewidth]{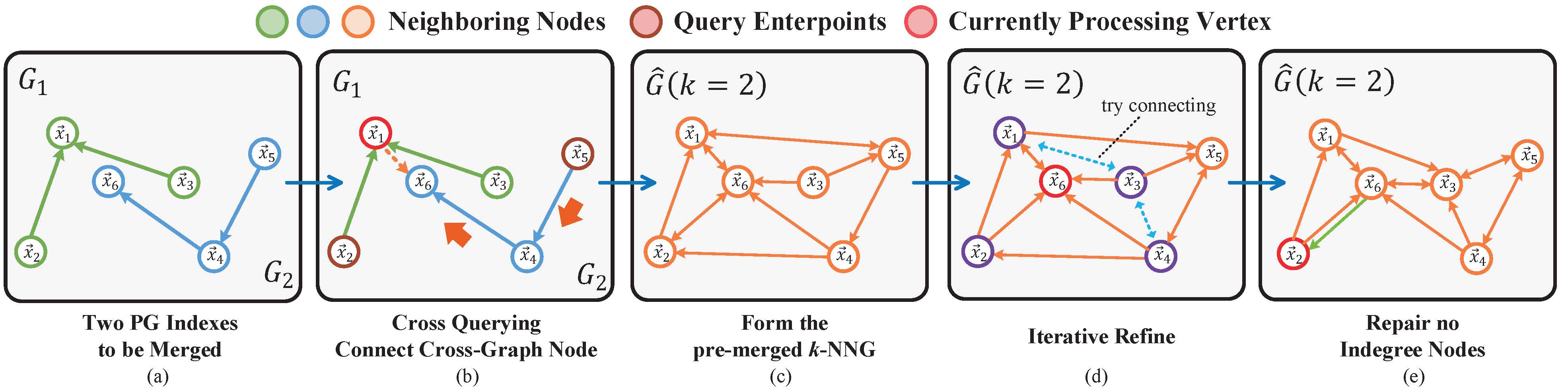}
	\vspace{-3ex}
	\figureBelowMargin
	\caption{An example of \textit{PGs to $k$-NNG transformation} and \textit{$k$-NNG refinement}.}
	\vspace{-2ex}
	\label{fig:pg2knng}
\end{figure*}

\begin{definition}[Local Candidate Neighbors] Given a vertex $u$ in $G_i$ in a set of graph-based indexes $\{G_1, G_2, \ldots, G_h\}$, the local candidate neighbors can be obtained from the original graph $G_i$:
	\begin{equation}
		C_{i}^{+}(u) = \{v \in V_i \mid (u, v) \in E_i\}
	\end{equation}
	where $V_i$ and $E_i$ are the vertex set and edge set of $G_i$, respectively.
\end{definition}

To obtain local candidate neighbors, we retrieve the neighbors of vertex $u$ directly from $G_i$. For cross candidate neighbors, we adopt a simple yet effective approach: treat each vertex $u$ in $G_i$ as a query point $\vec{x}_u$ and perform Algorithm~\ref{alg:anns} in the other indexes $\{G_1, G_2, \ldots, G_h\} \setminus G_i$. This \textit{cross-querying} process yields candidate neighbors across different graph-based indexes.

\begin{definition}[Cross Candidate Neighbors] Given a vertex $u$ in $G_i$ in a set of graph-based indexes $\{G_1, G_2, \ldots, G_h\}$ with enterpoints $\{ep_1, ep_2, \ldots, ep_h\}$, and a search pool size $L$ (i.e., the \textit{beam width} used in beam search), the cross candidate neighbors can be obtained by querying the other indexes $\{G_1, G_2, \ldots, G_h\} \setminus G_i$:
	\begin{equation}
		C_{i}^{-}(u) = \bigcup_{j \ne i} \textit{KNNSearch}(\vec{x}_u, G_j, L, L, ep_j)
	\end{equation}
\end{definition}

The \textit{cross-querying} technique acquires cross candidate neighbors from other PGs. In Figure~\ref{fig:pg2knng}, starting from $\vec{x}_1$ in $G_1$, we select $\vec{x}_5$ as the entry point in $G_2$ and follow the search path to obtain $\vec{x}_6$ as a candidate neighbor, creating a cross-graph connection between $\vec{x}_1$ and $\vec{x}_6$. This process is repeated for all vertices in $G_1$ and $G_2$, producing a set of cross-graph candidate neighbors for each vertex.

\subsubsection{Minimum Querying Strategy.}\label{subsubsec:min_querying} However, searching for cross-candidate neighbors in other indexes can be computationally expensive.  For example, the query time complexity of SOTA methods such as HNSW and NSG is $O(d m n^{\frac{2}{d}}\log n)$ with a probability of at least $1 - (1/e)^{\frac{d}{4}(1-\frac{3}{e^2})}$ \cite{wang2021comprehensive,fu2017fast,peng2023efficient}, where $d$ is the cost of a single distance computation in a $d$-dimensional space, $m$ is the number of neighbors evaluated at each vertex (i.e., the graph's outdegree constraint), and $n^{\frac{2}{d}}\log n$ represents the search path length (i.e., the number of hops). Given $h$ indexes, since vertices do not need to query their own index, the total cost of the \textit{cross-querying} technique is $O\big(dm \sum_{i=1}^{h} n_i \sum_{j \neq i} n_{j}^{\frac{2}{d}} \log n_j\big)$. Under typical conditions in the industry, where each index has roughly the same size ($n_i \approx n'$), this introduces a substantial $h^2$ term, as each index queries all others, yielding $h \times (h-1) \approx h^2$ pairwise interactions. To mitigate this, we propose a compensatory strategy: rather than retrieving precise neighbors at this stage, we quickly establish a set of inexpensive yet informative cross-graph connections. These serve as low-quality but \textit{pivotal links} that provide essential cross-graph information, enabling efficient updates in subsequent iterative refinement.

The \textit{Minimum Querying Strategy} ensures that each vertex $u$ has at least $k$ candidate neighbors in the initialized set. The minimum required search pool size $L$ used in the \textit{cross-querying} process must satisfy: $L \geq \left\lceil \frac{k - \min_i |C_{i}^{+}(u)|}{|\mathcal{G}| - 1} \right\rceil$. In the worst case (i.e., vertex $u$ has no outgoing neighbors in its own index), this simplifies to $L \geq \lceil \frac{k}{|\mathcal{G}| - 1} \rceil$.

\begin{algorithm}[t]\small
	\DontPrintSemicolon
	\KwIn{a set of graph-based indexes $\mathcal{G} = \{G_1, G_2, \ldots, G_n\}$, outdegree bound $k$, ID mapping function $\phi(\cdot)$}
	\KwOut{$k$-NNG $G$}
	
	Initialize $G \gets \emptyset$, candidate set $C \gets \emptyset$\;
	$L \gets \tfrac{k}{|\mathcal{G}| - 1}$\;
	\ForEach{$G_i \in \mathcal{G}$}{
		\ForEach{$u \in V_i$}{
			$C \gets \{v \in V_i \mid (u, v) \in E_i\}$\;
			\ForEach{$G_j \in \mathcal{G} \setminus \{G_i\}$}{
				$C \gets C \cup \textit{KNNSearch}(\vec{x}_u, G_j, L, L, ep_j)$\;
			}
			sort $C$ in ascending order of the distance to $\vec{x}_u$\;
			map each $v \in C$ to global ID using $\phi(\cdot)$\;
			$N_G(\phi(u)) \gets \textit{resize}(C, k)$\;
		}
	}
	\Return{$G$}\;
	\caption{PGs to $k$-NNG Transformation}
	\label{algo:pg2knng}
\end{algorithm}

Previous studies \cite{yang2024revisiting} have shown that a larger search pool size $L$ improves $k$-NNG accuracy, leading to better PG quality and ANNS performance. However, our approach uses minimal search parameters, making the routing process more susceptible to local optima \cite{prokhorenkova2020graph}. This will reduce $k$-NNG accuracy and degrade the merged PG quality. To address this, we optimize the $k$-NNG accuracy via a refinement method to ensure a high-quality merged index (\textsection\ref{sec:knngrefinement}). Once the $k$-NNG is sufficiently accurate, the final PG is generated through \textit{Neighbor Selection} (\textsection\ref{sec:knng2pg}), completing the merging process.

\subsubsection{Mapping onto $k$-NNG}\label{subsubsec:mapping} We obtain the candidate neighbor set of vertex $u$ in $G_i$ by combining local and cross neighbors: $C_i(u) = C_i^{+}(u) \cup C_i^{-}(u)$. Given a parameter $k$ (the maximum outdegree of the merged graph), candidates exceeding $k$ are ranked by distance to $u$, and only the top-$k$ are retained. Analogous to $M$ in HNSW and $R$ in Vamana, $k$ controls the trade-off between accuracy and efficiency and can be set to match the outdegree of the original graphs.

Moreover, each sub-index maintains a local identifier space, typically assigning vertex IDs starting from zero. To avoid identifier conflicts during merging, these local spaces must be aligned. We define an \emph{ID mapping function} $\phi(\cdot)$ that maps each valid local vertex ID in a sub-index to a unique global ID in the merged index. For each local vertex identifier $u_{\text{src}}$, the function is defined as:
\[
\phi(u_{\text{src}}) \rightarrow 
\begin{cases}
	(\texttt{true}, u_{\text{dst}}), & \text{if } u_{\text{src}} \text{ is retained in the merged index},\\[4pt]
	(\texttt{false}, \text{--}), & \text{otherwise,}
\end{cases}
\]
where the boolean flag indicates whether the source vertex is considered valid in the context of the merged index. A value of $\texttt{false}$ denotes an invalid or obsolete vertex that should be excluded, while $\texttt{true}$ associates $u_{\text{src}}$ with a unique global identifier $u_{\text{dst}}$ in the destination index. In practice, $\phi(\cdot)$ assigns disjoint global ID ranges to different sub-indexes by applying appropriate offsets to their local identifier spaces, while filtering out invalid vertices, thereby guaranteeing global uniqueness and conflict-free merging. Since $\phi(\cdot)$ is user-defined, it provides flexibility in ID management to accommodate system-specific requirements. For example, in systems \cite{chen2024singlestore,niu2025blendhouse,wang2021milvus} based on Log-Structured Merge (LSM) trees \cite{o1996log}, vertex validity can be determined by record existence or associated metadata such as timestamps.

Algorithm~\ref{algo:pg2knng} outlines the \textit{PG-to-$k$NNG transformation}. The algorithm first initializes and determines the search candidate size $L$ using the \textit{minimum querying strategy} (Lines 1-2). For each vertex $u$ in each PG $G_i$, it retrieves \textit{local} candidates from $G_i$ and \textit{cross} candidates from the other PGs (Lines 3-7). Finally, the top-$k$ neighbors are selected and assigned to $N_G(\phi(u))$ in the $k$-NNG (Lines 8-10).

\subsection{$k$-NNG Refinement}
\label{sec:knngrefinement}

After \textit{PGs to $k$-NNG transformation}, the candidate neighbors are retrieved from the existing graph-based indexes. However, the proposed \textit{minimum querying strategy} presents a critical challenge: the strategy may cause the \textit{cross-querying} process to become trapped in a local optimum \cite{prokhorenkova2020graph}, resulting in low-quality cross candidate neighbors in an inaccurate $k$-NNG. To address this issue, we propose an \textit{{Indegree-Aware $k$-NNG Refinement}} to improve the quality of the merged graph by iteratively refining the obtained $k$-NNG with \textit{streamlined refinement} process while improving the graph connectivity via an \textit{indegree repair mechanism}.

\subsubsection{Streamlined Refinement.}
\label{subsubsec:streamlined_refinement}

Since we have already established a sufficient number of cross-graph connections in \S\ref{sec:pg2knng}, these pivotal vertices can be leveraged to introduce more precise cross-graph links in the $k$-NNG, which is achieved through the refinement method of the $k$-NNG \cite{dong2011efficient}. This refinement operates by iteratively performing two batched operations, \textit{Visit} and \textit{Update}, over the entire graph to progressively improve neighbor quality.

Specifically, the \textit{Visit} procedure organizes the neighbors and reverse neighbors of each vertex into candidate sets of newly discovered and previously processed vertices, while maintaining corresponding reverse graphs. A flag-based mechanism is employed to distinguish \texttt{new} and \texttt{old} neighbors, reducing redundant distance computations without affecting the final $k$-NNG structure. Based on this classification, vertices are inserted into the candidate sets $C_n$ and $C_o$, and reverse links are recorded accordingly.

Given the candidate sets produced by \textit{Visit}, the \textit{Update} procedure examines all valid vertex pairs drawn from $C_n \times C_n$ and $C_n \times C_o$, where the two vertices originate from different graph indexes. For each such pair, the pairwise distance is computed and compared against the current farthest neighbors in their respective $k$-NN lists. If the new distance is smaller, the farthest neighbor is replaced, thereby establishing a direct and more accurate cross-graph connection. By iteratively applying this mutual update process, neighboring vertices gradually recognize each other, resulting in a $k$-NNG of higher quality.

\begin{algorithm}[t]\small
	\DontPrintSemicolon
	\KwIn{$k$-NNG $G$, maximum iteration $I_{max}$}
	\KwOut{refined $k$-NNG $G$}
	
	Initialize the reverse graph $\overline{G}_{\texttt{n}}$, $\overline{G}_{\texttt{o}}$ $\gets$ $\emptyset$\;
	
	\Repeat{$iter = I_{max}$}{
		$\mathcal{V}, \mathcal{S}$ $\gets$ $\emptyset$\;
		\ForEach{$u \in V$, $u \notin \mathcal{V}$}{
			push $u$ into $\mathcal{S}$ and $\mathcal{V}$\;
			\While{$\mathcal{S} \ne \emptyset$}{
				$v$ $\gets$ pop from $\mathcal{S}$\;
				$C_{n}$, $C_{o}$ $\gets$ perform \textit{Visit} for $v$\;
				push $\forall w \in N(v)$ into $\mathcal{S}$ and $\mathcal{V}$ if $w \notin \mathcal{V}$\;
				move $\forall v_1 \in \overline{G}_{\texttt{n}}[v]$ into $C_{n}$, $\forall v_2 \in \overline{G}_{\texttt{o}}[v]$ into $C_{o}$\;
				perform \textit{Update} on all $(v_1, v_2) \in C_{n}^2$ $\cup$ $(C_{n} \times C_{o})$ s.t. $v_1 \ne v_2$, $v_1$, $v_2$ from different $G_i$\;
			}
		}
		perform \textit{Indegree Repair} for $G$\;
		$iter$ $\gets$ $iter + 1$\;
	}
	\Return{$G$}\;
	
	\caption{$k$-NNG Refinement}
	\label{alg:refine}
\end{algorithm}

Then, we first introduce a \textit{cache-friendly merging strategy} based on a DFS-style traversal. In conventional approaches~\cite{dong2011efficient}, vertices are accessed in ascending ID order, causing random memory access and poor cache efficiency. To address this, we adopt a DFS-style access pattern, where after processing a vertex, we immediately visit its neighbors to perform the \textit{Update} operation. Following the ANNS principle that “a neighbor of a neighbor may also be a neighbor”~\cite{wang2021comprehensive}, this leverages spatial locality, so data already in cache (for memory indexes) or in memory (for SSD-based indexes) can be reused, thereby reducing access overhead and speeding up merging.\label{rev:cachefriendly}

Additionally, we propose a \textit{streamlined refinement} approach that combines \textit{Visit} and \textit{Update} into a single pass per vertex. This eliminates the need for separate passes for visiting neighbors and updating, as required in conventional methods \cite{dong2011efficient}. By integrating both operations within one traversal, we reduce overhead, access the latest graph snapshot, and continuously improve graph quality, ultimately enhancing the efficiency of the refinement process.

\begin{algorithm}[t]\small
	\DontPrintSemicolon
	\KwIn{graph $G$}
	\KwOut{repaired $G$, indegree recorder $\mathcal{T}$ of $G$}
	
	$\mathcal{T}$ $\gets$ calculate indegrees for $\forall u \in V$\;
	\ForEach{$\forall v^* \in V$ where $\mathcal{T}(v^*) = 0$}{
		\Repeat{$\mathcal{T}(v^*) > 0$ or all neighbors in $N_G(v^*)$ are visited}{
			$v$ $\gets$ the closest unvisited neighbor in $N_G(v^*)$\;
			\ForEach{$\forall v' \in N(v)$ in descending order of the distance to $\vec{x}_v$}{
				\If{$v'$ is not \texttt{replaced} and $\mathcal{T}(v') > 1$}{
					Replace $v'$ with $v^*$ in $N_G(v)$\;
					Mark $v^*$ in $N_G(v)$ as \texttt{replaced}\;
					$\mathcal{T}(v')$ $\gets$ $\mathcal{T}(v') - 1$\;
					$\mathcal{T}(v^*)$ $\gets$ $\mathcal{T}(v^*) + 1$\;
					break\;
				}
			}
		}
	}
	\Return{$G$, $\mathcal{T}$}\;
	
	\caption{Indegree Repair Heuristic}
	\label{alg:repair}
\end{algorithm}

With the two optimizations, the refinement process converges quickly, as confirmed in our experiments (\textsection\ref{subsec:ablation}). Given a $k$-NNG $G$, we initialize two auxiliary reverse graphs, $\overline{G}_{\texttt{n}}$ and $\overline{G}_{\texttt{o}}$, to track neighbors labeled as \textit{new} and \textit{old}, respectively (Algorithm~\ref{alg:refine} Line 1). The refinement proceeds for up to $I_{\text{max}}$ iterations. In each iteration, vertices are traversed in DFS order using a stack $\mathcal{S}$ and a visited set $\mathcal{V}$ (Lines 3-6). For each vertex $v$ popped from the stack, the \textit{Visit} operation accesses its neighbors $N_G(v)$ and classifies them into $C_n$ (new) and $C_o$ (old) (Lines 7-8). Reverse neighbors recorded in $\overline{G}_{\texttt{n}}$ and $\overline{G}_{\texttt{o}}$ are merged into these sets (Line 10). We then perform the \textit{Update} operation on all valid \textit{cross-graph} vertex pairs in $C_n^2 \cup (C_n \times C_o)$ (Line 11), ensuring pairs come from different subgraphs.

\subsubsection{Indegree Repair Mechanism.}

$k$-NNG does not inherently guarantee full connectivity or reachability among all vertices \cite{navarro2002searching}. For instance, a vertex with no indegree cannot be accessed during the query process, rendering it unreachable. Moreover, such vertices are unlikely to participate in the iterative process, which can degrade the overall graph quality. To address this issue, we propose a \textit{connectivity enhancement heuristic} that links such disconnected vertices, denoted as $v^*$, to the current graph $G$. Specifically, we try to connect each vertex $v^*$ to its nearest neighbor $v \in N_G(v^*)$. To ensure that other vertices with low indegrees are not adversely affected, we traverse the neighbors of $v$ to identify replaceable vertices that have not yet been replaced (i.e., not marked as \textit{replaced}) and have an indegree greater than 1 (Line 6 in Algorithm~\ref{alg:repair}). If $v'$ in $N_G(v)$ meets these conditions, we replace $v'$ with $v^*$ in $N_G(v)$, and mark $v^*$ as \textit{replaced} in $N_G(v)$ (Lines 7-11). Otherwise, we continue to visit the next closest neighbor $v \in N_G(v^*)$, following the ascending order of distance to $\vec{x}_{v^*}$, until no neighbors are left. Importantly, these supplementary edges maintain the outdegree constraint while significantly reducing the number of connected components, thus improving reachability to disconnected vertices for query routing.

\subsection{$k$-NNG to PG Transformation}
\label{sec:knng2pg}

Despite the refined $k$-NNG being already of high quality, it still falls short of performing ANNS tasks with both precision and efficiency because of weak navigability and longer search paths~\cite{li2019approximate, wang2021comprehensive}, compared to the SOTA graph-based ANNS methods~\cite{malkov2018efficient,fu2017fast,jayaram2019diskann}. To bridge this gap, we propose a process to transform $k$-NNG back into PG, which consists of two main steps: 1) \textit{neighbor selection}, 2) \textit{connectivity enhancement}, and 3) \textit{HNSW-adaptive merging}. Algorithm \ref{alg:prune} shows the process of the \textit{$k$-NNG to PG transformation}.

\subsubsection{Neighbor Selection.}

Given a $k$-NNG $G$, we perform \textit{neighbor selection} for each vertex $u \in V$. For each $u$, neighbors in $N_G(u)$ are sorted by distance to $\vec{x}_u$, and up to $k$ neighbors are selected (Lines 2–8). A neighbor is kept if it is closer to $u$ than to any previously selected neighbor (Lines 4–6), ensuring local diversity. This criterion is effective under common distance metrics (e.g., Euclidean, cosine) and can be extended to MIPS using standard $\ell_2$ transformations \cite{chen2025maximum}. To maintain reachability, we also ensure that pruning does not remove the only incoming edge of any vertex (Line 5).

This selection procedure removes distant neighbors that have closer alternatives in the graph space, establishing a sparser graph that enhances efficiency in ANNS tasks. Furthermore, the framework supports the integration of alternative pruning strategies (e.g., $\alpha$-RNG~\cite{jayaram2019diskann} and $\tau$-MNG~\cite{peng2023efficient}) by reimplementing the \textit{neighbor selection} process, allowing for flexibility in the choice of methods.

\begin{algorithm}[t]\small
	\DontPrintSemicolon
	\KwIn{$k$-NNG $G$, indegree recorder $\mathcal{T}$, outdegree bound $k$}
	\KwOut{merged graph index $\hat{G}$}
	
	Initialize the graph $\hat{G}(V', E')$ $\gets$ $\emptyset$\;
	
	\ForEach{$\forall u \in V$}{
		$\mathcal{R}$ $\gets$ $\emptyset$\;
		\ForEach{$\forall v$ $\in$ $N_G(u)$ in ascending order of $\delta(u, v)$}{
			\If{$\mathcal{T}(v)=1$ or $\forall w \in \mathcal{R}$, $\delta(u, v) < \delta(w, v)$}{
				$\mathcal{R}$ $\gets$ $\mathcal{R}$ $\cup$ $\{v\}$ \;
			}
			break if $\lvert \mathcal{R} \rvert = k$\;
		}
		$N_{\hat{G}}(u)$ $\gets$ $\mathcal{R}$\;
	}
	\ForEach{$\forall u \in V'$}{
		$N_{\hat{G}}(u)$ $\gets$ $N_{\hat{G}}(u)$ $\cup$ $\{v \in V' \mid (v, u) \in E'\}$\;
		sort $N_{\hat{G}}(u)$ in ascending order of the distance to $\vec{x}_u$\;
		resize $N_{\hat{G}}(u)$ to $k$\;
	}
	\Return{$\hat{G}$}\;
	
	\caption{$k$-NNG to PG Transformation}
	\label{alg:prune}
\end{algorithm}

\subsubsection{Connectivity Enhancement.}

Although limiting vertex outdegree simplifies graph management, it can weaken global connectivity. To reduce strongly connected components, we perform \textit{connectivity enhancement} on the pruned graph $\hat{G}$. Following the principle that \textit{one tends to regard as important those who regard oneself as important} \cite{ootomo2024cagra}, we add reverse neighbors to each vertex’s neighbor set $N_{\hat{G}}(u)$, then sort $N_{\hat{G}}(u)$ by distance to $\vec{x}_u$ (Lines 9–11). Finally, we retain only the top-$k$ neighbors to avoid hubness (Line 12). These added edges improve long-range routing and graph navigability \cite{prokhorenkova2020graph} while making better use of the degree budget.

\begin{figure}[t]
	\centering
	\vspace{-2ex}
	
	\begin{minipage}{0.65\textwidth}
		\centering
		\subfigure[][\scriptsize Sift1M]{
			\includegraphics[width=0.47\linewidth]{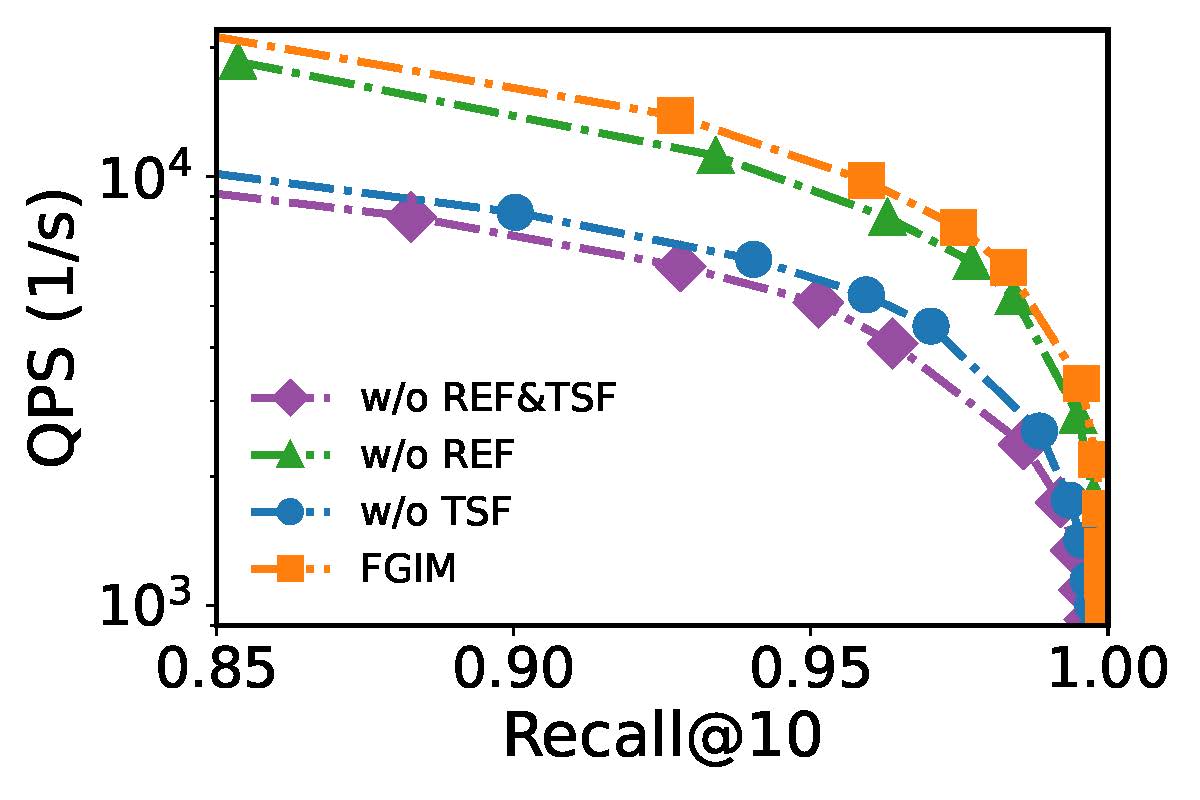}
			\label{fig:sift_effectiveness}
		}
		\hspace{-1em}
		\subfigure[][\scriptsize Gist1M]{
			\includegraphics[width=0.47\linewidth]{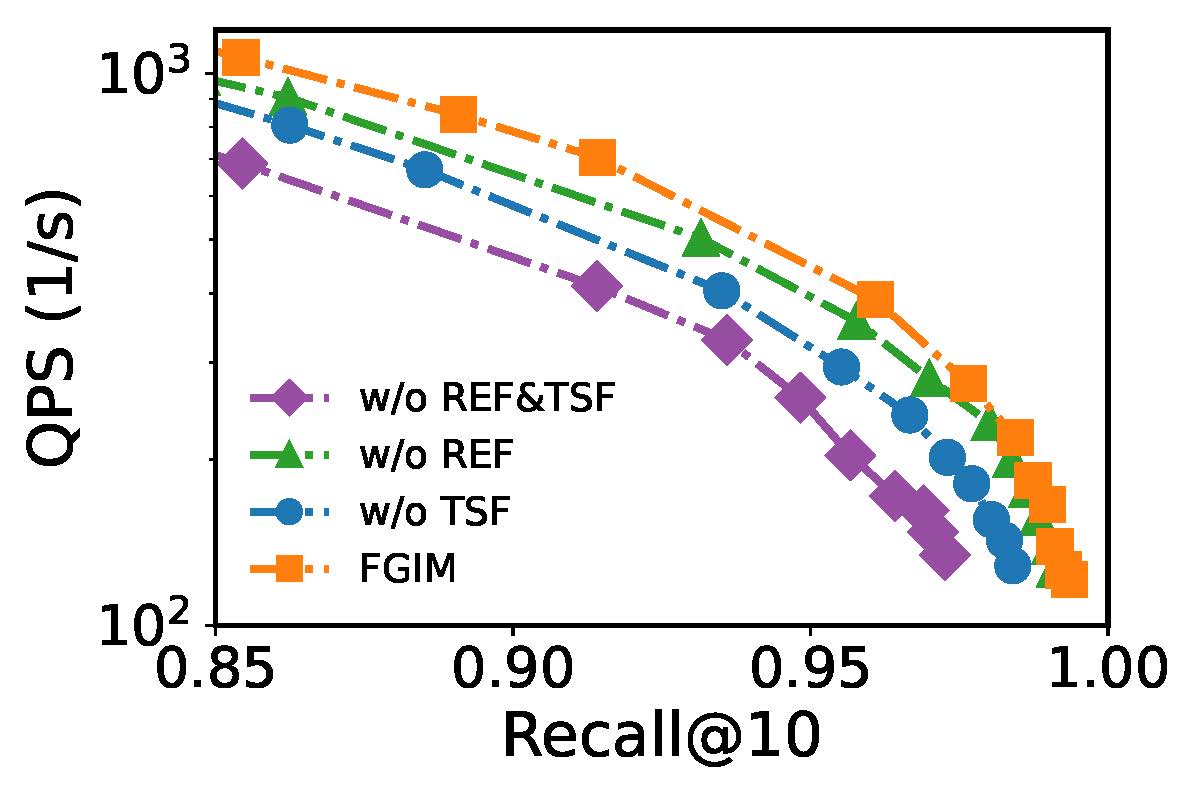}
			\label{fig:gist_effectiveness}
		}
	\end{minipage}
	
	\figureCaptionMargin
	\vspace{-1ex}
	\caption{Effectiveness of $k$-NNG refinement (REF, \textsection\ref{sec:knngrefinement}) and $k$-NNG to PG transformation (TSF, \textsection\ref{sec:knng2pg}).}
	\label{fig:effectiveness}
	\figureBelowMargin
\end{figure}

We validate the effectiveness of the proposed \textit{$k$-NNG refinement} (\textsection\ref{sec:knngrefinement}) and \textit{$k$-NNG to PG transformation} (\textsection\ref{sec:knng2pg}) on the Sift1M and Gist1M datasets. Figure~\ref{fig:effectiveness} demonstrates that both components effectively improve ANNS performance by enhancing graph quality.

\subsubsection{HNSW-Adaptive Merging}

Many mainstream graph-based ANNS methods typically construct a single-layer index, where we can directly apply our FGIM framework to merge the indexes. However, the widely used HNSW \cite{malkov2018efficient} method builds a multi-layer graph structure, where each layer contains vertices of varying density, thereby enabling superior search performance. Notably, like skip lists \cite{pugh1990skip}, the number of vertices in each layer of HNSW decreases exponentially with increasing layer depth, i.e., $P(level \ge k) \propto \exp(-k)$. In real-world applications, this indicates that the number of non-base layer vertices is significantly smaller than that of the base layer. Therefore, we tried a simplistic yet effective strategy that first merges the base layer of HNSW using the proposed FGIM framework and subsequently reconstructs the higher layers accordingly.

Specifically, we reconstruct the upper layers following the original HNSW design principles. Each vertex in the merged base graph is first assigned a random level according to the HNSW level distribution, and the hierarchical graph-based index is initialized with a number of layers equal to the maximum assigned level. For a vertex with level $l>1$, we perform greedy navigation to the closest vertex at level $l+1$, and then search for candidate neighbors from level $l$ down to level $1$ using Algorithm~\ref{alg:anns}. At each level, candidate neighbors are pruned using the RNG pruning heuristic~\cite{toussaint1980relative} introduced in HNSW~\cite{malkov2018efficient}, ensuring that the outdegree constraint is satisfied. Bidirectional edges are then added, and the same pruning strategy is applied to affected neighbors.

\subsection{Complexity Analysis}
\label{subsec:complexity}

In this part, we analyze the complexity of our FGIM framework.

\noindent\underline{\textit{PGs to $k$-NNG transformation.}} The complexity of this phase depends on the search complexity of the underlying graph-based index. For SOTA methods such as HNSW~\cite{malkov2018efficient} and NSG~\cite{fu2017fast}, the expected search complexity is $O(d m n^{\frac{2}{d}} \log n)$ with probability at least $1 - (1/e)^{\frac{d}{4}(1 - \frac{3}{e^2})}$ \cite{wang2021comprehensive,peng2023efficient}, where $d$ is the cost of distance computation, $m$ is the outdegree (i.e., the number of neighbors visited per vertex), and $n^{\frac{2}{d}} \log n$ corresponds to the search path length. For the merged graph, we set the outdegree constraint $k$ equal to that of the original graphs (i.e., $k = m$). Under our \textit{Minimum Querying Strategy}, only $m' = \lceil k / (h - 1) \rceil$ neighbors are visited at each hop, where $m' < m$. Given $h$ PGs $\mathcal{G} = \{G_1, G_2, \dots, G_h\}$ with $\{n_1, n_2, \dots, n_h\}$ vertices, the complexity of this phase is $O\big(\frac{kd}{h - 1} \sum_{i=1}^{h} n_i \sum_{j \neq i} n_{j}^{\frac{2}{d}} \log n_j \big)$. When all $n_i = n'$, this simplifies to $O(kdhn'^{\frac{2 + d}{d}} \log n')$, which provides a good approximation for the general case.

\noindent\underline{\textit{$k$-NNG refinement.}} This step involves at most $I_{\max}$ iterations, where each iteration refines the graph using the \textit{streamlined refinement} technique with complexity $O(I_{\max} \cdot n \cdot k^2 d)$. Additionally, the \textit{indegree repair mechanism} incurs a complexity of $O(I_{\max} \cdot n \cdot k)$, which is subsumed by the dominant term. Therefore, the total complexity of the \textit{$k$-NNG refinement} phase is $O(I_{\max} \cdot n \cdot k^2 d)$.

\noindent\underline{\textit{$k$-NNG to PG transformation.}} In this phase, the complexity of \textit{neighbor selection} is $O(n \cdot k^2 d)$, while for \textit{connectivity enhancement} is $O(n \cdot k)$. Hence, the total complexity of this phase is $O(n \cdot k^2 d)$. Combining both the \textit{$k$-NNG refinement} and the \textit{PG transformation} phases, the overall complexity becomes $O(I_{\max} \cdot n \cdot k^2 d)$.

\noindent\underline{\textit{HNSW-adaptive merging.}} The reconstruction of the upper-layer graph follows the original HNSW design, where elements are inserted sequentially via greedy search with $O(\log n)$ complexity per insertion \cite{malkov2018efficient, wang2021comprehensive}. The probability that a node is assigned to level $l$ or higher is given by $\mathbb{P}(\text{level}_i \geq l) = \int_0^{e^{-l \cdot \log(k)}} dx = k^{-l}$, which implies that the expected number of nodes in non-zero layers is $n/k$. As a result, the time complexity of this procedure is $O\left(\frac{n}{k} \log \frac{n}{k}\right)$. For space, the number of nodes at level $l$ is $n_l = n \cdot k^{-l}(1 - 1/k)$, and each maintains $O(k \cdot (l+1))$ neighbors. Summing across levels, the total space requirement becomes $\sum_{l=0}^\infty n \cdot k^{-l} \left(1 - \frac{1}{k} \right) \cdot (l + 1) \cdot k = O\left(n \cdot \frac{k^2}{k - 1}\right)=O(nk)$.

\noindent\underline{\textit{Total complexity.}} Summarizing all components, the time complexity of FGIM framework is: $$O\big(\frac{kd}{h - 1} \sum_{i=1}^{h} n_i \sum_{j \neq i} n_{j}^{\frac{2}{d}} \log n_j + I_{\max} \cdot n \cdot k^2d + \frac{n}{k} \log \frac{n}{k}\big).$$ The space complexity of the framework is $O(nk)$, accounting for the candidate neighbor set of each vertex in the merged graph, the two auxiliary reverse-neighbor graphs, the stack and visited set used for DFS traversal during refinement, and the upper-layer graphs. Since each vertex has at most $k$ positive neighbors, the reverse graphs require $O(nk)$ space.

\section{Experimental Study}
\label{sec:experimental}

In this section, we present experimental results of our FGIM framework on seven real-world datasets. Our evaluation seeks to answer the following research questions:

\begin{itemize}[leftmargin=0pt,label={}, topsep=0pt, itemsep=0pt]
	\item \textbf{RQ1:} How does FGIM compare to incremental methods in terms of merging efficiency and search performance? (\textsection\ref{subsec:efficiency})
	\item \textbf{RQ2:} How does FGIM perform in merging multiple indexes? (\textsection\ref{subsec:multiple})
	\item \textbf{RQ3:} How does FGIM compare to other graph-based indexes, and is it applicable in scenarios involving deletion and overlap? (\textsection\ref{subsec:applicability})
	\item \textbf{RQ4:} How do different strategies contribute to FGIM? (\textsection\ref{subsec:ablation})
	\item \textbf{RQ5:} How does FGIM scale w.r.t. data size and thread count? (\textsection\ref{subsec:scalability})
\end{itemize}

\begin{table}[t]
	\centering\small
	\caption{The properties of the datasets.}
	\vspace{-1ex}
	\begin{tabular}{l|l|l|l|l}
		\toprule
		\textbf{Dataset} & \textbf{Dim} & \#\textbf{Base} & \#\textbf{Query} & \textbf{Metric} \\
		\midrule
		Sift1M \cite{jegou2010product} & 128 & 1,000,000  & 10,000 & Euclidean \\
		Sift2M \cite{jegou2010product} & 128 & 2,000,000  & 10,000 & Euclidean \\
		Sift5M \cite{jegou2010product} & 128 & 5,000,000  & 10,000 & Euclidean \\
		Gist1M \cite{jegou2010product} & 960 & 1,000,000  & 1,000 & Euclidean \\
		Deep1M \cite{babenko2016efficient} & 96 & 1,000,000 & 10,000 & cosine \\
		GloVe \cite{pennington2014glove} & 100 & 1,183,513 & 10,000 & cosine \\
		MSong \cite{bertin2011million} & 420 & 994,185 & 1,000 & Euclidean  \\
		Crawl \cite{commoncrawl} & 300 & 1,989,995 & 10,000 & cosine \\
		Internet Search \cite{Alipay2025} & 768 & 9,991,307 & 172 & Euclidean \\
		\bottomrule
	\end{tabular}
	\label{tab:dataset}
\end{table}

\subsection{Experimental Settings} 
\label{subsec:settings}

\noindent\textbf{Datasets.} The experiments are conducted on several popular benchmarking datasets. All of them are real-world datasets and have been widely used in the literature \cite{wang2021comprehensive,aumuller2020ann}. The datasets cover various applications such as image (Sift \cite{jegou2010product}, Gist \cite{jegou2010product}, Deep \cite{babenko2016efficient}), text (Glove \cite{pennington2014glove}, Crawl \cite{commoncrawl}, Internet Search \cite{Alipay2025}), and audio (MSong \cite{bertin2011million}). The datasets' properties are summarized in Table \ref{tab:dataset}.

\noindent\textbf{Compared algorithms.} We evaluate 9 representative graph-based ANNS methods: (1) HNSW~\cite{malkov2018efficient}, a widely used hierarchical graph index; (2) Vamana~\cite{jayaram2019diskann}, which optimizes HNSW’s neighbor selection for improved search performance; (3) $\tau$-MNG~\cite{peng2023efficient}, which constructs a monotonic neighborhood graph from an existing index; (4) NSW \cite{malkov2014approximate}, a classical navigable small-world graph; (5) NNDescent \cite{dong2011efficient}, a representative $k$-NNG construction method; (6) NNMerge \cite{zhao2021merge}, a merging method for $k$-NNGs; (7) DiskANN \cite{jayaram2019diskann}, which introduces a merging strategy for large-scale datasets; (8) FreshDiskANN~\cite{singh2021freshdiskann}, a representative graph-based index designed for SSDs; and (9) Lucene \cite{xian2024vector}, which implements a merging strategy for HNSW in its vector search system.
We did not apply PQ~\cite{jegou2010product} or SQ~\cite{aguerrebere2023similarity}, as these are orthogonal techniques applicable to any index. All baseline methods use their original pruning strategies, while our FGIM employs the algorithm in Algorithm~\ref{alg:prune} for neighbor selection during merging.

\begin{table}[t]
	\small
	\centering
	\caption{Parameter variations for different algorithms.}
	\vspace{-1ex}
	\label{tab:parameter}
	\begin{tabular}{l|p{7.0cm}}
		\toprule
		\textbf{Algorithm} & \textbf{Parameter Settings} \\
		\midrule
		\textit{FGIM} &
		$\text{max\_degree} \in \{8, 12, 16, 20, 24, 28, 32, 36, 48, 64, 96, 128\}$ \\
		\midrule
		\textit{NSW}, \textit{HNSW} &
		\begin{tabular}[t]{@{}l@{}}
			$\text{maximum\_neighbors} \in \{4, 8, 12, 16, 20, 24, 36, 48, 64, 96\}$, \\
			$\text{ef\_construction} = 200$
		\end{tabular} \\
		\midrule
		\textit{Vamana}, \textit{FreshDiskANN} &
		\begin{tabular}[t]{@{}l@{}}
			$\text{R} \in \{ 8, 12, 16, 20, 24, 28, 32, 36, 40, 56, 64, 72, 80 \},\ \text{L} = 200$ \\
			$\alpha = 1.2$
		\end{tabular} \\
		\midrule
		\textit{$\tau$-MNG} &
		$\text{b} \in \{50, 100, 150, 200, 250, 300, 350, 400, 450, 500\},\ \text{h} = 200$ \\
		\midrule
		\textit{NNDescent} &
		$\text{k} \in \{8, 12, 16, 20, 24, 28, 32, 36, 48, 64, 96, 128\}$ \\
		\bottomrule
	\end{tabular}
\end{table}

\noindent\textbf{Parameters.} We follow the approach of benchmarking papers \cite{aumuller2020ann,li2019approximate,wang2021comprehensive} by employing grid search to determine the optimal parameter values for each index, ensuring that the algorithm achieves its best search performance. For building efficiency evaluations with parameter variations, we report the specific configurations in Table~\ref{tab:parameter}. For both the subgraphs to be merged and the merged graph, we consistently use the same construction parameters.

\noindent\textbf{Computing Environment.} We implemented our method in C++11 and compiled the code using CMake 3.30.2 with GCC 11.4.0 as the compiler. Almost all experiments were conducted on a machine equipped with an Intel Core i7-12700H CPU and 32 GB of RAM, running WSL2 Ubuntu 22.04 as the operating system. The only exception is that Exp.12 (\S\ref{subsec:scalability}) was conducted on 
a server equipped with an Intel Xeon Platinum 8163 CPU and 256 GB of RAM.

\noindent\textbf{Measurements.} 
To measure the accuracy of the search results, we use the $Recall$ metric defined in \textsection\ref{sec:problemDefinition}. Additionally, Queries Per Second (QPS) and Number of Distance Computation (NDC) are used to measure the search efficiency. We measure the search performance of each method by plotting the QPS vs. $Recall@10$ and $Recall@100$ curves while varying the search parameters. Besides, the building efficiency is also evaluated by plotting the building time (BT) vs. $Recall@10$ curves while varying the construction parameters.

\subsection{Efficiency Evaluation}
\label{subsec:efficiency}

\noindent\textbf{Exp.1 \& 2: Comparison with the Incremental Method in Merging Efficiency and Search Performance.} We evaluate the merge efficiency and search performance of our FGIM framework against the incremental construction methods of HNSW, NSW, and FreshDiskANN. Each dataset is evenly split into two subsets, and separate indexes are built on them. For the baselines, the index on one subset is incrementally expanded by inserting all data from the other. HNSW and NSW are constructed single-threaded in memory, while FreshDiskANN operates in a disk-based setting using 20 threads. In contrast, FGIM directly merges two pre-built indexes.

\begin{figure}[t]
	\figureTopMargin
	\centering
	\includegraphics[width=0.65\textwidth]{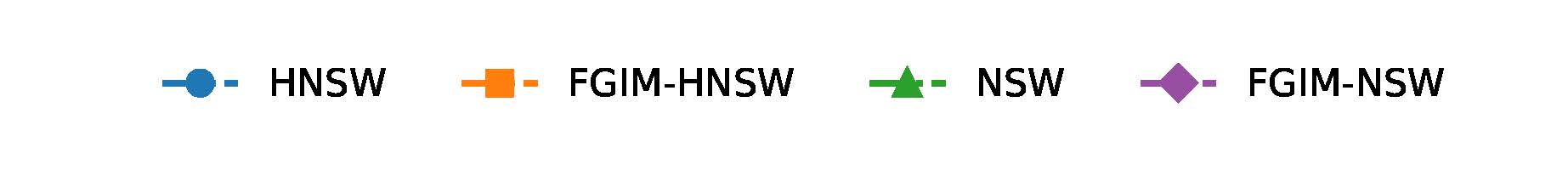}
	
	\vspace{-3ex}
	\subfigure[][{\scriptsize Sift1M}]{
		\hspace{-0.73em}
		\includegraphics[width=0.173\textwidth]{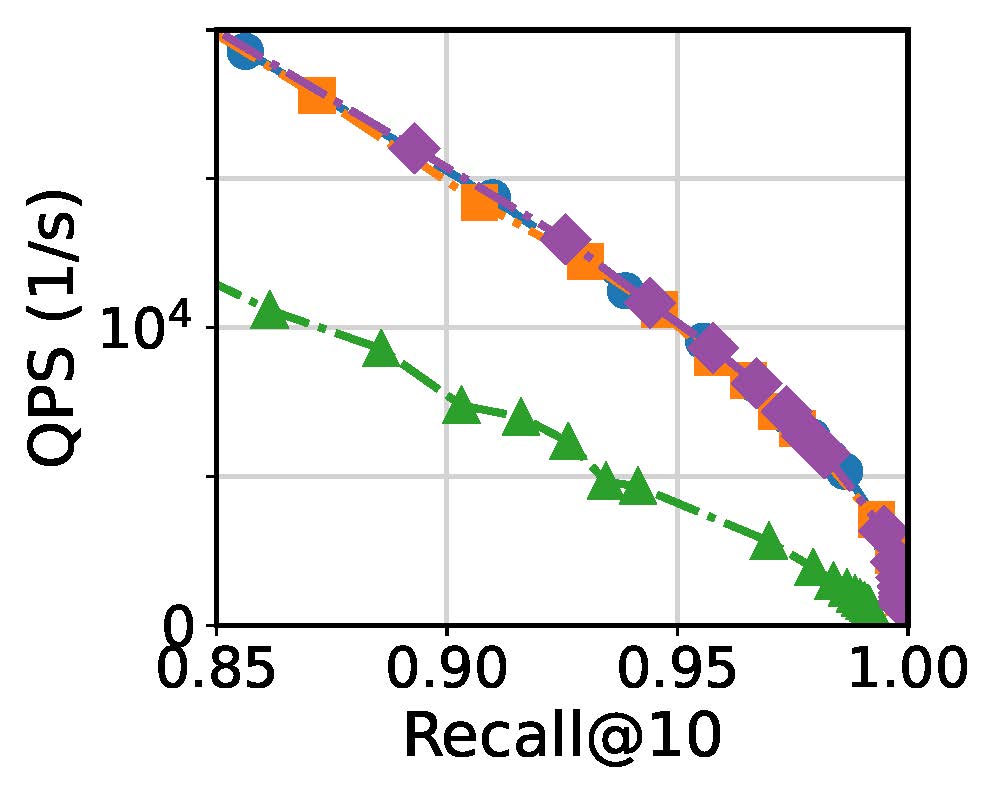}
		\hspace{-0.73em}
	}
	\subfigure[][{\scriptsize Deep1M}]{
		\hspace{-0.73em}
		\includegraphics[width=0.173\textwidth]{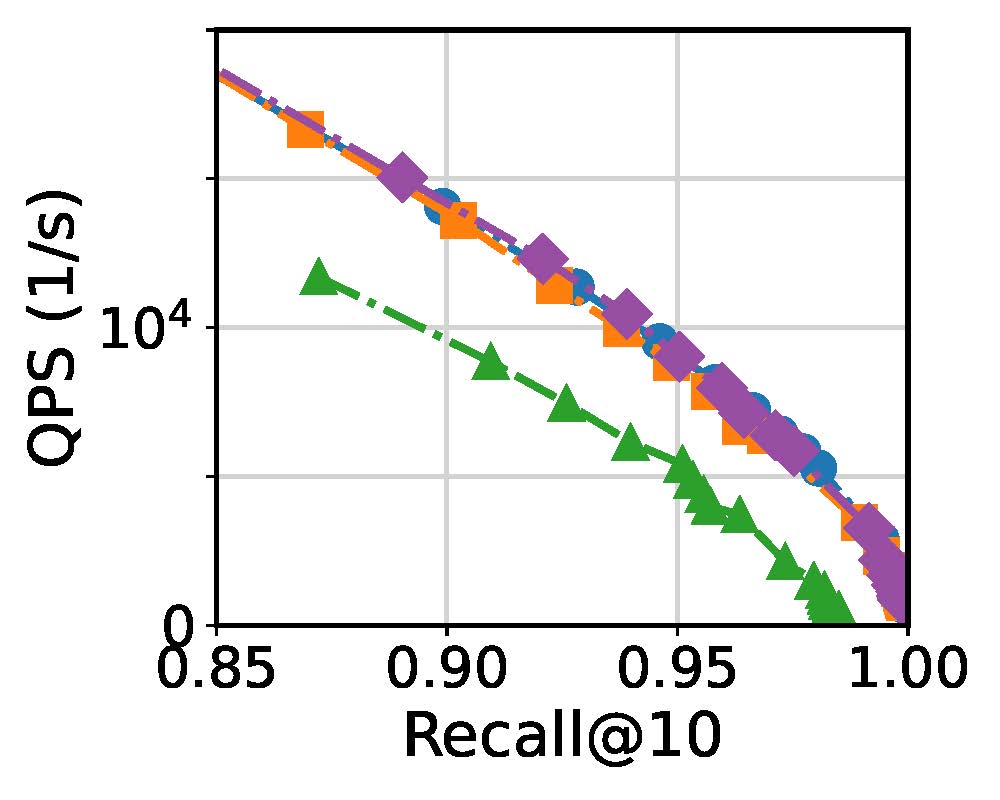}
		\hspace{-0.73em}
	}
	\subfigure[][{\scriptsize MSong}]{
		\hspace{-0.73em}
		\includegraphics[width=0.173\textwidth]{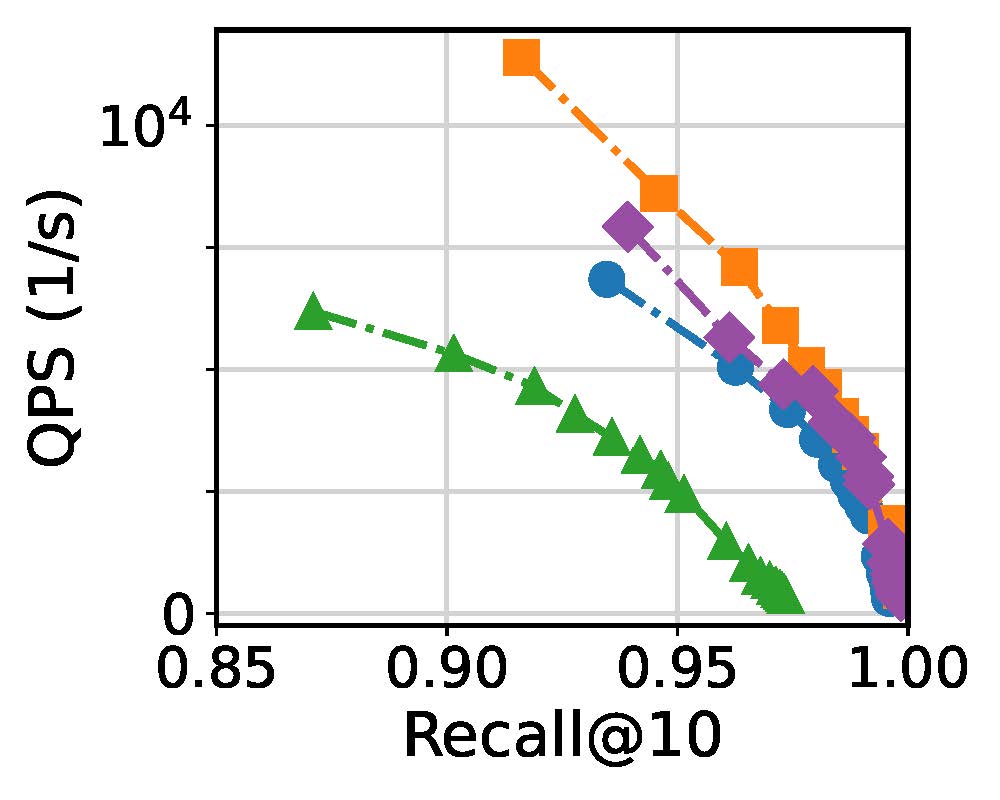}
		\hspace{-0.73em}
	}
	\subfigure[][{\scriptsize GloVe}]{
		\hspace{-0.73em}
		\includegraphics[width=0.173\textwidth]{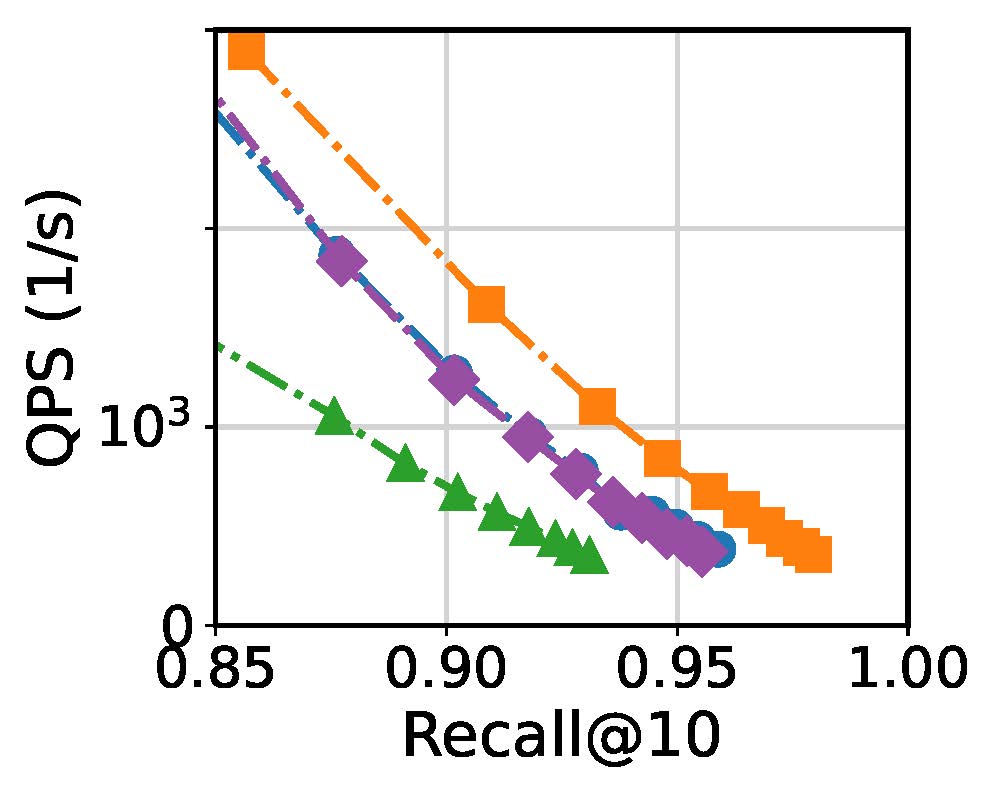}
		\hspace{-0.73em}
	}
	\subfigure[][{\scriptsize Gist1M}]{
		\hspace{-0.73em}
		\includegraphics[width=0.173\textwidth]{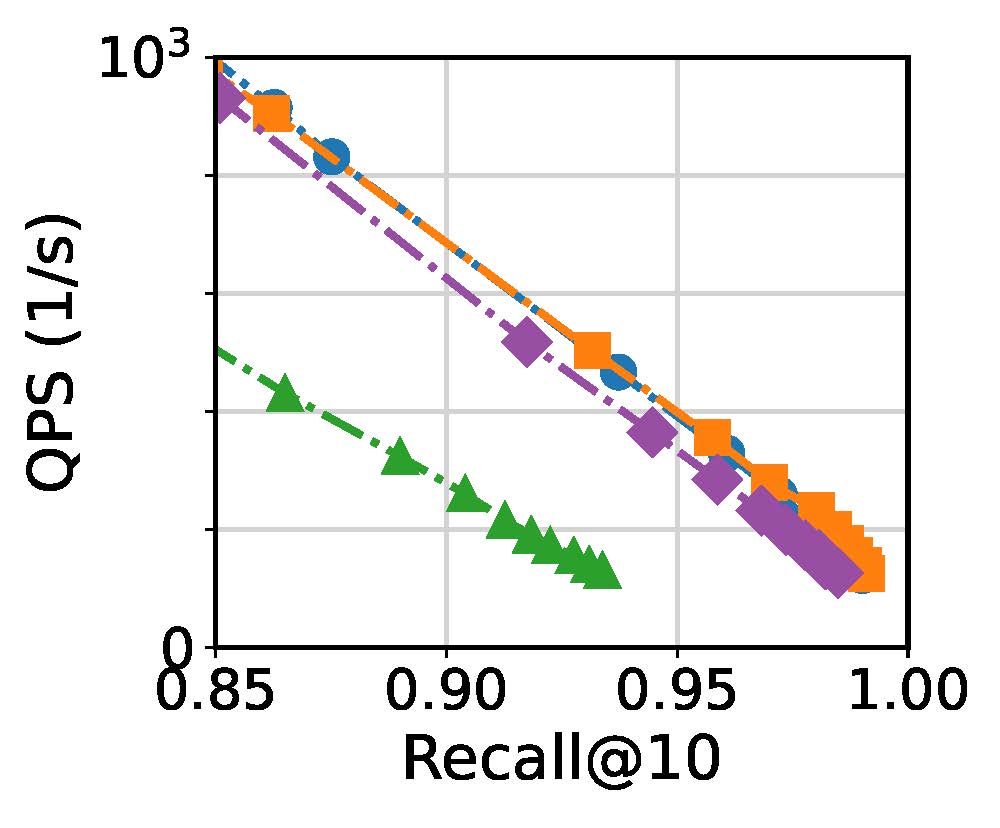}
		\hspace{-0.73em}
	}
	\subfigure[][{\scriptsize Crawl}]{
		\hspace{-0.73em}
		\includegraphics[width=0.173\textwidth]{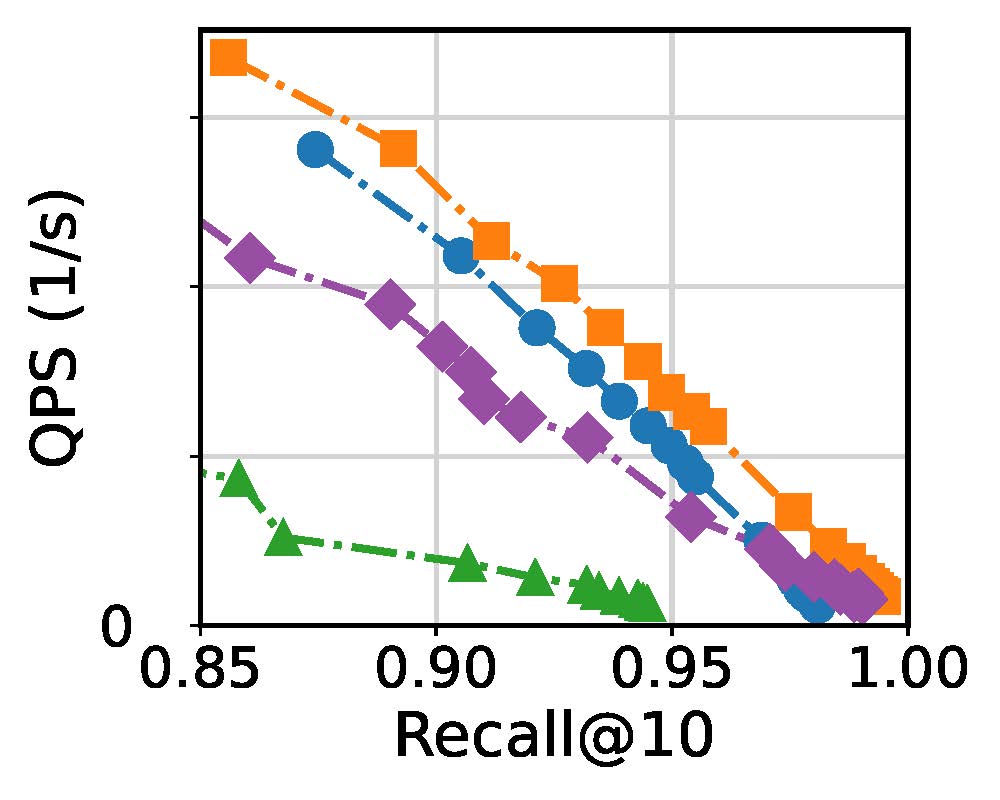}
		\hspace{-0.73em}
	}\hfill\\
	
	\vspace{-2ex}
	
	\subfigure[][{\scriptsize Sift1M}]{
		\hspace{-0.73em}
		\includegraphics[width=0.173\textwidth]{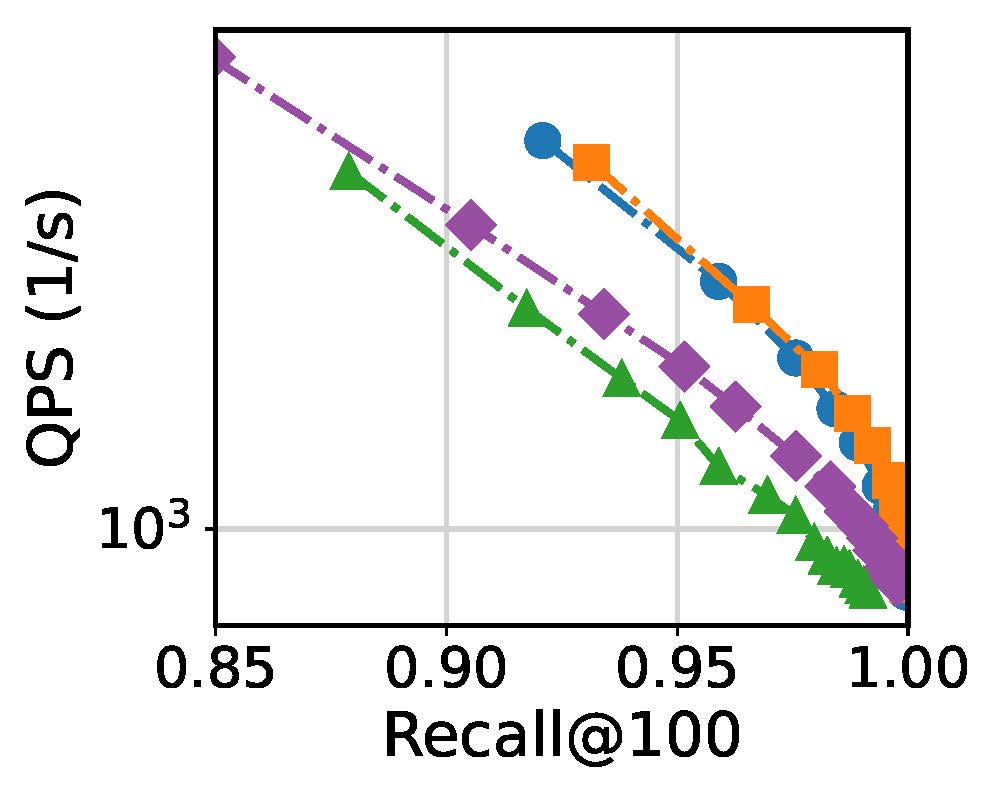}
		\hspace{-0.73em}
	}
	\subfigure[][{\scriptsize Deep1M}]{
		\hspace{-0.73em}
		\includegraphics[width=0.173\textwidth]{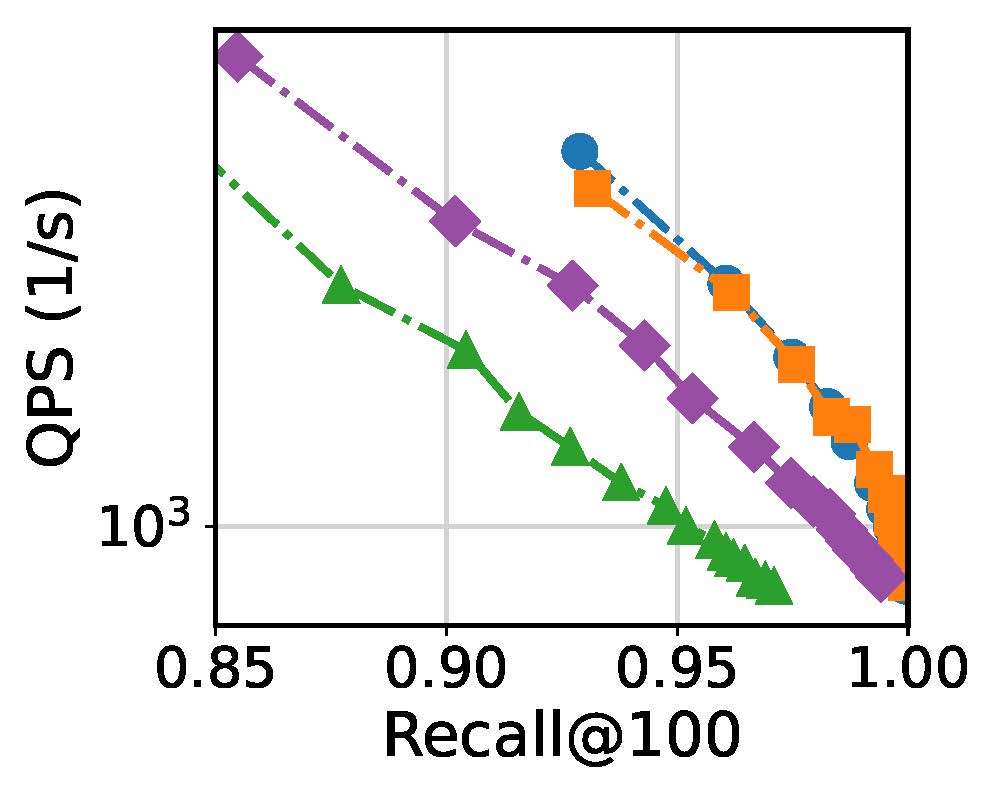}
		\hspace{-0.73em}
	}
	\subfigure[][{\scriptsize MSong}]{
		\hspace{-0.73em}
		\includegraphics[width=0.173\textwidth]{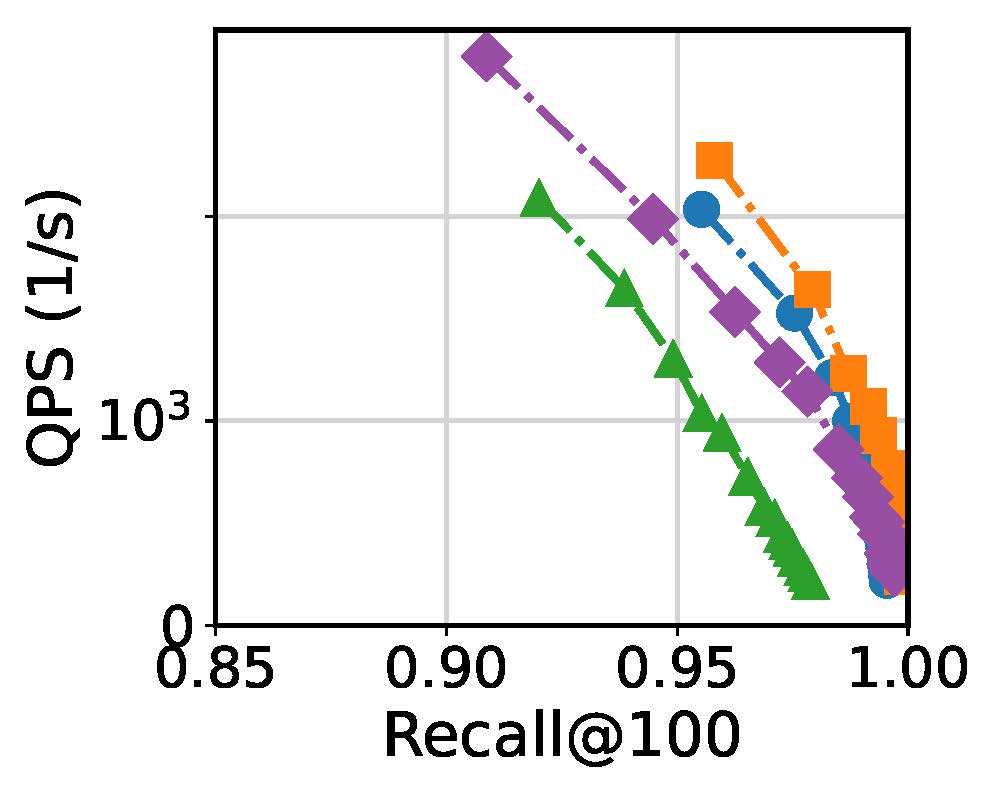}
		\hspace{-0.73em}
	}
	\subfigure[][{\scriptsize GloVe}]{
		\hspace{-0.73em}
		\includegraphics[width=0.173\textwidth]{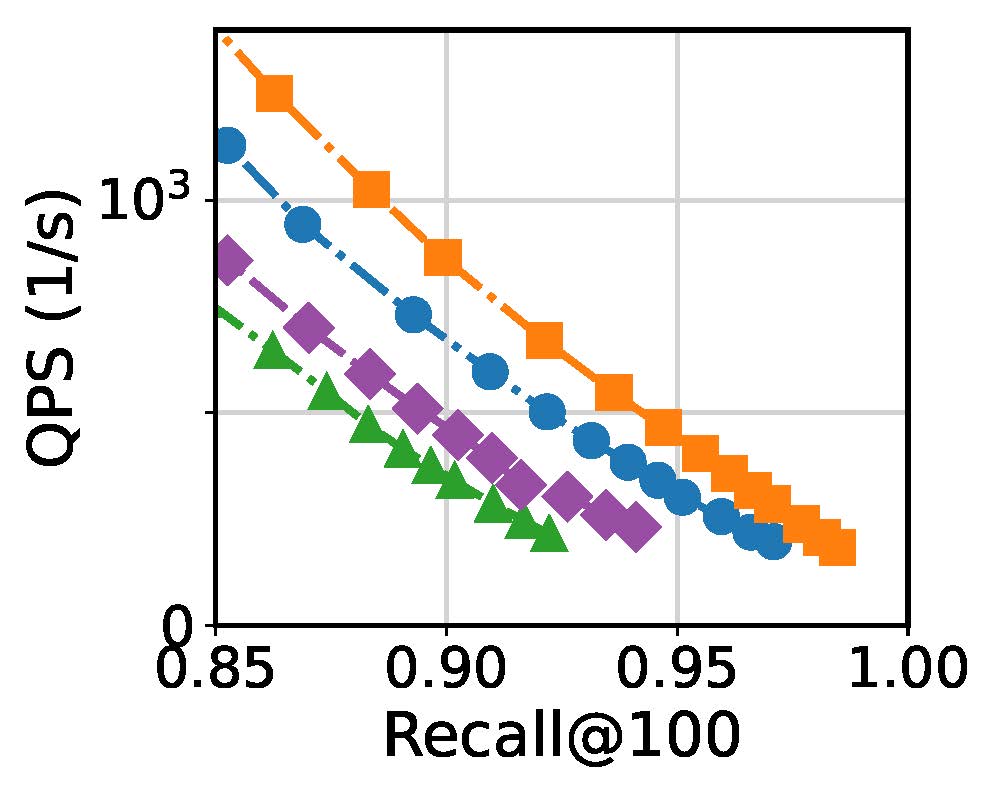}
		\hspace{-0.73em}
	}
	\subfigure[][{\scriptsize Gist1M}]{
		\hspace{-0.73em}
		\includegraphics[width=0.173\textwidth]{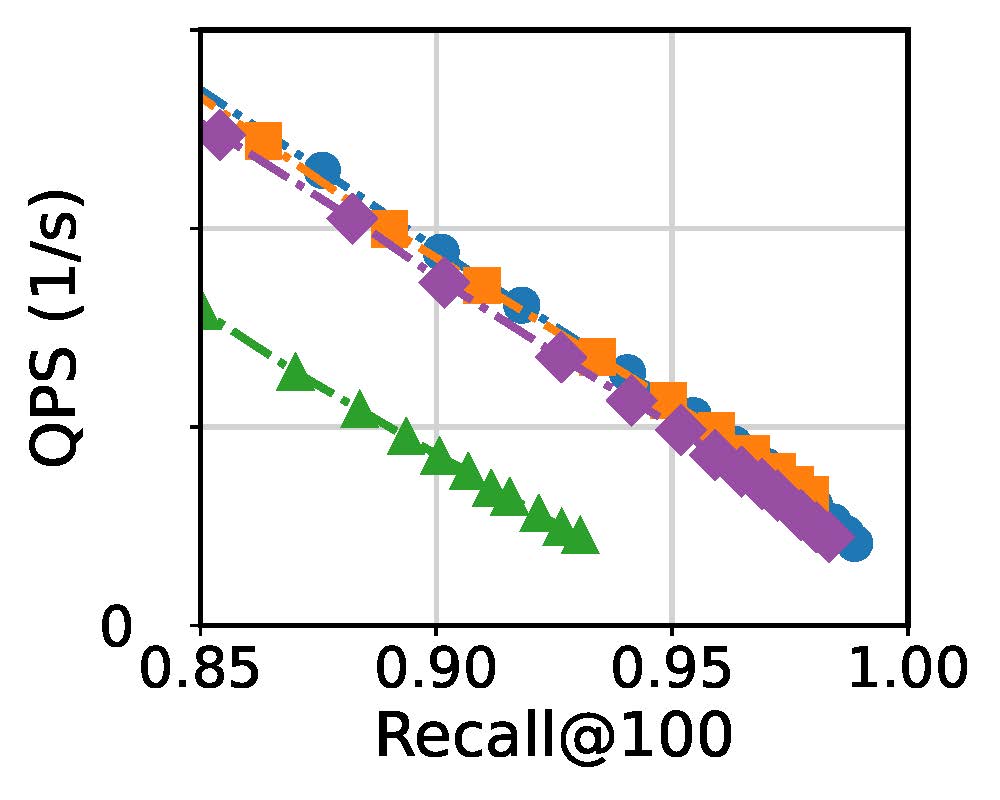}
		\hspace{-0.73em}
	}
	\subfigure[][{\scriptsize Crawl}]{
		\hspace{-0.73em}
		\includegraphics[width=0.173\textwidth]{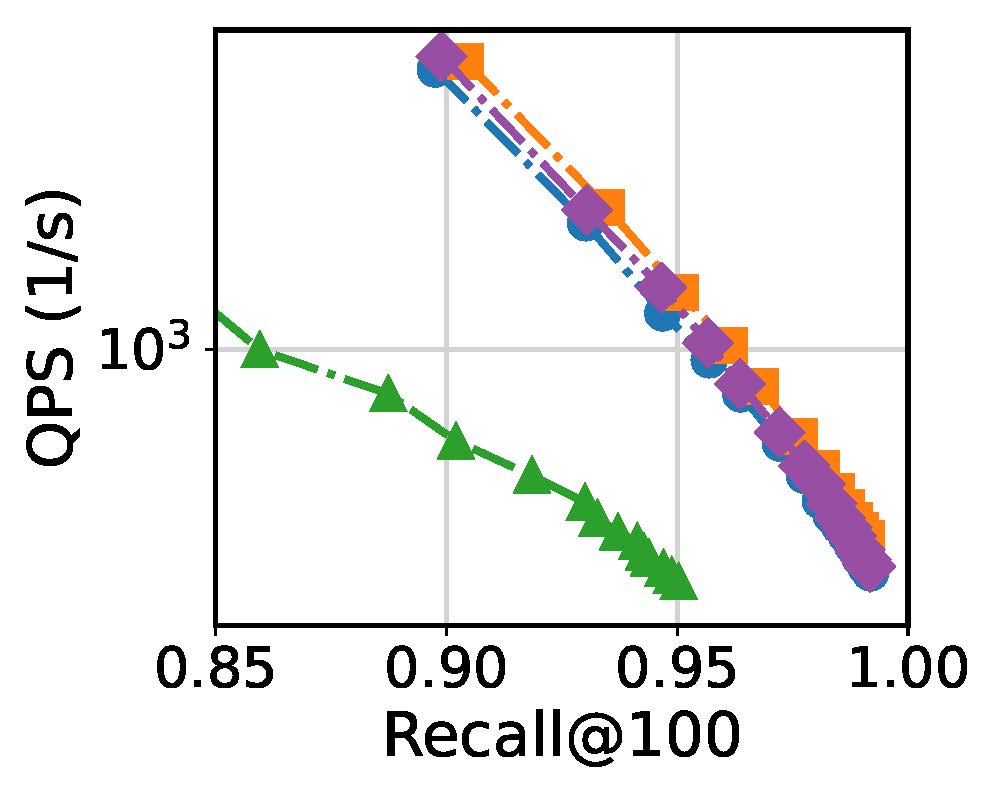}
		\hspace{-0.73em}
	}
	\hfill\\
	
	\vspace{-2ex}
	\subfigure[][{\scriptsize Sift1M}]{
		\hspace{-0.73em}
		\includegraphics[width=0.173\textwidth]{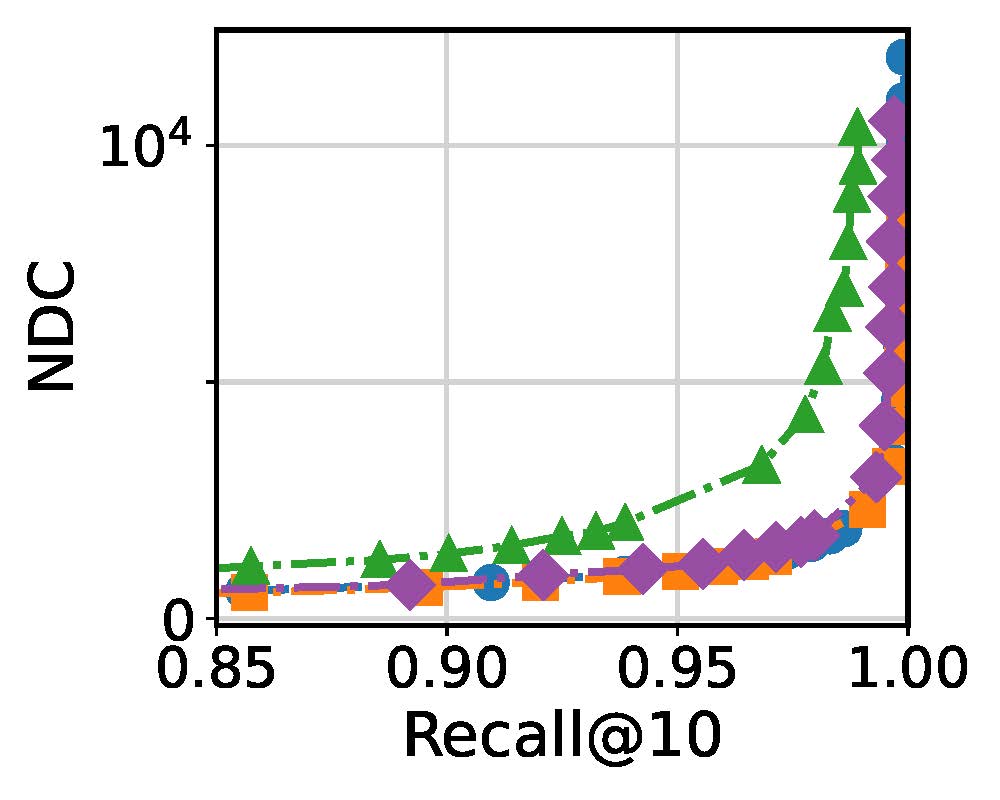}
		\hspace{-0.73em}
	}
	\subfigure[][{\scriptsize Deep1M}]{
		\hspace{-0.73em}
		\includegraphics[width=0.173\textwidth]{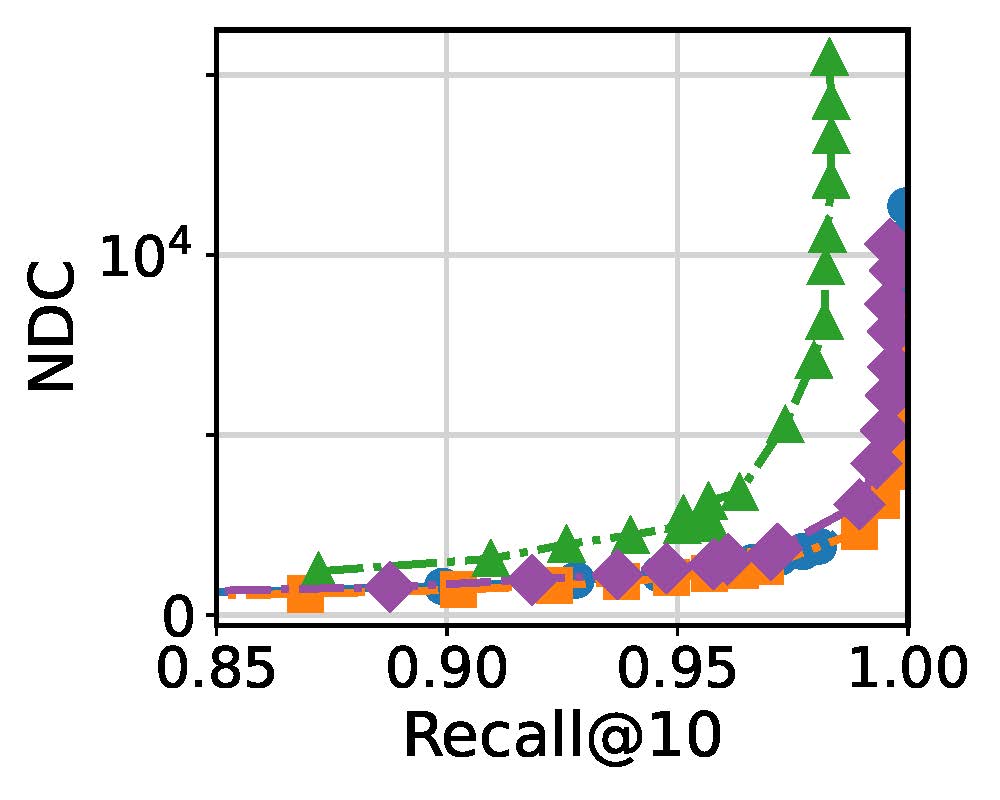}
		\hspace{-0.73em}
	}
	\subfigure[][{\scriptsize MSong}]{
		\hspace{-0.73em}
		\includegraphics[width=0.173\textwidth]{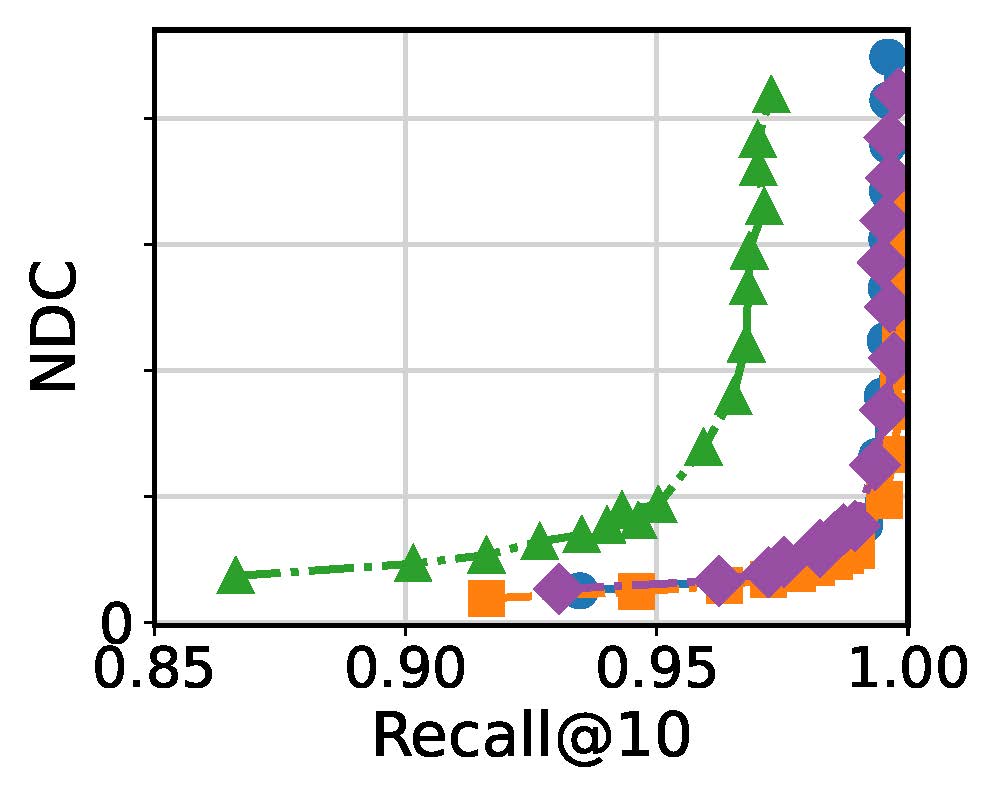}
		\hspace{-0.73em}
	}
	\subfigure[][{\scriptsize GloVe}]{
		\hspace{-0.73em}
		\includegraphics[width=0.173\textwidth]{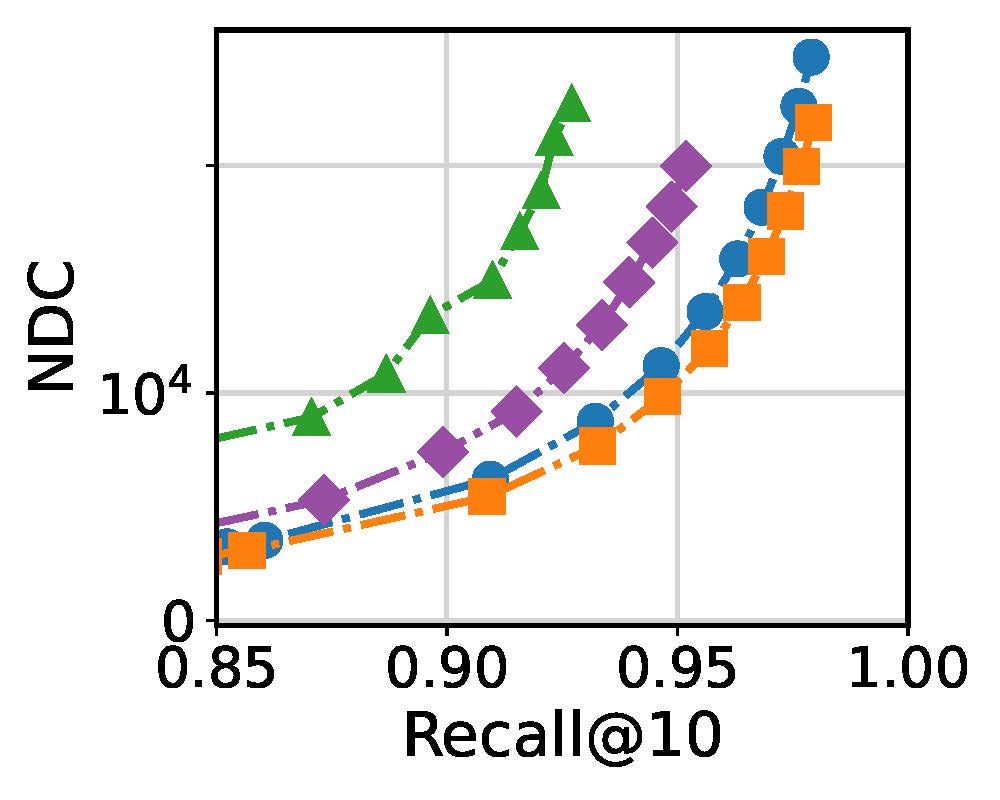}
		\hspace{-0.73em}
	}
	\subfigure[][{\scriptsize Gist1M}]{
		\hspace{-0.73em}
		\includegraphics[width=0.173\textwidth]{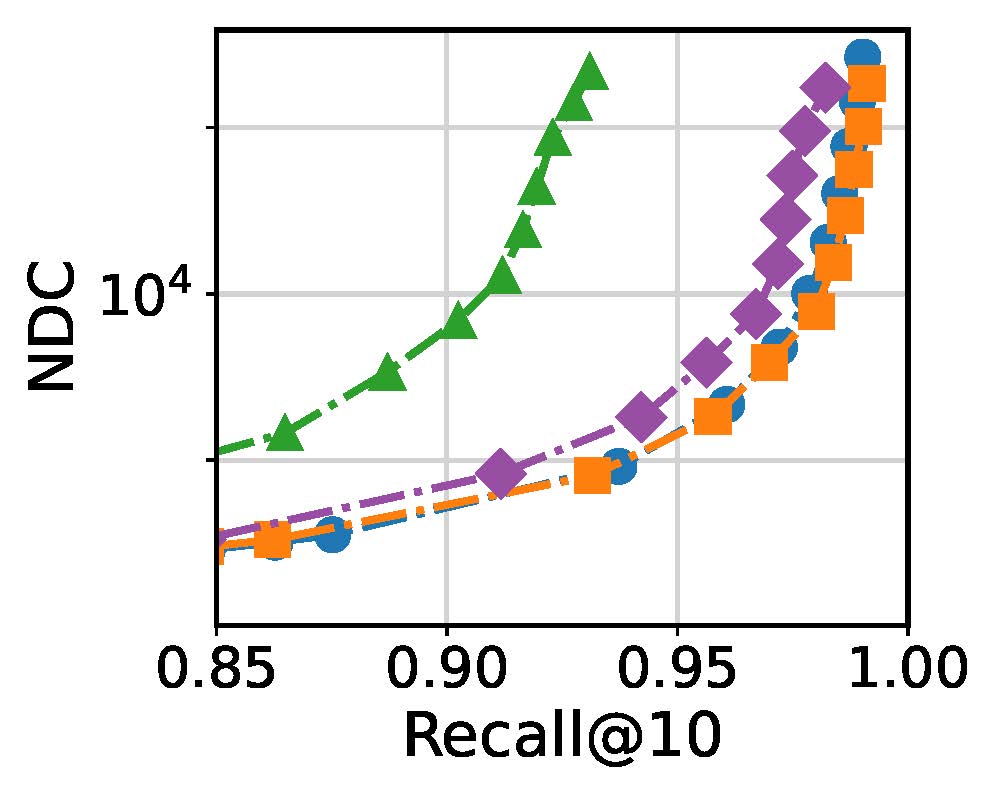}
		\hspace{-0.73em}
	}
	\subfigure[][{\scriptsize Crawl}]{
		\hspace{-0.73em}
		\includegraphics[width=0.173\textwidth]{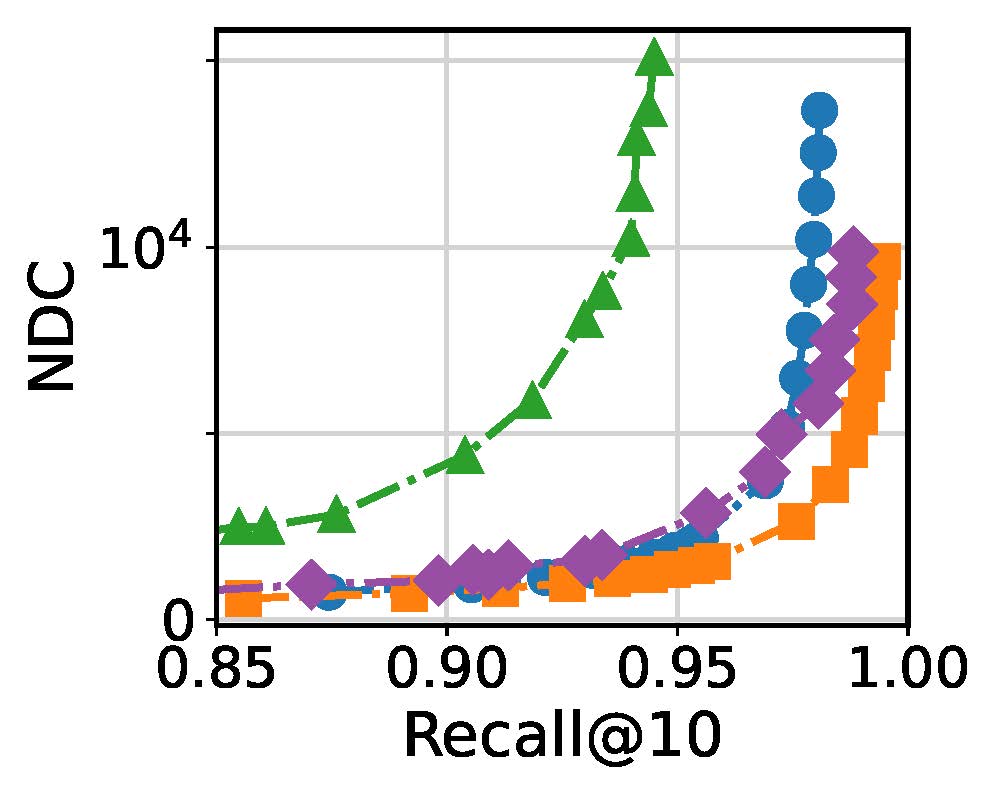}
		\hspace{-0.73em}
	}
	\figureCaptionMargin\vspace{-2ex}
	\caption{Search performance of the merged index using FGIM and incremental methods. (a--f: $Recall@10$-QPS, g--l: $Recall@100$-QPS, m--r: $Recall@10$-NDC; merging 2 indexes, Exp.~2)}
	\label{fig:hnsw_search}
\end{figure}

\begin{figure}[t]
	\centering
	\begin{minipage}{0.7\textwidth}
		\centering
		
		\begin{minipage}{0.48\linewidth}
			\centering
			\includegraphics[width=\linewidth]{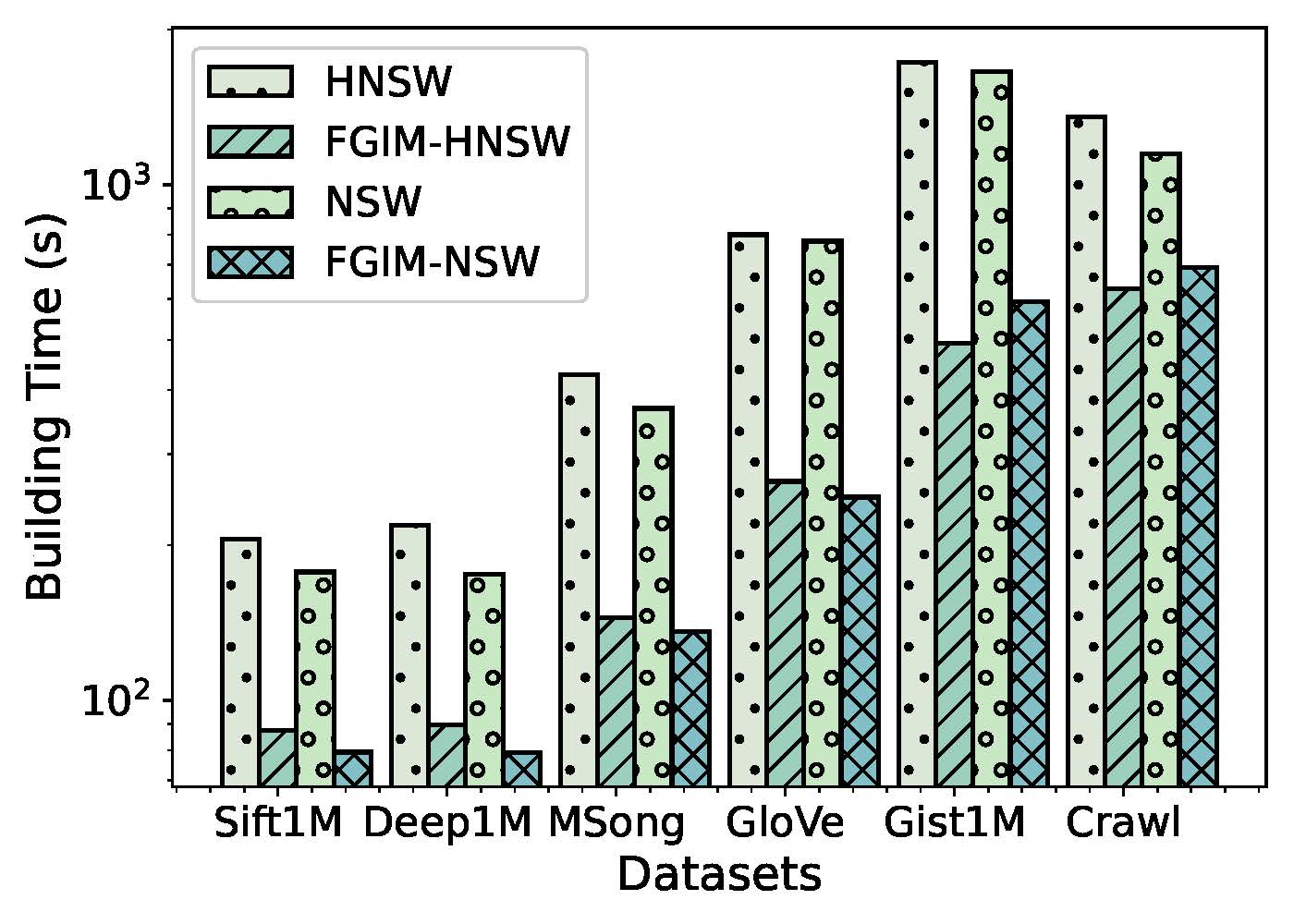}
			\vspace{-3ex}
			\captionof{figure}{Merging efficiency. (Exp.~1)}
			\label{fig:bhnsw}
		\end{minipage}
		\hfill
		\begin{minipage}{0.48\linewidth}
			\centering
			\includegraphics[width=\linewidth]{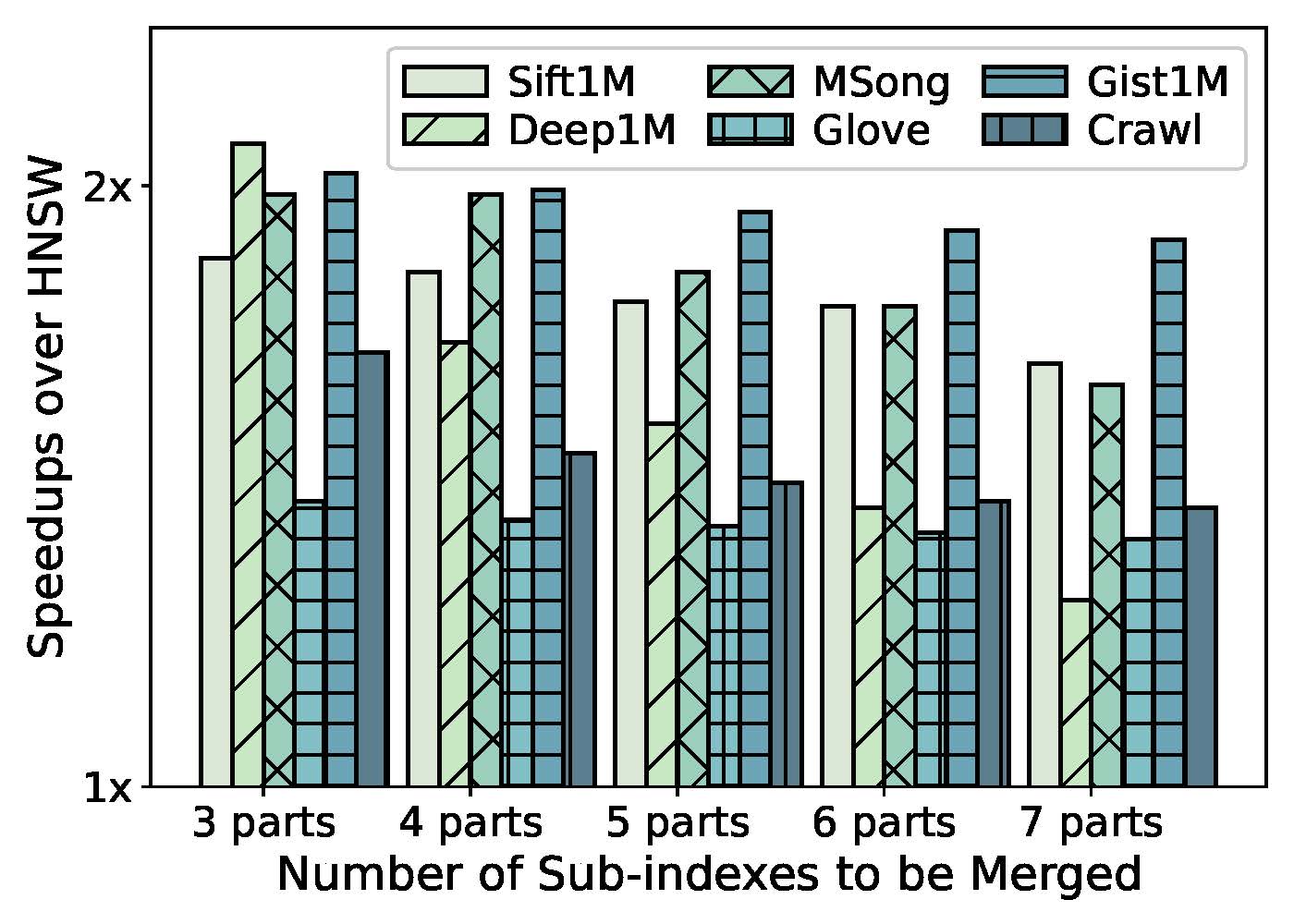}
			\vspace{-4ex}
			\captionof{figure}{Merging efficiency in multiple indexes (Exp.~4)}
			\label{fig:multiple}
		\end{minipage}
		
	\end{minipage}
	\figureBelowMargin
\end{figure}

Figure~\ref{fig:bhnsw} reports the construction cost of the merged index using the FGIM framework and the incremental method of HNSW, NSW in the memory-based setting. Our methods achieve substantially higher efficiency, demonstrating speedups of 2.35$\times$ and 2.24$\times$ on Sift, 2.44$\times$ and 2.22$\times$ on Deep, 2.96$\times$ and 2.71$\times$ on MSong, 3.01$\times$ and 3.14$\times$ on GloVe, 3.51$\times$ and 2.79$\times$ on Gist, 2.16$\times$ and 1.66$\times$ on Crawl, respectively, compared to HNSW and NSW. We also report the build time of HNSW and NSW in Table~\ref{tab:baseline_build_time} for reference. In the disk-based setting, Table~\ref{tab:disk_merge_time} shows that FGIM achieves speedups of 9.07$\times$, 9.48$\times$, 11.52$\times$, 8.33$\times$, 5.61$\times$, and 5.87$\times$ over FreshDiskANN on the corresponding datasets.
This improvement can be attributed to FGIM's ability to leverage both search mechanisms and iterative optimization strategies, thereby minimizing computational overhead. Overall, FGIM delivers higher efficiency than incremental methods in both in-memory and disk-based settings.

Figure~\ref{fig:hnsw_search} and Table~\ref{tab:disk_qps} compare the search performance of incremental and our methods. Across all datasets, FGIM achieves comparable or better performance, showing that it can effectively merge indexes while preserving or improving search accuracy.

\begin{table}[t]
	\small
	\centering
	\vspace{-1ex}
	\caption{Comparison of FreshDiskANN (FDA) and the proposed method in terms of merge time in the disk-based setting using 20 threads. (Exp.~1)}
	\vspace{-2ex}
	\begin{tabular}{lcccccc}
		\toprule
		\textbf{Methods} & Sift1M & Deep1M & MSong & GloVe & Gist1M & Crawl \\
		\midrule
		FDA & 623.50 & 654.67 & 598.36 & 1588.05 & 965.59 & 1678.53 \\
		FGIM-FDA & 68.78 & 69.08 & 51.92 & 190.67 & 172.18 & 286.11 \\
		\bottomrule
	\end{tabular}
	\vspace{-1ex}
	\label{tab:disk_merge_time}
\end{table}

\begin{table}[t]
	\small
	\centering
	\caption{Comparison of build time (in seconds) for baseline methods in the memory-based setting. (Exp.~1)}
	\vspace{-1ex}
	\begin{tabular}{lcccccc}
		\toprule
		\multirow{2}{*}{\textbf{Methods}} & \multicolumn{3}{c}{\textbf{Half Build Time}} & \multicolumn{3}{c}{\textbf{Total Build Time}} \\
		\cmidrule(lr){2-4} \cmidrule(lr){5-7}
		& {Sift1M} & {Deep1M} & {MSong} & {Sift1M} & {Deep1M} & {MSong} \\
		\midrule
		HNSW & 140.02 & 136.60 & 257.53 & 345.37 & 355.13 & 685.59 \\
		NSW  & 158.99 & 158.54 & 275.93 & 343.54 & 340.89 & 665.46 \\
		\bottomrule
	\end{tabular}
	\label{tab:baseline_build_time}
	\vspace{-1ex}
\end{table}

\begin{table}[t]
	\small
	\centering
	\caption{Search performance (QPS) of FreshDiskANN (FDA) and our method in the disk-based setting. (Exp.~2)}
	\vspace{-1ex}
	\begin{tabular}{lcccccccccccc}
		\toprule
		\multirow{2}{*}{\textbf{Dataset}} & \multicolumn{2}{c}{\textbf{$Recall@10=95\%$}} & \multicolumn{2}{c}{\textbf{$Recall@10=97\%$}} & \multicolumn{2}{c}{\textbf{$Recall@10=99\%$}} \\
		\cmidrule(lr){2-3} \cmidrule(lr){4-5} \cmidrule(lr){6-7}
		& {FDA} & {FGIM-FDA} & {FDA} & {FGIM-FDA} & {FDA} & {FGIM-FDA} \\
		\midrule
		Sift1M & 3841 & 3857 & 3054 & 2947 & 1571 & 1646 \\
		Deep1M & 3840 & 3782 & 2894 & 2746 & 1584 & 1559 \\
		MSong & 6709 & 6524 & 4999 & 4942 & 2146 & 2433 \\
		\bottomrule
	\end{tabular}
	\label{tab:disk_qps}
	\vspace{-1ex}
\end{table}

\begin{table}[t]
	\small
	\centering
	\caption{Comparison of existing strategies. (Exp.~3)}
	\vspace{-1ex}
	\begin{tabular}{lccccccc}
		\toprule
		\textbf{Method} & \textbf{Metric} & \textbf{Sift1M} & \textbf{Deep1M} & \textbf{MSong} & \textbf{GloVe} & \textbf{Gist1M} & \textbf{Crawl} \\
		\midrule
		\multirow{3}{*}{HNSW} 
		& Indexing Time & 3.42m & 3.64m & 7.13m & 13.3m & 28.8m & 22.6m \\
		& $Recall@10$ & 99.0\% & 98.7\% & 97.8\% & 93.9\% & 87.6\% & 96.9\% \\
		& Index Size & 164.2M & 173.2M & 124.2M & 292.2M & 96.7M & 348.0M \\
		\midrule
		\multirow{3}{*}{NNMerge}
		& Indexing Time & 3.24m & 3.23m & 6.31m & 10.2m & 28.3m & 32.0m \\
		& $Recall@10$ & 94.4\% & 92.6\% & 91.6\% & 89.9\% & 74.5\% & 90.6\% \\
		& Index Size & 220.5M & 220.6M & 219.3M & 522.5M & 430.6M & 925.6M \\
		\midrule
		\multirow{3}{*}{DiskANN}
		& Indexing Time & -- & -- & -- & -- & -- & -- \\
		& $Recall@10$ & 48.6\% & 49.7\% & 49.9\% & 48.7\% & 46.4\% & 49.7\% \\
		& Index Size & 189.4M & 210.0M & 137.2M & 430.0M & 120.8M & 474.4M \\
		\midrule
		\multirow{3}{*}{Lucene}
		& Indexing Time & 1.88m & 2.13m & 3.98m & 5.37m & 9.53m & 10.1m \\
		& $Recall@10$ & 98.8\% & 98.3\% & 97.6\% & 94.3\% & 86.7\% & 98.3\% \\
		& Index Size & 119.3M & 103.8M & 78.8M & 181.1M & 66.2M & 207.2M \\
		\midrule
		\multirow{3}{*}{FGIM (Ours)}
		& Indexing Time & \cellcolor{gray!30}{1.46m} & \cellcolor{gray!30}{1.49m} & \cellcolor{gray!30}{2.41m} & \cellcolor{gray!30}{4.43m} & \cellcolor{gray!30}{8.20m} & \cellcolor{gray!30}{10.5m} \\
		& $Recall@10$ & \cellcolor{gray!30}{99.3\%} & \cellcolor{gray!30}{99.1\%} & \cellcolor{gray!30}{98.8\%} & \cellcolor{gray!30}{96.5\%} & \cellcolor{gray!30}{91.3\%} & \cellcolor{gray!30}{99.0\%} \\
		& Index Size & \cellcolor{gray!30}{123.3M} & \cellcolor{gray!30}{134.3M} & \cellcolor{gray!30}{92.3M} & \cellcolor{gray!30}{324.3M} & \cellcolor{gray!30}{100.7M} & \cellcolor{gray!30}{277.2M} \\
		\bottomrule
	\end{tabular}
	\label{tab:existing_method}
	\vspace{-1ex}
\end{table}

\noindent\textbf{Exp.3: Comparison with other methods.} As discussed in \textsection\ref{subsec:currentStudies}, many existing methods exhibit significant limitations. In this part, we evaluate these methods alongside our approach to demonstrate their shortcomings. Table~\ref{tab:existing_method} presents a comparison of different methods, focusing on merging time, $Recall@10$ with fixed QPS (4000 for Sift, Deep, MSong, and 800 for GloVe, Gist, and Crawl), and index size. NNMerge achieves relatively fast merging but suffers from a lower $Recall@10$, consistent with findings in previous benchmark studies \cite{li2019approximate,wang2021comprehensive}, as $k$-NNG is not optimized for efficient ANNS. Moreover, the merging strategy employed by DiskANN leads to a noticeable drop in search accuracy, primarily due to the random spatial distribution of real-world datasets. Without a spatial clustering step during preprocessing, the merging process reduces to a naive structural union, which compromises global navigability and connectivity of the merged graph.
Compared to HNSW, Lucene’s merge strategy shows a modest decline in recall on several datasets, caused by reducing $ef_{construction}$ to improve merge efficiency at the cost of graph quality.
	In contrast, our FGIM framework outperforms other methods in terms of merging efficiency on most datasets, achieving consistently superior search performance. Overall, existing approaches remain ineffective for merging graph-based indexes.

\subsection{Multiple Indexes Merging}
\label{subsec:multiple}

As noted in \textsection\ref{sec:introduction}, efficiently merging multiple graph-based indexes is crucial for real-time systems that generate frequent small indexes. We vary the number of indexes to merge from 3 to 7, each constructed by randomly selecting an equal number of vectors from the original dataset. We then compare our merging cost with HNSW's incremental construction method.

\noindent\textbf{Exp.4: Multi-Index Merging Efficiency.} As shown in Figure~\ref{fig:multiple}, our merging method consistently outperforms the incremental approach as the number of indexes increases. Notably, although the acceleration gain of our method gradually diminishes with a growing number of indexes, it still achieves a noticeable speedup even when merging seven indexes. However, in real-world scenarios, the generated index fragments are merged offline in the background, preventing an excessive number of indexes. Therefore, our method remains effective in this context, ensuring its applicability and scalability to dynamic and evolving data environments.

\begin{figure}[t]
	\centering
	
	\includegraphics[width=0.7\textwidth]{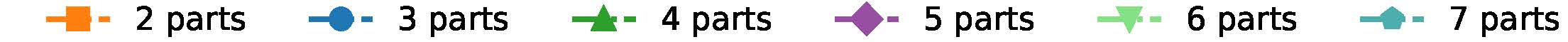}
	
	\subfigure[][{\scriptsize Sift1M}]{
		\hspace{-0.73em}
		\includegraphics[width=0.173\textwidth]{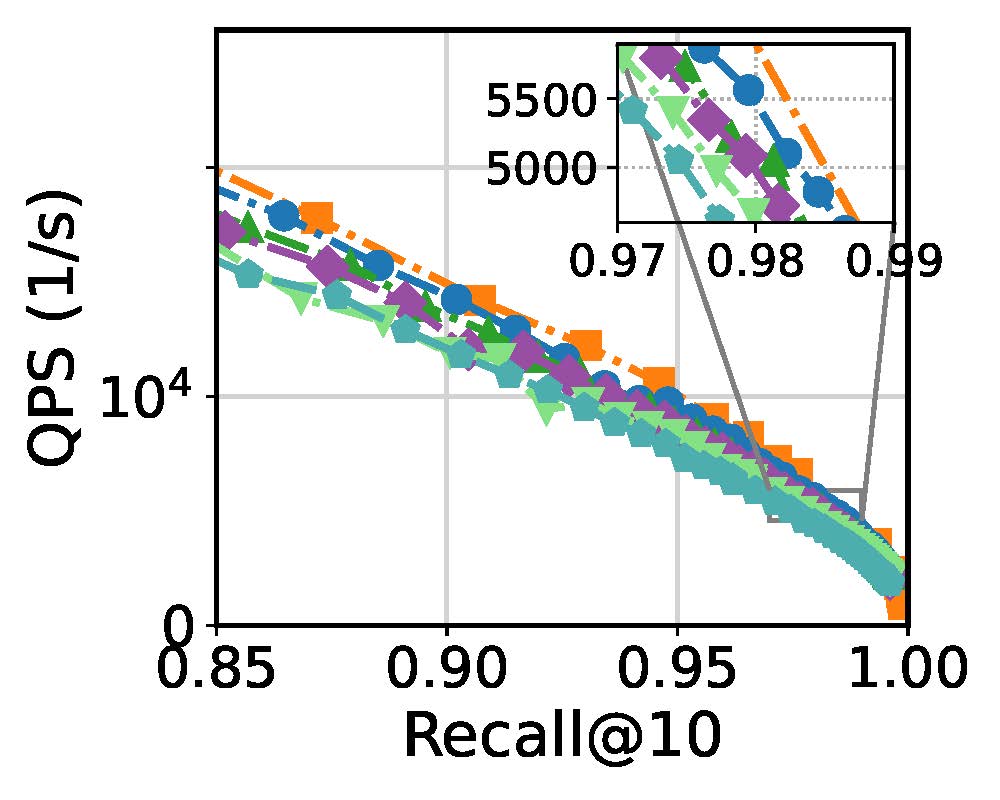}
		\hspace{-0.73em}
	}
	\subfigure[][{\scriptsize Deep1M}]{
		\hspace{-0.73em}
		\includegraphics[width=0.173\textwidth]{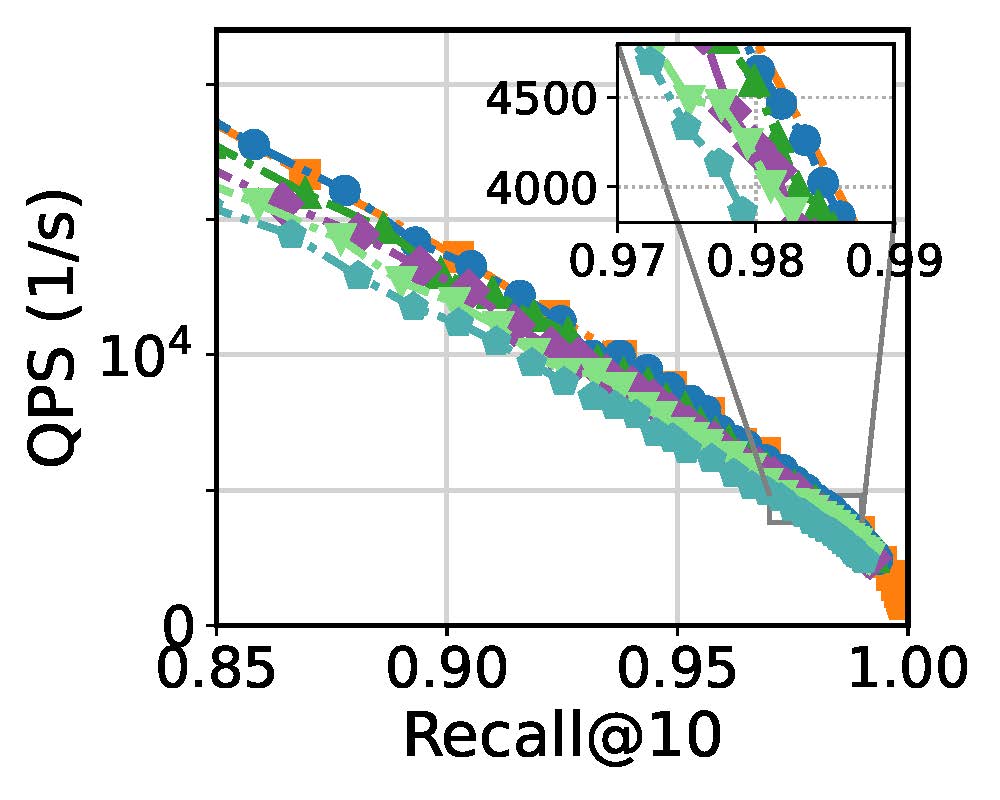}
		\hspace{-0.73em}
	}
	\subfigure[][{\scriptsize MSong}]{
		\hspace{-0.73em}
		\includegraphics[width=0.173\textwidth]{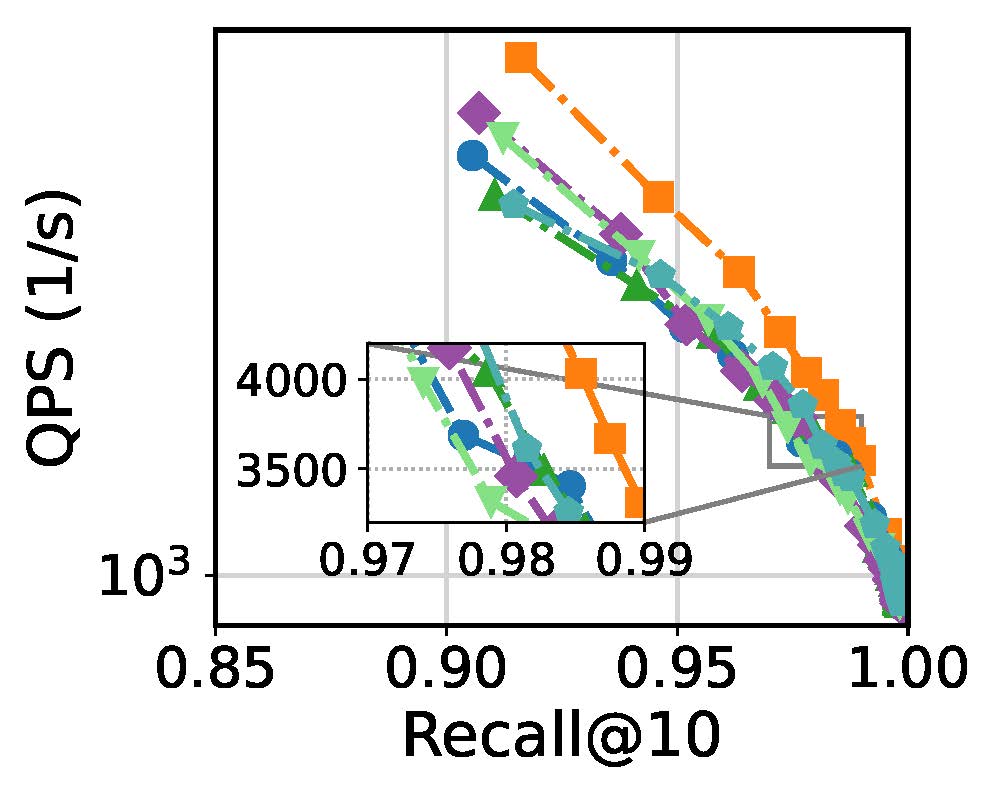}
		\hspace{-0.73em}
	}
	\subfigure[][{\scriptsize GloVe}]{
		\hspace{-0.73em}
		\includegraphics[width=0.173\textwidth]{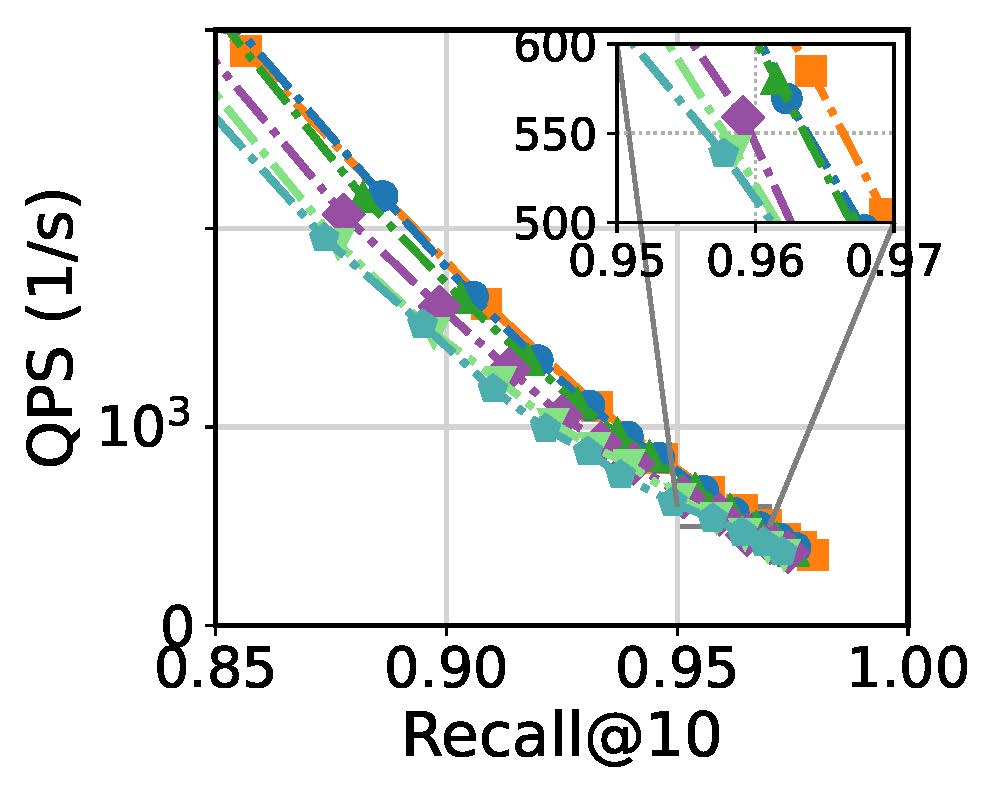}
		\hspace{-0.73em}
	}
	\subfigure[][{\scriptsize Gist1M}]{
		\hspace{-0.73em}
		\includegraphics[width=0.173\textwidth]{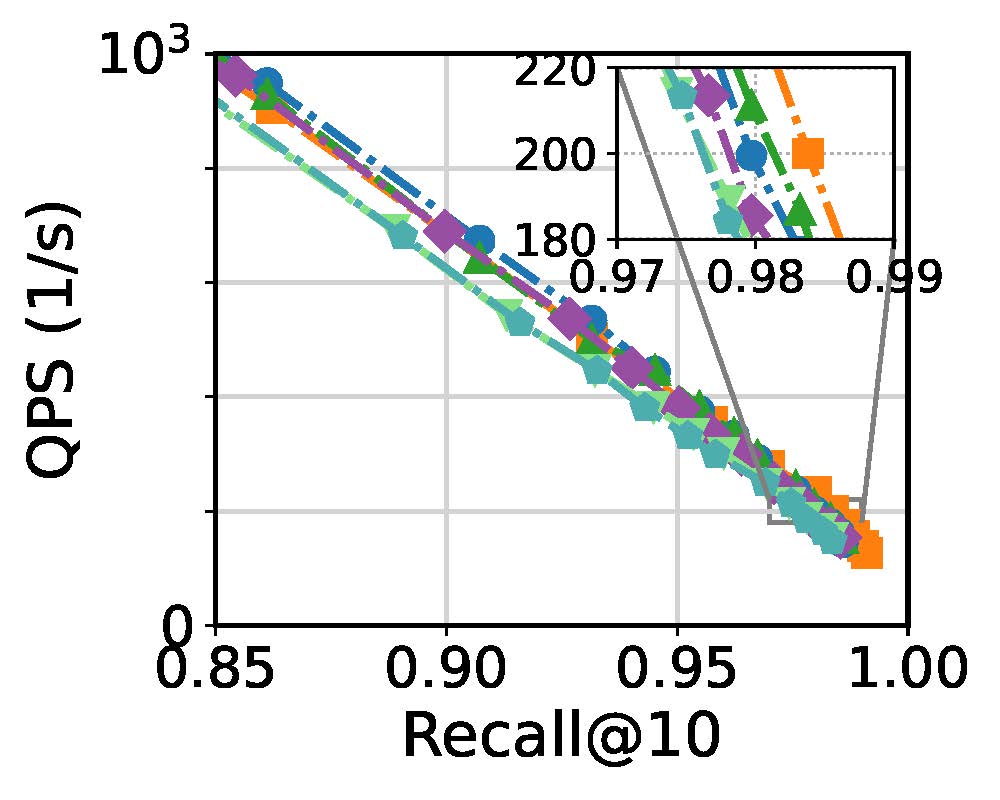}
		\hspace{-0.73em}
	}
	\subfigure[][{\scriptsize Crawl}]{
		\hspace{-0.73em}
		\includegraphics[width=0.173\textwidth]{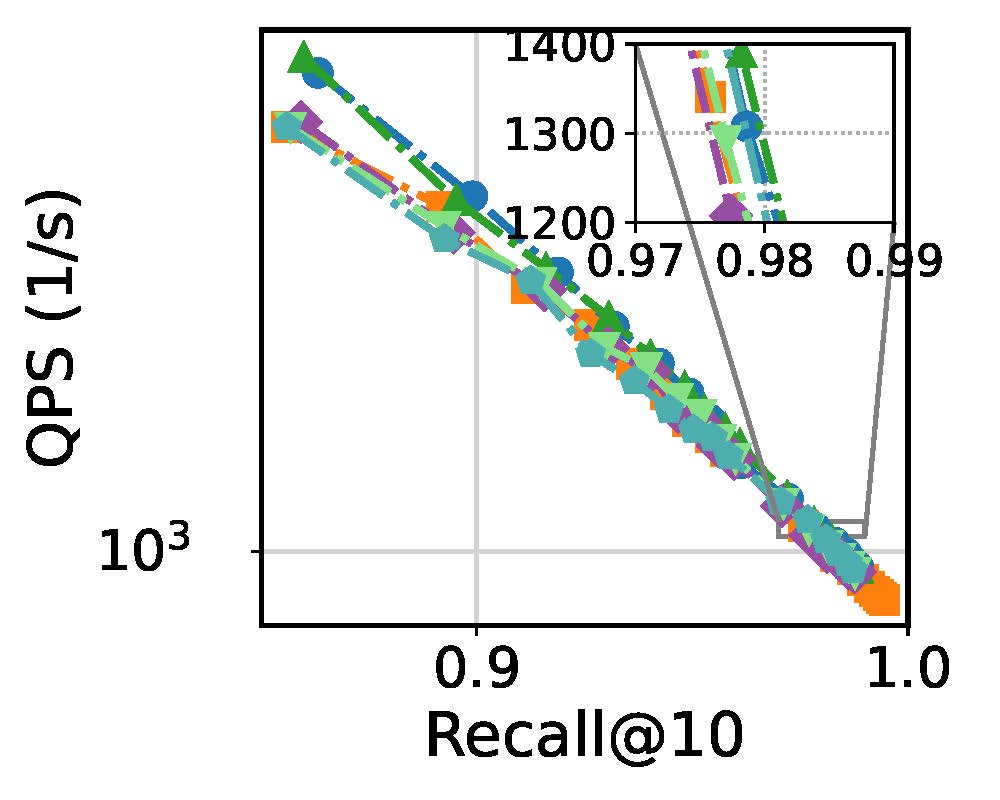}
		\hspace{-0.73em}
	}
	
	\vspace{-3ex}
	\caption{Search performance of the merged index with the number of merged indexes ranging from 2 to 7. (a-f: $Recall@10$-QPS; Exp.~5)}
	\label{fig:mul_qps}
\end{figure}

\noindent\textbf{Exp.5: Search Performance.} Figure~\ref{fig:mul_qps} reports the search performance of the merged index with the number of merged indexes ranging from 2 to 7. The results demonstrate that our method maintains consistent search performance across varying numbers of merged indexes. Across all datasets, the $Recall@10$ drops by at most only \textasciitilde1\% under the same QPS as the number of merged indexes increases. This demonstrates that our method can effectively merge multiple indexes while preserving high search accuracy.

\subsection{Framework Applicability}
\label{subsec:applicability}

In this part, we apply FGIM to merge four representative graph-based indexes, i.e., Vamana, $\tau$-MNG, NSW, and NNDescent, further exploring the generality and effectiveness of our framework.

\begin{figure}[t]
	\vspace{-1ex}
	\centering
	\includegraphics[width=0.9\textwidth]{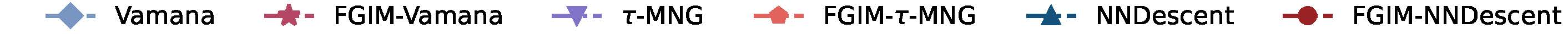}
	\subfigure[][{\scriptsize Sift1M}]{
		\hspace{-0.73em}
		\includegraphics[width=0.173\textwidth]{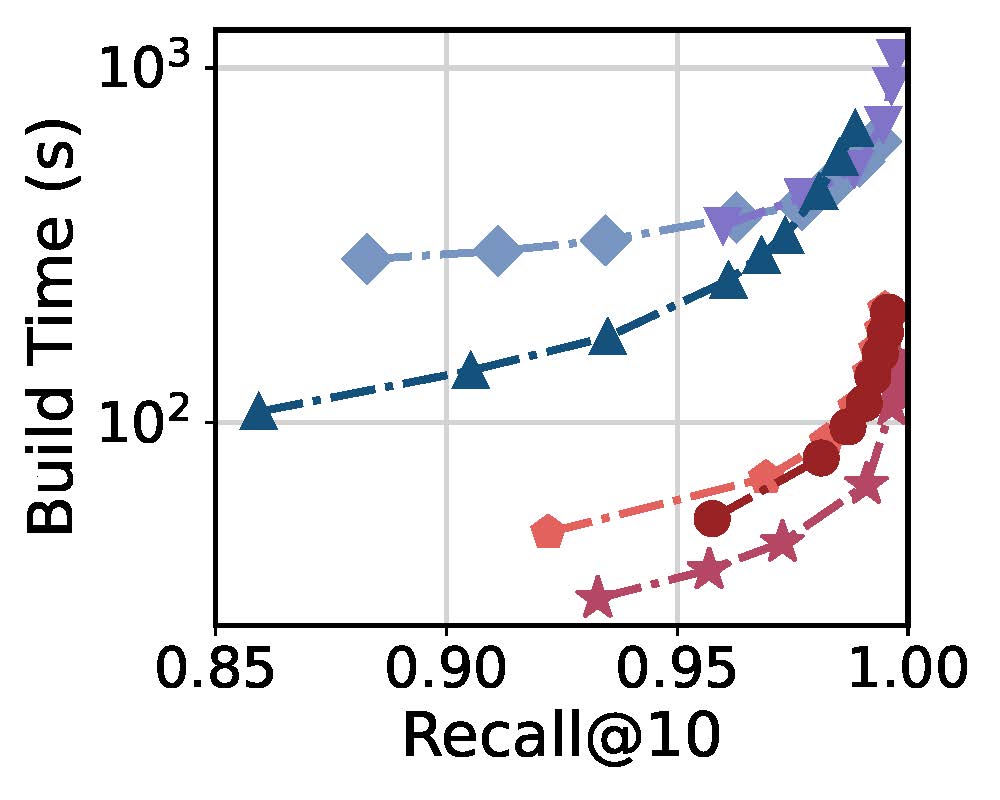}
		\hspace{-0.73em}
	}
	\subfigure[][{\scriptsize Deep1M}]{
		\hspace{-0.73em}
		\includegraphics[width=0.173\textwidth]{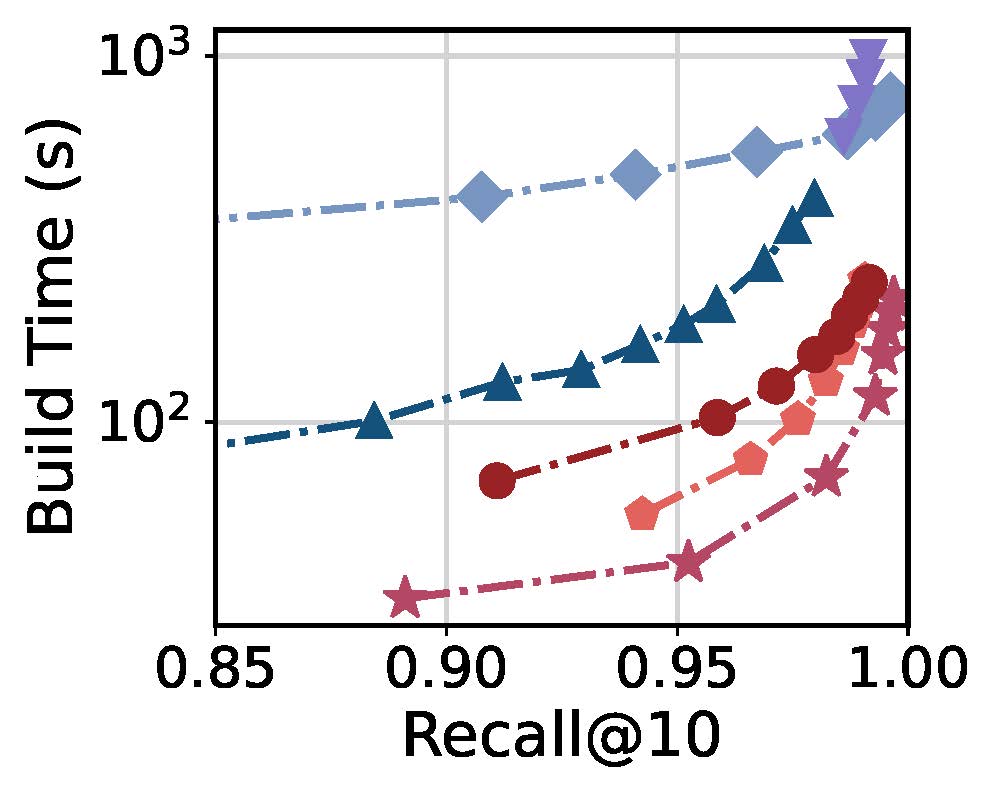}
		\hspace{-0.73em}
	}
	\subfigure[][{\scriptsize MSong}]{
		\hspace{-0.73em}
		\includegraphics[width=0.173\textwidth]{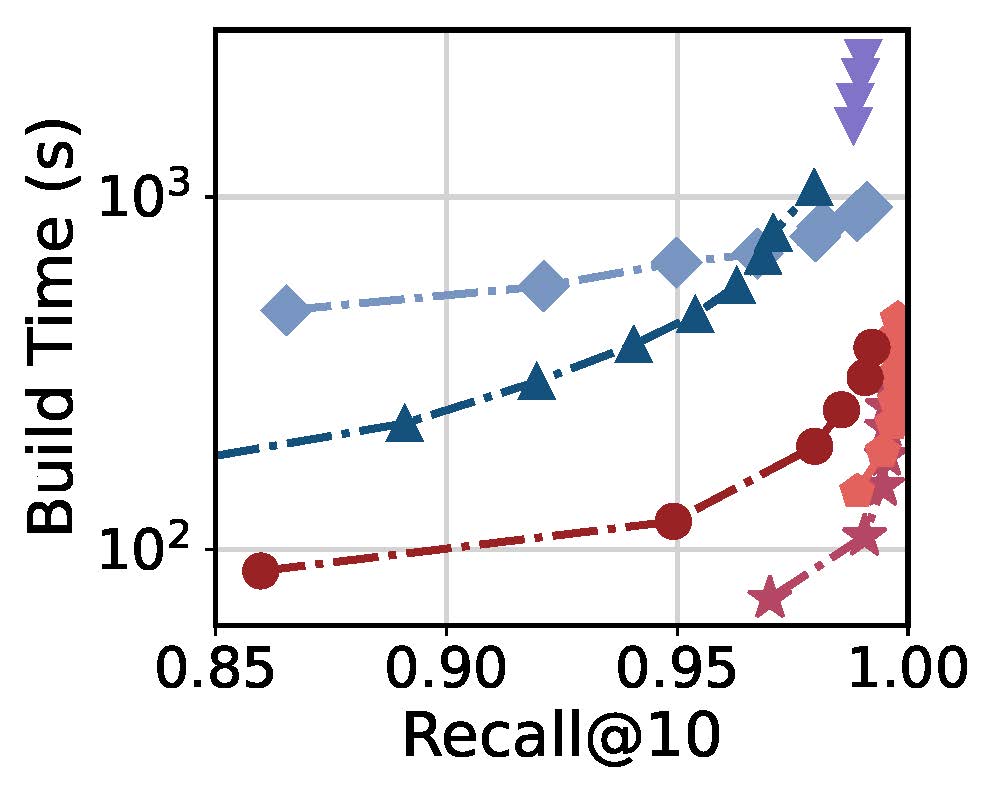}
		\hspace{-0.73em}
	}
	\subfigure[][{\scriptsize GloVe}]{
		\hspace{-0.73em}
		\includegraphics[width=0.173\textwidth]{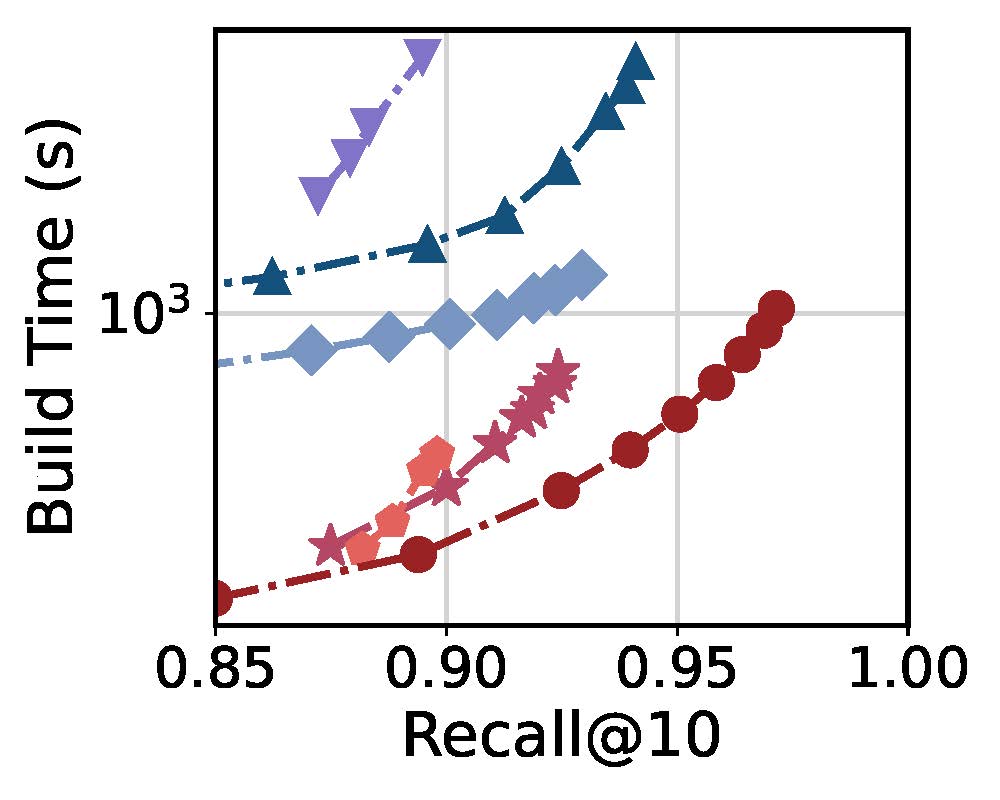}
		\hspace{-0.73em}
	}
	\subfigure[][{\scriptsize Gist1M}]{
		\hspace{-0.73em}
		\includegraphics[width=0.173\textwidth]{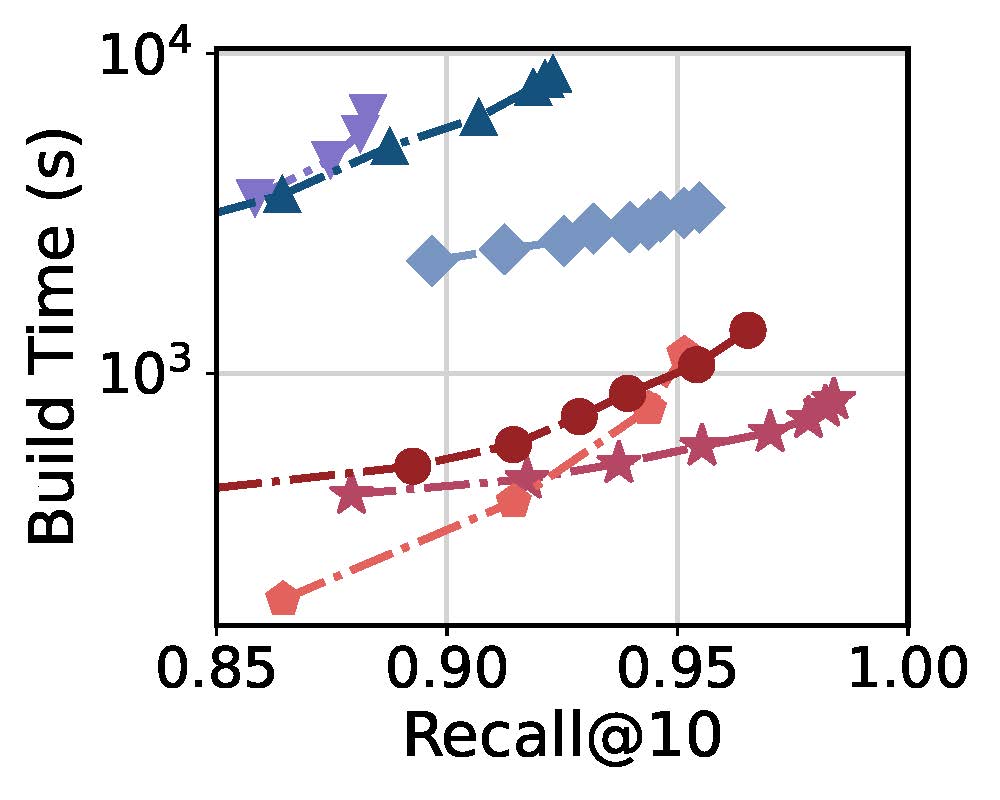}
		\hspace{-0.73em}
	}
	\subfigure[][{\scriptsize Crawl}]{
		\hspace{-0.73em}
		\includegraphics[width=0.173\textwidth]{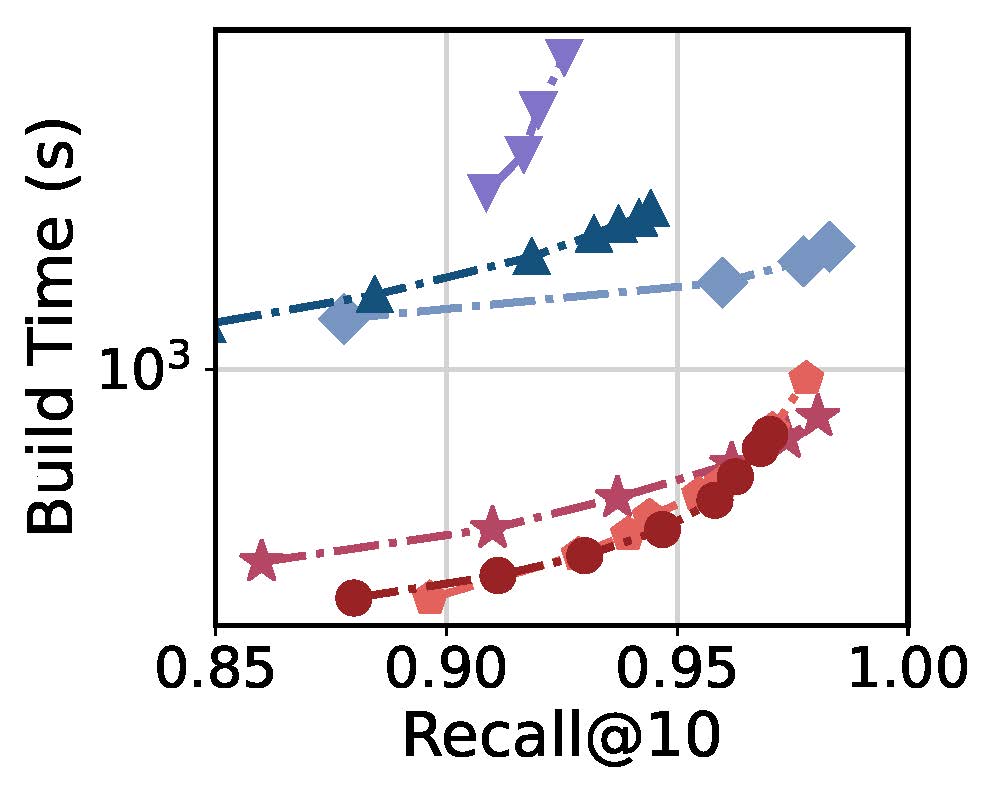}
		\hspace{-0.73em}
	}
	
	\figureCaptionMargin\vspace{-2ex}
	\caption{Construction time for our merging methods and the original methods rebuilt from scratch. (a--f: $Recall@10$-BT; Exp.~6)}
	\label{fig:app_bd}
\end{figure}

\noindent\textbf{Exp.6: Comparison with the Original Methods Rebuilt from Scratch.} In this experiment, we employ building time and $Recall@10$ curves that keep search parameters fixed (i.e., $L=200$). For each method, we systematically vary its construction parameters to reconstruct indexes on the entire dataset from scratch, ensuring that each method has at least four different parameter configurations. Figure~\ref{fig:app_bd} presents the experimental results. We can find out that our methods fulfilled the merging task with high efficiency. Specifically, at the same Recall$@$10, (1) FGIM-Vamana achieved 3.4$\sim$ 6.9$\times$ speedups over Vamana; (2) FGIM-$\tau$-MNG achieved 4.3$\sim$ 23.2$\times$ speedups over $\tau$-MNG; (3) FGIM-NNDescent achieved 3.3$\sim$ 6.7$\times$ speedups over NNDescent. On average, our method achieves a 7.4$\times$ speedup, demonstrating its efficiency in merging graph-based indexes over reconstructing an entirely new index from scratch.

\begin{figure}[t]
	\centering
	\includegraphics[width=0.9\textwidth]{app_legend-eps-converted-to.jpg}
	\subfigure[][{\scriptsize Sift1M}]{
		\hspace{-0.73em}
		\includegraphics[width=0.173\textwidth]{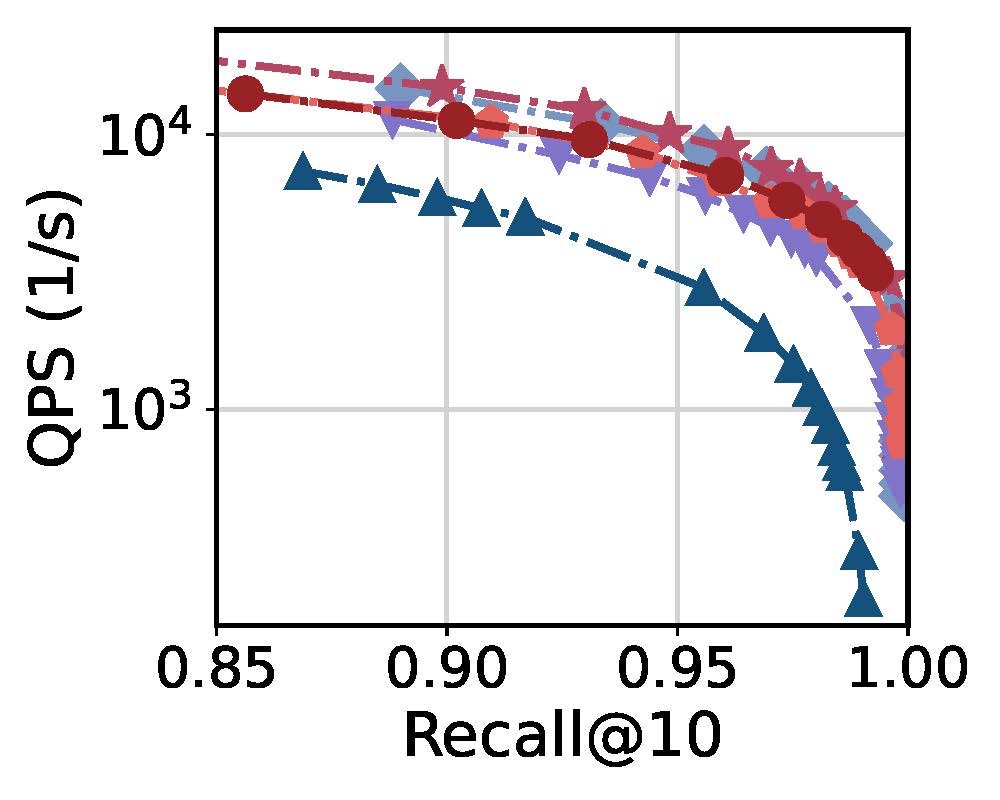}
		\hspace{-0.73em}
	}
	\subfigure[][{\scriptsize Deep1M}]{
		\hspace{-0.73em}
		\includegraphics[width=0.173\textwidth]{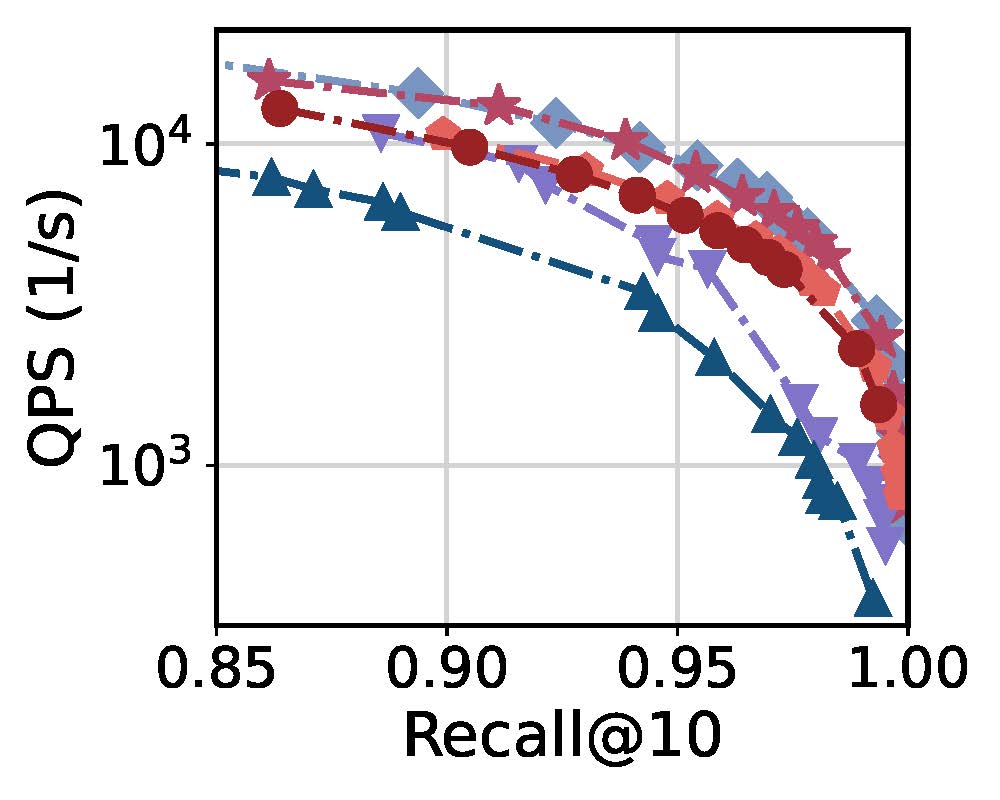}
		\hspace{-0.73em}
	}
	\subfigure[][{\scriptsize MSong}]{
		\hspace{-0.73em}
		\includegraphics[width=0.173\textwidth]{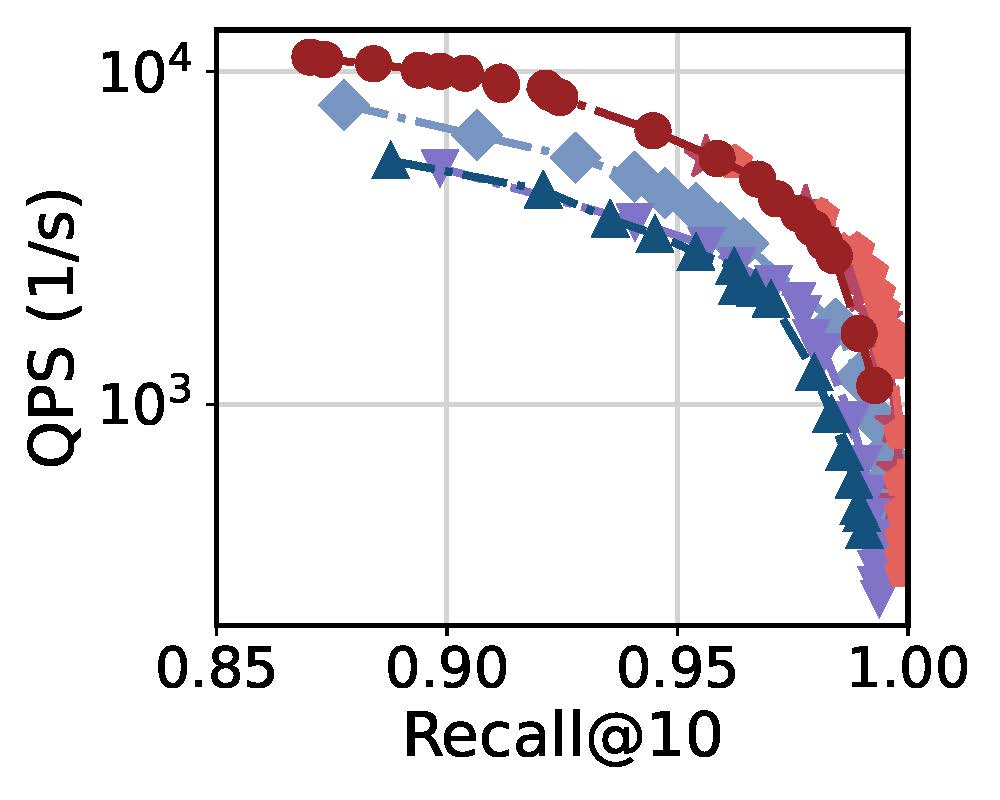}
		\hspace{-0.73em}
	}
	\subfigure[][{\scriptsize GloVe}]{
		\hspace{-0.73em}
		\includegraphics[width=0.173\textwidth]{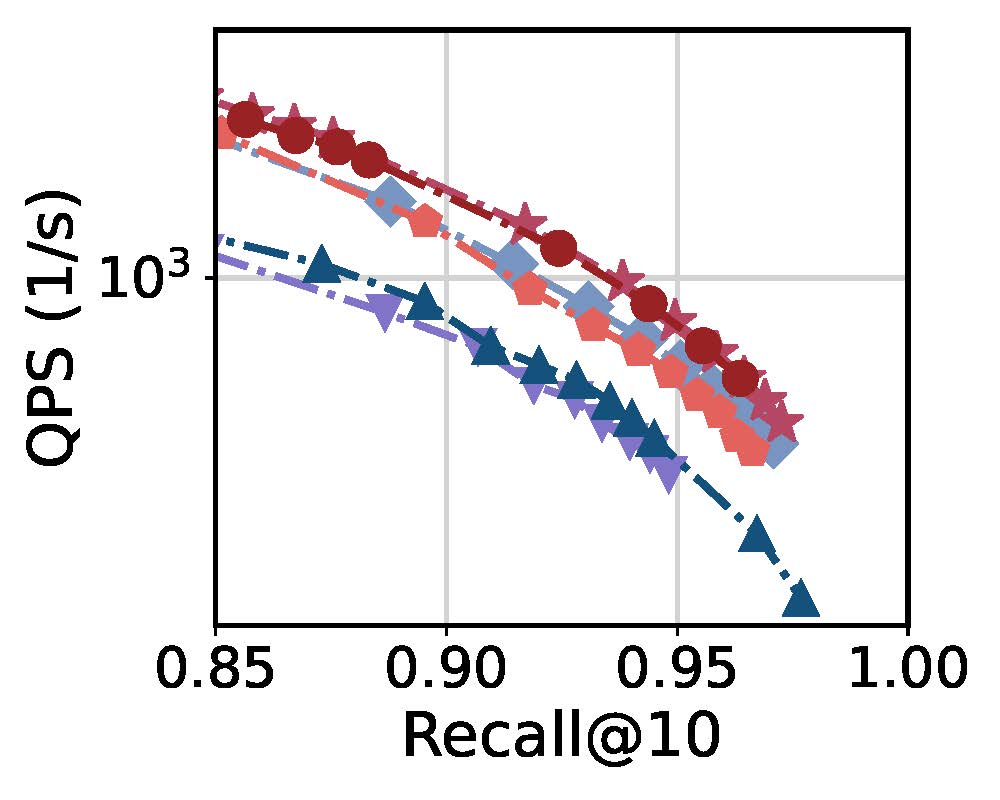}
		\hspace{-0.73em}
	}
	\subfigure[][{\scriptsize Gist1M}]{
		\hspace{-0.73em}
		\includegraphics[width=0.173\textwidth]{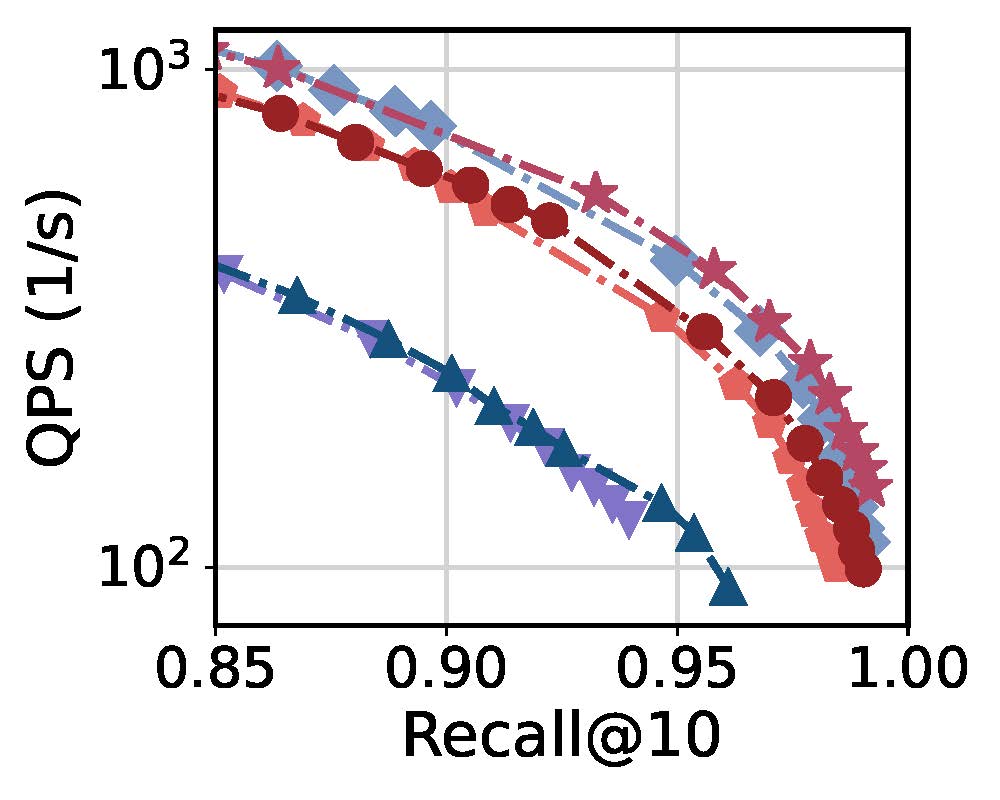}
		\hspace{-0.73em}
	}
	\subfigure[][{\scriptsize Crawl}]{
		\hspace{-0.73em}
		\includegraphics[width=0.173\textwidth]{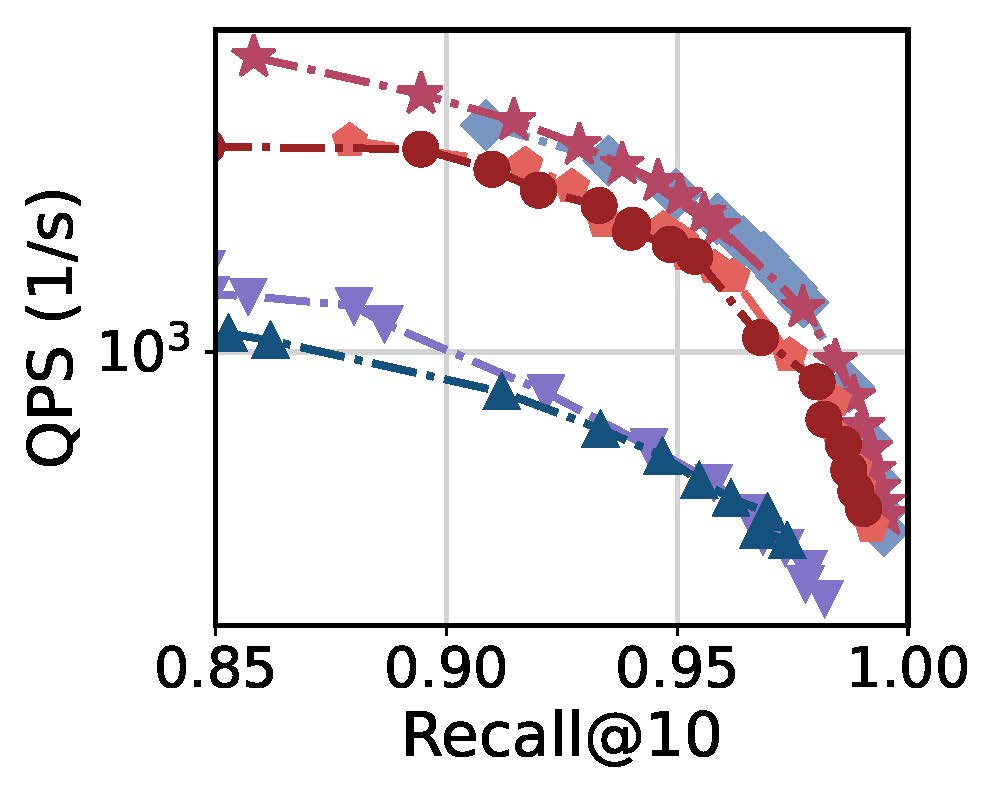}
		\hspace{-0.73em}
	}
	
	\figureCaptionMargin\vspace{-2ex}
	\caption{Search performance for our merging methods and the original methods rebuilt from scratch. (a--f: $Recall@10$-QPS; Exp.~7)}
	\label{fig:app_search}
\end{figure}

\noindent\textbf{Exp.7: Search Performance.} As depicted in Figure~\ref{fig:app_search}, FGIM-Vamana, FGIM-$\tau$-MNG, FGIM-NSW, and FGIM-NNDescent consistently achieved comparable or superior performance compared to their original approaches. This improvement stems from the more accurate initialization of the $k$-NNG in our method. For instance, while Vamana starts from a randomly initialized graph, FGIM-Vamana begins with an approximate $k$-NNG obtained via the PG-to-$k$-NNG transformation. This better starting graph enables our method to identify more accurate $k$-nearest neighbors, improving graph quality and explaining FGIM-Vamana’s superior performance on some datasets.

\begin{table}[t]
	\small
	\centering
	\caption{Effectiveness of FGIM under the deletion and overlap settings on Sift1M. (Exp.~8)}
	\begin{tabular}{lcccc}
		\toprule
		\multirow{2}{*}{\textbf{Ratio}} 
		& \multicolumn{2}{c}{\textbf{Deletion Setting}} 
		& \multicolumn{2}{c}{\textbf{Overlap Setting}} \\
		\cmidrule(lr){2-3} \cmidrule(lr){4-5}
		& \textbf{Recall@10} & \textbf{QPS} & \textbf{Recall@10} & \textbf{QPS} \\
		\midrule
		1/4 & 99.50\% & 3354 & 99.49\% & 3216 \\
		1/3 & 99.49\% & 3365 & 99.48\% & 3341 \\
		1/2 & 99.43\% & 3466 & 99.43\% & 3392 \\
		\bottomrule
	\end{tabular}
	\label{tab:deletion_overlap}
	\vspace{-2ex}
\end{table}

\noindent\textbf{Exp.8: Deletion and Overlap Handling.}  In dynamic scenarios of modern vector database systems, it is essential to verify that FGIM can robustly handle deletions and overlapping data. To evaluate this, we conduct experiments on the Sift1M dataset. For each case, we build an index on the full dataset and another on a subset determined by a given ratio. In the deletion scenario, the subset represents deleted data, while in the overlap scenario, it overlaps with the existing data. As shown in Table~\ref{tab:deletion_overlap}, FGIM effectively manages both deletions and overlaps, maintaining high search accuracy and efficiency across different ratios. These results demonstrate the robustness of our framework in dynamic settings.

Overall, these findings demonstrate the high applicability of our methods, which are capable of efficiently and effectively merging graph-based indexes while preserving search accuracy.

\begin{figure}[t]
	\centering
	\begin{minipage}{0.5\textwidth}
		\centering
		\includegraphics[width=0.65\textwidth]{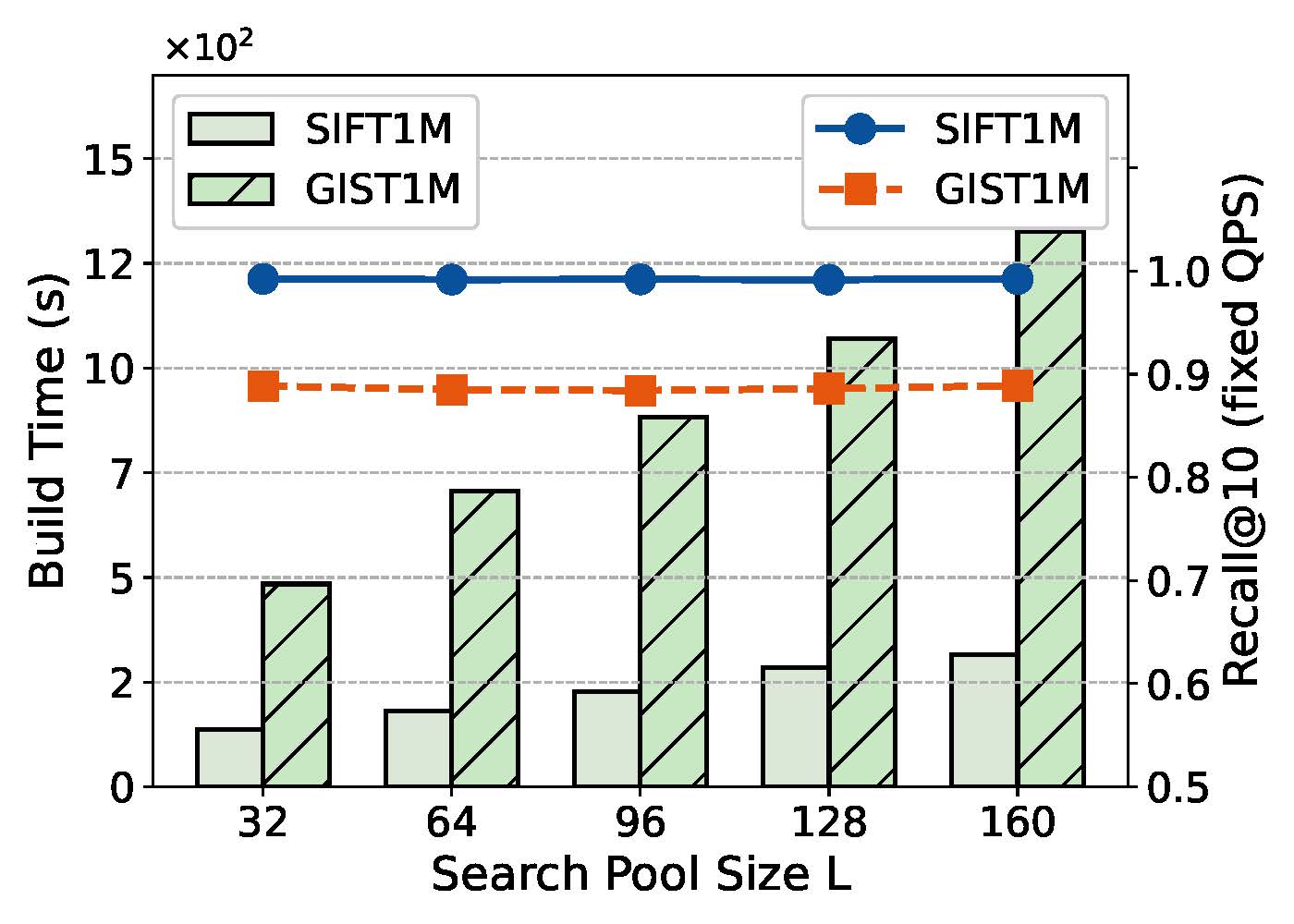}
		\vspace{-2ex}
		\captionof{figure}{Effects of search pool size. (Exp.~9)}
		\label{fig:ablation0}
	\end{minipage}
	\hspace{-3.5ex}
	\begin{minipage}{0.5\textwidth}
		\centering
		\includegraphics[width=0.65\textwidth]{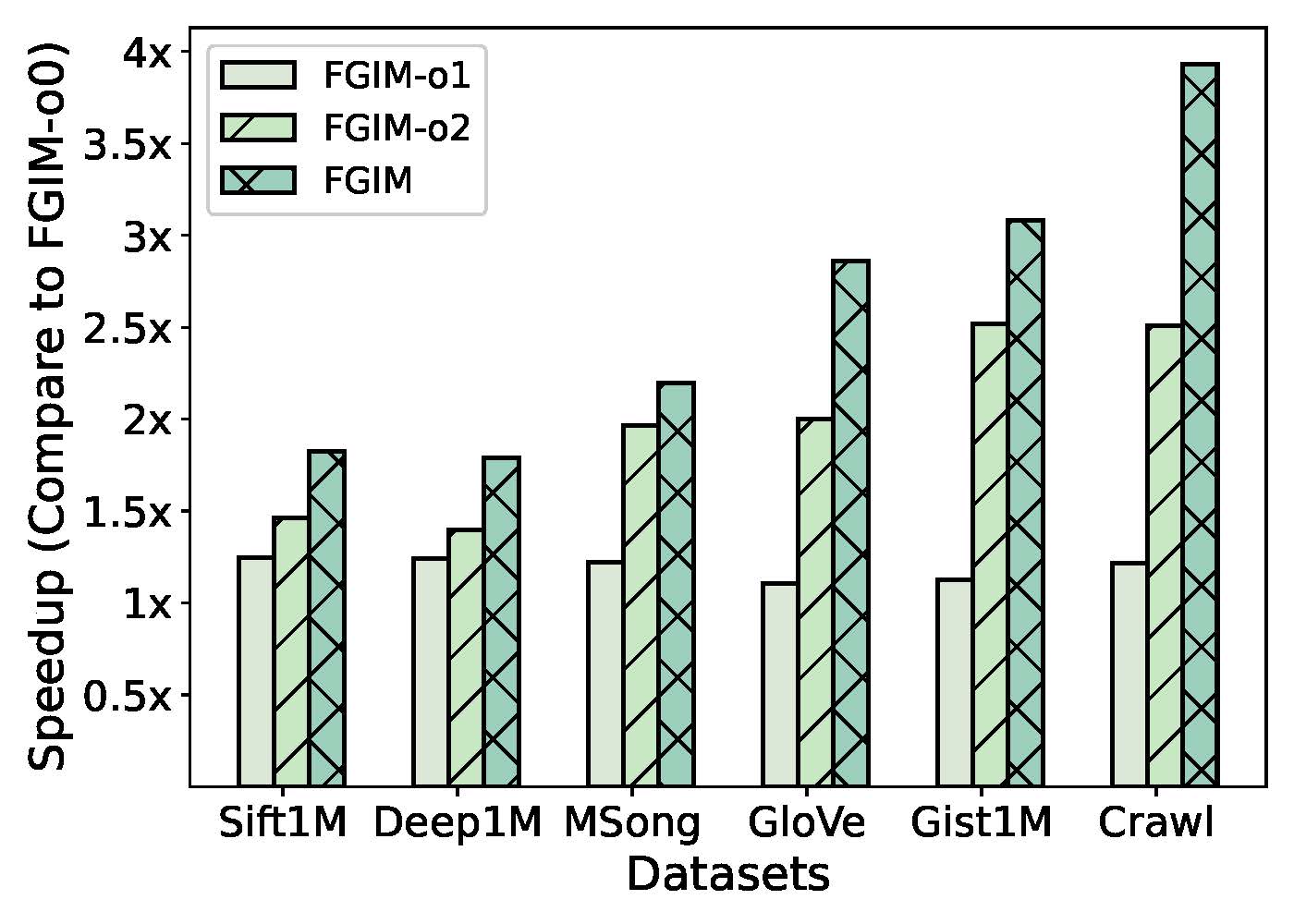}
		\vspace{-2ex}
		\captionof{figure}{Effects of acceleration techniques. (Exp.~10)}
		\label{fig:ablation1}
	\end{minipage}
\end{figure}

\begin{figure}[t]
	\centering
	\begin{minipage}{0.5\textwidth}
		\centering
		\includegraphics[width=0.65\textwidth]{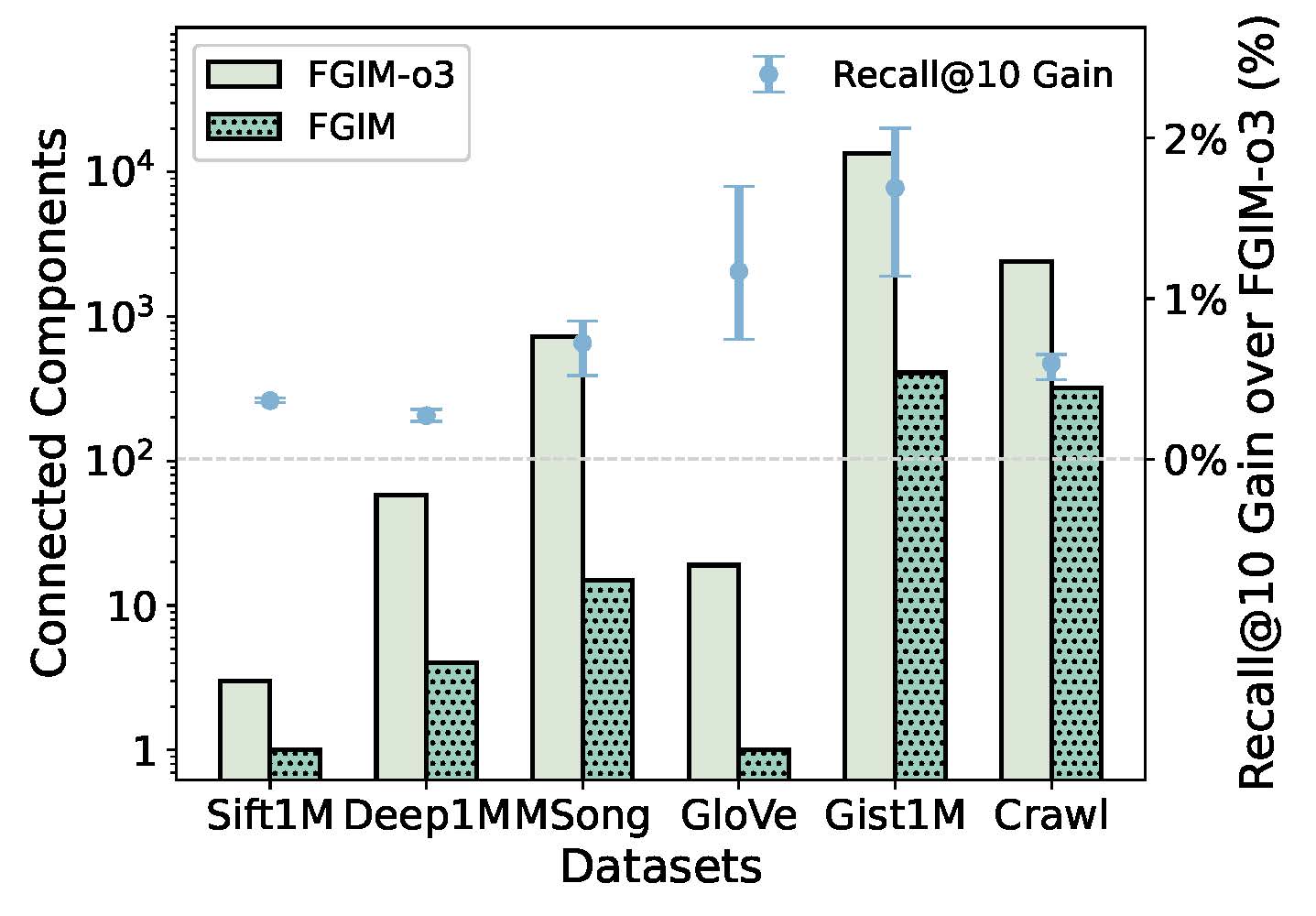}
		\vspace{-2ex}
		\captionof{figure}{Effects of indegree repair mechanism. (Exp.~11)}
		\label{fig:ablation2}
	\end{minipage}
	\hspace{-3.5ex}
	\begin{minipage}{0.5\textwidth}
		\vspace{-2ex}
		\centering
		\includegraphics[width=0.65\textwidth]{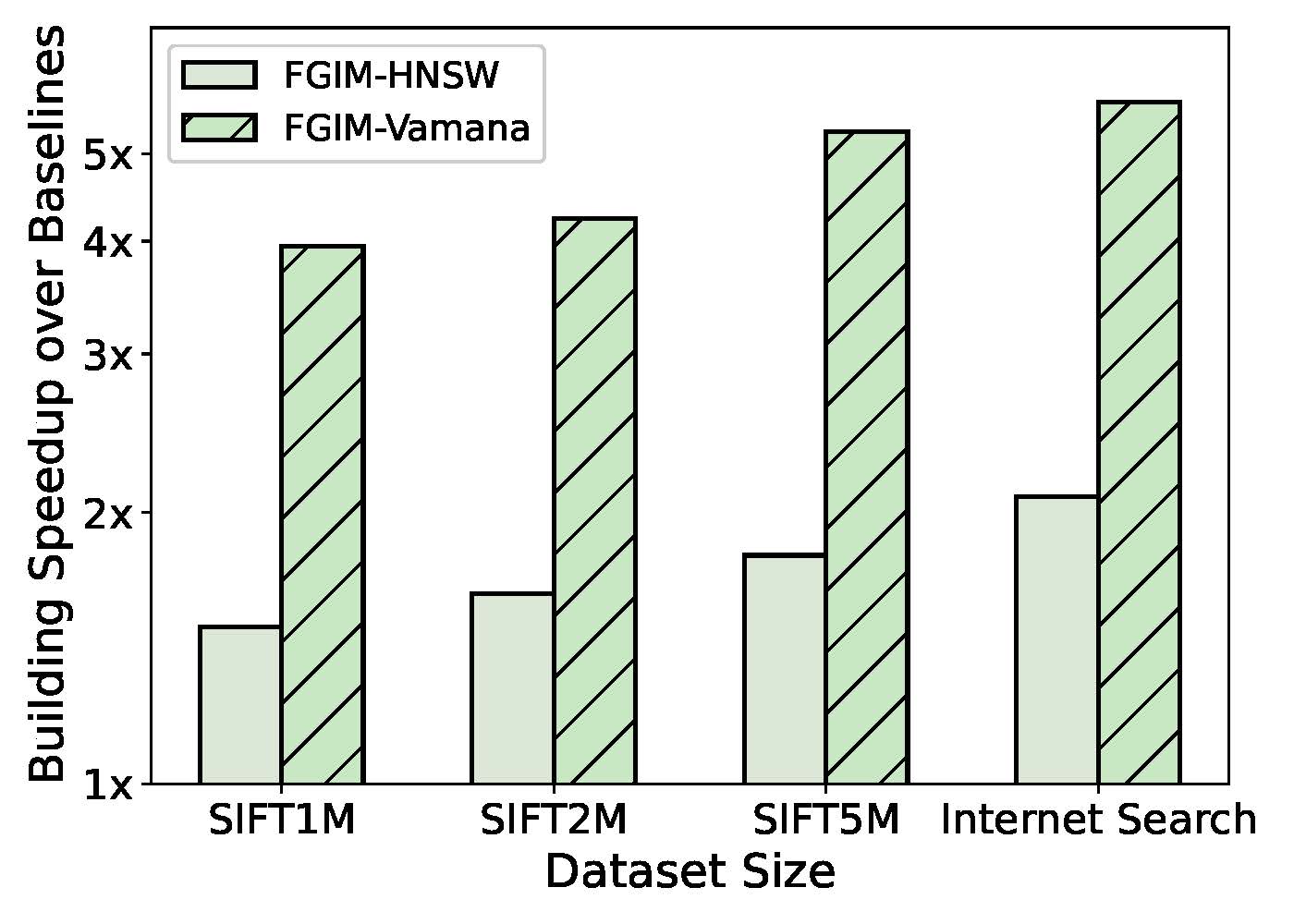}
		\vspace{-2ex}
		\captionof{figure}{Scalability study. (Exp.~12)}
		\label{fig:scala}
	\end{minipage}
	\vspace{-1ex}
\end{figure}

\subsection{Ablation Analysis}
\label{subsec:ablation}

In this part, we assess the effects of our techniques used in the framework, i.e., the \textit{Minimum Querying Strategy}, the \textit{cross-querying}, the \textit{streamlined refinement}, and the \textit{indegree repair mechanism}.

\noindent\textbf{Exp.9: Effects of Minimum Querying Strategy.} According to our strategy, when selecting cross-graph candidate neighbors, we set the search pool size $L$ to the minimum feasible value. A key question is whether this affects the quality of the graph. To investigate this, we increase $L$ while keeping QPS fixed (4000 for Sift1M and 800 for Gist) and observe the changes in merging time and $Recall@10$, as shown in Figure~\ref{fig:ablation0}. Our findings demonstrate that this strategy achieves the shortest index merging time without compromising graph quality, which can be explained that the iterative refinement allows the graph structure to converge to a similar structure \cite{dong2011efficient}, thus mitigating the impact of initial candidate selection.

\noindent\textbf{Exp.10: Effects of Optimization Technique.} In this part, we evaluate the effects of the \textit{cross-querying} and the \textit{streamlined refinement}. We define four variants for comparison: FGIM-o0 without any technique; FGIM-o1, which incorporates only \textit{cross-querying} and adopts the original refinement; FGIM-o2, which employs only \textit{streamlined refinement} and initializes the merged index by randomly introducing vertices from the other indexes. We report the speedup achieved by each method over FGIM-o0 in Figure~\ref{fig:ablation1}. Overall, each technique contributes to a significant acceleration of the merging process. Since these optimizations are mutually independent, their combined application FGIM yields the lowest construction cost.

\noindent\textbf{Exp.11: Effects of Indegree Repair Mechanism.} Figure~\ref{fig:ablation2} evaluates the effects of the \textit{indegree repair mechanism}, reporting the Strongly Connected Components (SCC) in the merged index and the $Recall@10$ improvement over the method without this mechanism (i.e., FGIM-o3). Overall, when this mechanism is applied, the number of strongly connected components decreases significantly, highlighting the effectiveness in improving graph connectivity within the degree budget. Besides, we observe that the $Recall@10$ is improved across all datasets, demonstrating the effectiveness of this mechanism in enhancing search performance.

\begin{figure}[t]
	\centering
	\includegraphics[width=0.6\textwidth]{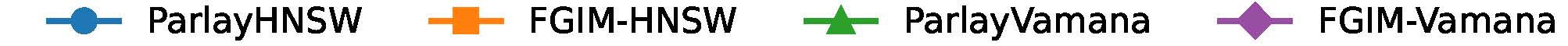}
	
	\subfigure[][{\scriptsize Sift1M}]{
		\includegraphics[width=0.28\textwidth]{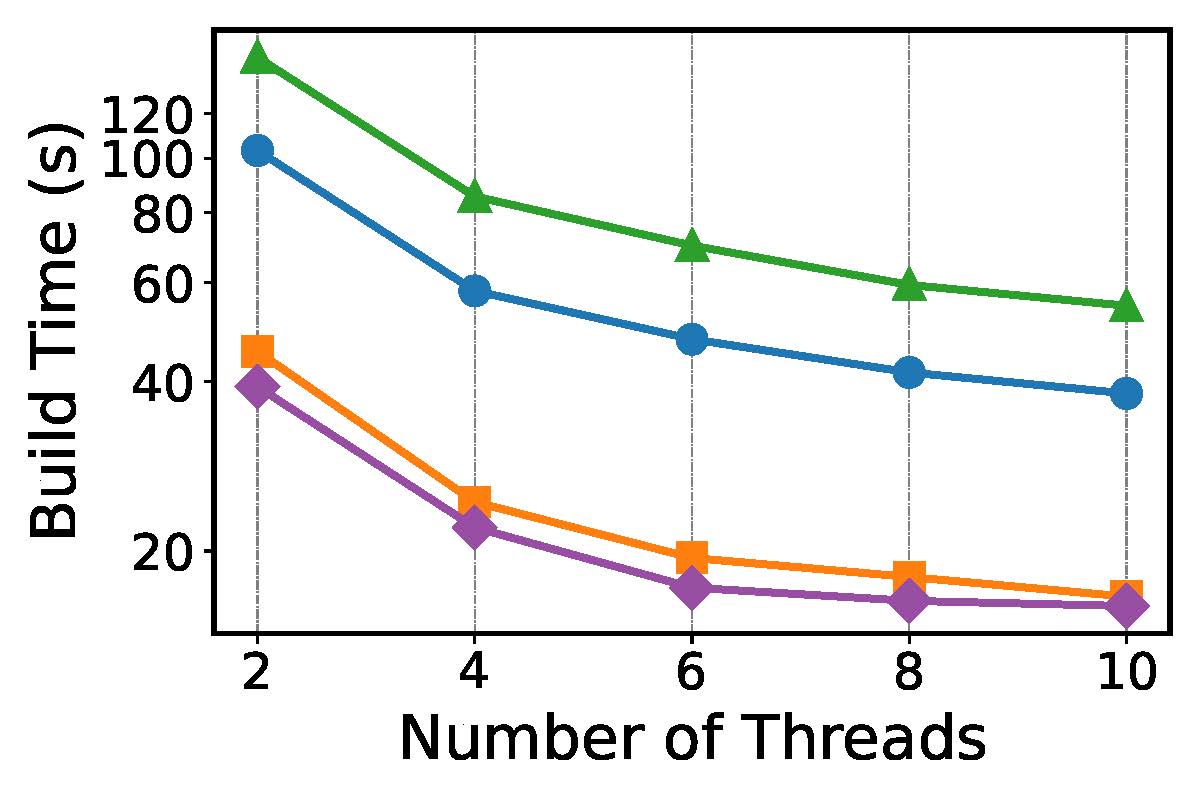}
		\label{fig:sift_multithread}
	}
	\hspace{3ex}
	\subfigure[][{\scriptsize Gist1M}]{
		\includegraphics[width=0.28\textwidth]{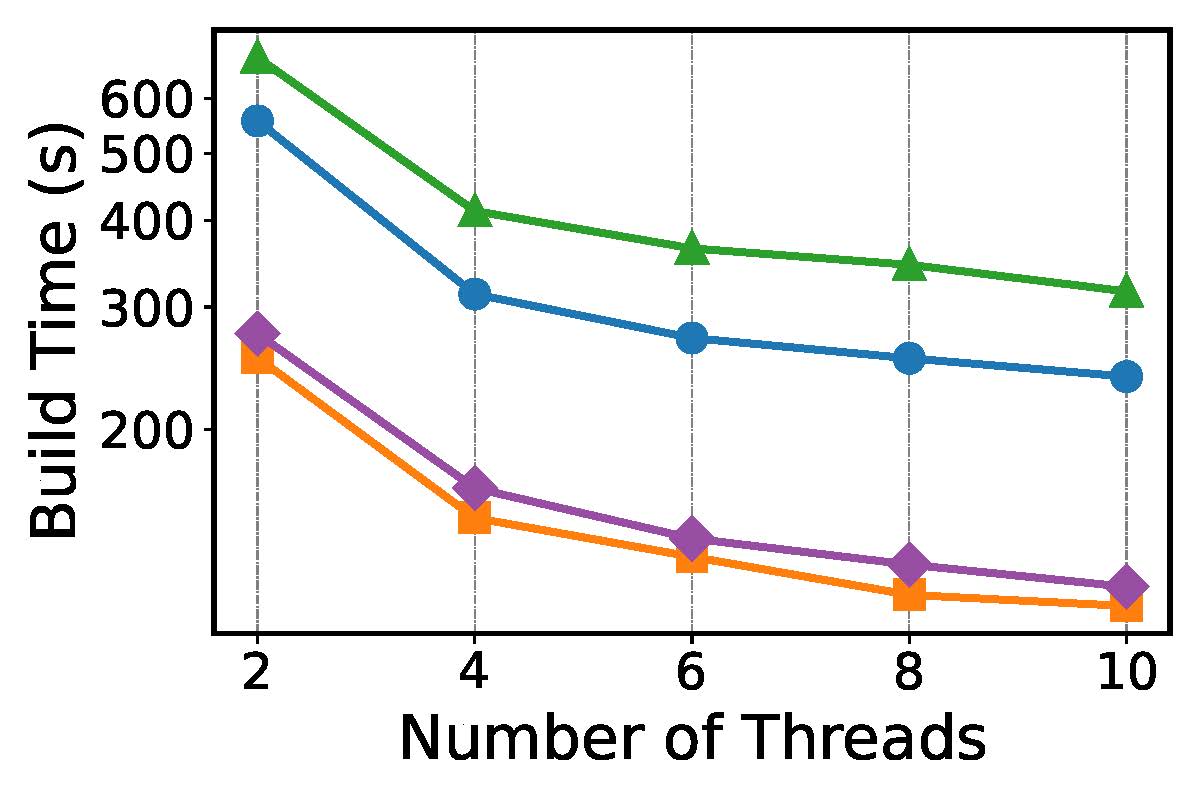}
		\label{fig:gist_multithread}
	}
	
	\vspace{-2ex}
	\caption{Comparison of multi-threading efficiency. (Exp.~13)}
	\label{fig:multithread}
	\vspace{-1ex}
\end{figure}

\subsection{Scalability Study}
\label{subsec:scalability}

\noindent\textbf{Exp.12: Scalability Study.} Figure~\ref{fig:scala} evaluates the scalability of our framework on the Sift and Internet Search dataset, with HNSW and Vamana as baselines. The results indicate that our method exhibits strong scalability as the number of vectors increases. Notably, the speedup achieved by our approach becomes amplified with larger dataset sizes, highlighting its superior efficiency in handling large-scale data. These findings demonstrate the robustness of our framework, making it well-suited for large-scale applications.

\noindent\textbf{Exp.13: Multi-threaded Merging Efficiency.} We use the parallel incremental insertion implementations of ParlayHNSW and ParlayVamana (based on ParlayANN) as baselines, and compare the merge efficiency of our framework under 2\textasciitilde10 threads, as shown in Figure~\ref{fig:multithread}. The results demonstrate that FGIM maintains high multi-thread efficiency across datasets and indexing methods, indicating that its efficiency can be further improved with increasing parallelism, while incurring only modest parallelization overhead.

\section{Related Work}
\label{sec:related}

Recently, with the rise of \textit{Large Language Models} (LLMs) and \textit{retrieval-augmented generation} (RAG), ANNS \cite{arya1993approximate, yang2025effective, yang2024fast} has attracted increasing attention, acting as a crucial component in these applications. Generally, ANNS methods are devoted to efficiently finding the most similar data points to a query in large-scale, high-dimensional datasets. Recent works \cite{naidan2015permutation,aumuller2020ann} have shown that graph-based methods \cite{fu2017fast,malkov2018efficient,jayaram2019diskann, zhong2025enhancegraph} outperform traditional methods such as hashing-based methods \cite{indyk1998approximate,sun2014srs,huang2015query}, tree-based methods \cite{bentley1975multidimensional,silpa2008optimised,muja2014scalable}, inverted index-based methods \cite{babenko2014inverted,babenko2016efficient,zhang2022bi, li2025sindi}, and quantization-based methods \cite{jegou2010product, ge2013optimized} in terms of search quality and efficiency.

The mainstream graph-based methods \cite{malkov2014approximate,malkov2018efficient,fu2017fast,fu2021high,peng2023efficient,jayaram2019diskann, cai2024navigating} statically build a proximity graph where each node is a base vector and edges connect several nearby vectors. Among them, HNSW \cite{malkov2018efficient} is a widely used graph-based method that is derived from small-world networks and is constructed incrementally in an online fashion. Other SOTA methods, such as NSG \cite{fu2017fast} and $\tau$-MNG \cite{peng2023efficient}, construct indexes by performing refinement operations on the existing graph (e.g., refining a pre-constructed $k$-NNG). Vamana\cite{jayaram2019diskann} achieves improved search performance by relaxing the neighbor selection policy to retain long-range edges. Due to their superior performance, these methods have been widely adopted in production \cite{wang2021milvus, Yang2024EffectiveAG}. Our proposed framework is compatible with the abovementioned methods and can be adapted to merge their indexes.

To meet the demands of real-time systems in practical applications, several vector search systems~\cite{singh2021freshdiskann,xu2023spfresh, zhong2025lsm, yu2025topology, xu2025place} support dynamic updates. SPFresh \cite{xu2023spfresh} utilizes a clustering-based index but demonstrates diminished performance in high-dimensional vector spaces. FreshDiskANN\cite{singh2021freshdiskann}, IP-DiskANN\cite{xu2025place}, and Greator \cite{yu2025topology} enable disk-based index updates. However, incremental methods fail to fully leverage the information from the existing indexes in index merging contexts, resulting in significant overhead.

\section{Conclusion}
\label{sec:conclusion}

In this paper, we study the problem of merging existing graph-based indexes into a single one for effective ANNS. We propose a general FGIM framework with three core techniques: 1) a \textit{PGs to $k$-NNG transformation} that consists of \textit{cross-querying} and \textit{minimum querying Strategy} to extract initial candidate neighbors from the existing graph-based indexes; 2) a streamlined and indegree-aware \textit{$k$-NNG refinement} method to improve candidate neighbors' quality; 3) a \textit{$k$-NNG to PG transformation} and HNSW-adaptive process are presented for better ANNS performance. Experimental results demonstrate the generality and effectiveness of our FGIM framework, yielding noticeable speedups over SOTA methods without compromising the search performance.

\section{Acknowledgment}
\label{sec:ack}

This work was supported by Ant Group. Peng Cheng, Jingkuan Song and Heng Tao Shen were supported by Fundamental Research Funds for the Central Universities, Shanghai Central and Local Science and Technology Development Fund Project (Grant No. YDZX20253100002004).
Yongxin Tong was partially supported by National Science Foundation of China (NSFC) (Grant Nos. 62425202, U21A20516, 62336003), the Beijing Natural Science Foundation (Z230001). 
Lei Chen’s work is partially supported by National Key Research and Development Program of China Grant No. 2023YFF0725100, National Science Foundation of China (NSFC) under Grant No. U22B2060, Guangdong-Hong Kong Technology Innovation Joint Funding Scheme Project No. 2024A0505040012, the Hong Kong RGC GRF Project 16213620, RIF Project R6020-19, AOE Project AoE/E-603/18, Theme-based project TRS T41-603/20R, CRF Project C2004-21G, Key Areas Special Project of Guangdong Provincial  Universities 2024ZDZX1006,  Guangdong Province Science and Technology Plan Project 2023A0505030011, Guangzhou municipality big data intelligence key lab, 2023A03J0012, Hong Kong ITC ITF grants MHX/078/21 and PRP/004/22FX, Zhujiang scholar program 2021JC02X170, Microsoft Research Asia Collaborative Research Grant, HKUST-Webank joint research lab and 2023 HKUST Shenzhen-Hong Kong Collaborative Innovation Institute Green Sustainability Special Fund from Shui On Xintiandi and the InnoSpace GBA.
Xuemin Lin was supported by NSFC U2241211.

\newpage
\bibliographystyle{ACM-Reference-Format}
\bibliography{add}

@inproceedings{dong2011efficient,
  title     = {Efficient k-nearest neighbor graph construction for
               generic similarity measures},
  author    = {Dong, Wei and Moses, Charikar and Li, Kai},
  booktitle = {Proceedings of the 20th international conference on World
               wide web},
  pages     = {577--586},
  year      = {2011}
}

@article{fu2016efanna,
  title   = {Efanna: An extremely fast approximate nearest neighbor
             search algorithm based on knn graph},
  author  = {Fu, Cong and Cai, Deng},
  journal = {arXiv preprint arXiv:1609.07228},
  year    = {2016}
}

@article{malkov2014approximate,
  title     = {Approximate nearest neighbor algorithm based on navigable
               small world graphs},
  author    = {Malkov, Yury and Ponomarenko, Alexander and Logvinov,
               Andrey and Krylov, Vladimir},
  journal   = {Information Systems},
  volume    = {45},
  pages     = {61--68},
  year      = {2014},
  publisher = {Elsevier}
}

@article{malkov2018efficient,
  title     = {Efficient and robust approximate nearest neighbor search
               using hierarchical navigable small world graphs},
  author    = {Malkov, Yu A and Yashunin, Dmitry A},
  journal   = {IEEE transactions on pattern analysis and machine
               intelligence},
  volume    = {42},
  number    = {4},
  pages     = {824--836},
  year      = {2018},
  publisher = {IEEE}
}

@article{zhao2021merge,
  title     = {On the Merge of k-NN Graph},
  author    = {Zhao, Wan-Lei and Wang, Hui and Lin, Peng-Cheng and Ngo,
               Chong-Wah},
  journal   = {IEEE Transactions on Big Data},
  volume    = {8},
  number    = {6},
  pages     = {1496--1510},
  year      = {2021},
  publisher = {IEEE}
}

@inproceedings{wang2012scalable,
  title        = {Scalable k-nn graph construction for visual descriptors},
  author       = {Wang, Jing and Wang, Jingdong and Zeng, Gang and Tu,
                  Zhuowen and Gan, Rui and Li, Shipeng},
  booktitle    = {2012 IEEE Conference on Computer Vision and Pattern
                  Recognition},
  pages        = {1106--1113},
  year         = {2012},
  organization = {IEEE}
}

@article{wang2020survey,
  title     = {A survey on large-scale machine learning},
  author    = {Wang, Meng and Fu, Weijie and He, Xiangnan and Hao, Shijie
               and Wu, Xindong},
  journal   = {IEEE Transactions on Knowledge and Data Engineering},
  volume    = {34},
  number    = {6},
  pages     = {2574--2594},
  year      = {2020},
  publisher = {IEEE}
}

@article{jegou2010product,
  title     = {Product quantization for nearest neighbor search},
  author    = {Jegou, Herve and Douze, Matthijs and Schmid, Cordelia},
  journal   = {IEEE transactions on pattern analysis and machine
               intelligence},
  volume    = {33},
  number    = {1},
  pages     = {117--128},
  year      = {2010},
  publisher = {IEEE}
}

@article{babenko2014inverted,
  title     = {The inverted multi-index},
  author    = {Babenko, Artem and Lempitsky, Victor},
  journal   = {IEEE transactions on pattern analysis and machine
               intelligence},
  volume    = {37},
  number    = {6},
  pages     = {1247--1260},
  year      = {2014},
  publisher = {IEEE}
}

@article{zhang2022bi,
  title   = {Bi-Phase Enhanced IVFPQ for Time-Efficient Ad-hoc
             Retrieval},
  author  = {Zhang, Peitian and Liu, Zheng},
  journal = {arXiv preprint arXiv:2210.05521},
  year    = {2022}
}

@article{toussaint1980relative,
  title     = {The relative neighbourhood graph of a finite planar set},
  author    = {Toussaint, Godfried T},
  journal   = {Pattern recognition},
  volume    = {12},
  number    = {4},
  pages     = {261--268},
  year      = {1980},
  publisher = {Elsevier}
}

@inproceedings{babenko2016efficient,
  title     = {Efficient indexing of billion-scale datasets of deep
               descriptors},
  author    = {Babenko, Artem and Lempitsky, Victor},
  booktitle = {Proceedings of the IEEE Conference on Computer Vision and
               Pattern Recognition},
  pages     = {2055--2063},
  year      = {2016}
}

@article{fu2017fast,
  title   = {Fast Approximate Nearest Neighbor Search With The
             Navigating Spreading-out Graph},
  author  = {Fu, Cong and Xiang, Chao and Wang, Changxu and Cai, Deng},
  journal = {Proceedings of the VLDB Endowment},
  volume  = {12},
  number  = {5},
  year    = {2017}
}

@article{jayaram2019diskann,
  title   = {Diskann: Fast accurate billion-point nearest neighbor
             search on a single node},
  author  = {Jayaram Subramanya, Suhas and Devvrit, Fnu and Simhadri,
             Harsha Vardhan and Krishnawamy, Ravishankar and Kadekodi,
             Rohan},
  journal = {Advances in Neural Information Processing Systems},
  volume  = {32},
  year    = {2019}
}

@article{fu2021high,
  title     = {High dimensional similarity search with satellite system
               graph: Efficiency, scalability, and unindexed query
               compatibility},
  author    = {Fu, Cong and Wang, Changxu and Cai, Deng},
  journal   = {IEEE Transactions on Pattern Analysis and Machine
               Intelligence},
  volume    = {44},
  number    = {8},
  pages     = {4139--4150},
  year      = {2021},
  publisher = {IEEE}
}

@article{yang2024revisiting,
  author    = {Yang, Shuo and Xie, Jiadong and Liu, Yingfan and Yu,
               Jeffrey Xu and Gao, Xiyue and Wang, Qianru and Peng, Yanguo
               and Cui, Jiangtao},
  title     = {Revisiting the Index Construction of Proximity Graph-Based
               Approximate Nearest Neighbor Search},
  year      = {2025},
  journal   = {Proceedings of the VLDB Endowment},
  volume    = {18},
  pages     = {1825--1838},
  publisher = {VLDB Endowment}
}

@inproceedings{indyk1998approximate,
  title     = {Approximate nearest neighbors: towards removing the curse
               of dimensionality},
  author    = {Indyk, Piotr and Motwani, Rajeev},
  booktitle = {Proceedings of the thirtieth annual ACM symposium on
               Theory of computing},
  pages     = {604--613},
  year      = {1998}
}

@article{sun2014srs,
  title   = {SRS: solving c-approximate nearest neighbor queries in
             high dimensional euclidean space with a tiny index},
  author  = {Sun, Yifang and Wang, Wei and Qin, Jianbin and Zhang, Ying
             and Lin, Xuemin},
  journal = {Proceedings of the VLDB Endowment},
  year    = {2014}
}

@article{huang2015query,
  title     = {Query-aware locality-sensitive hashing for approximate
               nearest neighbor search},
  author    = {Huang, Qiang and Feng, Jianlin and Zhang, Yikai and Fang,
               Qiong and Ng, Wilfred},
  journal   = {Proceedings of the VLDB Endowment},
  volume    = {9},
  number    = {1},
  pages     = {1--12},
  year      = {2015},
  publisher = {VLDB Endowment}
}

@article{naidan2015permutation,
  title   = {Permutation Search Methods are Efficient, Yet Faster
             Search is Possible},
  author  = {Naidan, Bilegsaikhan and Boytsov, Leonid and Nyberg,
             Eric},
  journal = {Proceedings of the VLDB Endowment},
  volume  = {8},
  number  = {12},
  year    = {2015}
}

@article{aumuller2020ann,
  title     = {ANN-Benchmarks: A benchmarking tool for approximate
               nearest neighbor algorithms},
  author    = {Aum{\"u}ller, Martin and Bernhardsson, Erik and Faithfull,
               Alexander},
  journal   = {Information Systems},
  volume    = {87},
  pages     = {101374},
  year      = {2020},
  publisher = {Elsevier}
}

@article{bentley1975multidimensional,
  title     = {Multidimensional binary search trees used for associative
               searching},
  author    = {Bentley, Jon Louis},
  journal   = {Communications of the ACM},
  volume    = {18},
  number    = {9},
  pages     = {509--517},
  year      = {1975},
  publisher = {ACM New York, NY, USA}
}

@inproceedings{silpa2008optimised,
  title        = {Optimised KD-trees for fast image descriptor matching},
  author       = {Silpa-Anan, Chanop and Hartley, Richard},
  booktitle    = {2008 IEEE Conference on Computer Vision and Pattern
                  Recognition},
  pages        = {1--8},
  year         = {2008},
  organization = {IEEE}
}

@article{muja2014scalable,
  title     = {Scalable nearest neighbor algorithms for high dimensional
               data},
  author    = {Muja, Marius and Lowe, David G},
  journal   = {IEEE transactions on pattern analysis and machine
               intelligence},
  volume    = {36},
  number    = {11},
  pages     = {2227--2240},
  year      = {2014},
  publisher = {IEEE}
}

@article{ge2013optimized,
  title     = {Optimized product quantization},
  author    = {Ge, Tiezheng and He, Kaiming and Ke, Qifa and Sun, Jian},
  journal   = {IEEE transactions on pattern analysis and machine
               intelligence},
  volume    = {36},
  number    = {4},
  pages     = {744--755},
  year      = {2013},
  publisher = {IEEE}
}

@inproceedings{ootomo2024cagra,
  title        = {Cagra: Highly parallel graph construction and approximate
                  nearest neighbor search for gpus},
  author       = {Ootomo, Hiroyuki and Naruse, Akira and Nolet, Corey and
                  Wang, Ray and Feher, Tamas and Wang, Yong},
  booktitle    = {2024 IEEE 40th International Conference on Data
                  Engineering (ICDE)},
  pages        = {4236--4247},
  year         = {2024},
  organization = {IEEE}
}

@inproceedings{wang2021milvus,
  title     = {Milvus: A purpose-built vector data management system},
  author    = {Wang, Jianguo and Yi, Xiaomeng and Guo, Rentong and Jin,
               Hai and Xu, Peng and Li, Shengjun and Wang, Xiangyu and
               Guo, Xiangzhou and Li, Chengming and Xu, Xiaohai and
               others},
  booktitle = {Proceedings of the 2021 International Conference on
               Management of Data},
  pages     = {2614--2627},
  year      = {2021}
}

@article{cost1993weighted,
  title     = {A weighted nearest neighbor algorithm for learning with
               symbolic features},
  author    = {Cost, Scott and Salzberg, Steven},
  journal   = {Machine learning},
  volume    = {10},
  pages     = {57--78},
  year      = {1993},
  publisher = {Springer}
}

@article{wang2021comprehensive,
  title     = {A comprehensive survey and experimental comparison of
               graph-based approximate nearest neighbor search},
  author    = {Wang, Mengzhao and Xu, Xiaoliang and Yue, Qiang and Wang,
               Yuxiang},
  journal   = {Proceedings of the VLDB Endowment},
  volume    = {14},
  number    = {11},
  pages     = {1964--1978},
  year      = {2021},
  publisher = {VLDB Endowment}
}

@article{li2019approximate,
  title     = {Approximate nearest neighbor search on high dimensional
               data—experiments, analyses, and improvement},
  author    = {Li, Wen and Zhang, Ying and Sun, Yifang and Wang, Wei and
               Li, Mingjie and Zhang, Wenjie and Lin, Xuemin},
  journal   = {IEEE Transactions on Knowledge and Data Engineering},
  volume    = {32},
  number    = {8},
  pages     = {1475--1488},
  year      = {2019},
  publisher = {IEEE}
}

@inproceedings{bertin2011million,
  author    = {Thierry Bertin-Mahieux and Daniel P.W. Ellis and Brian
               Whitman and Paul Lamere},
  title     = {The Million Song Dataset},
  booktitle = {{Proceedings of the 12th International Conference on Music
               Information Retrieval ({ISMIR} 2011)}},
  year      = {2011}
}

@inproceedings{pennington2014glove,
  title     = {Glove: Global vectors for word representation},
  author    = {Pennington, Jeffrey and Socher, Richard and Manning,
               Christopher D},
  booktitle = {Proceedings of the 2014 conference on empirical methods in
               natural language processing (EMNLP)},
  pages     = {1532--1543},
  year      = {2014}
}

@misc{commoncrawl,
  author       = {Anon.},
  title        = {Common Crawl},
  year         = {Retrieved April 15, 2020},
  howpublished = {\url{http://commoncrawl.org/}}
}

@inproceedings{meng2020pmd,
  title        = {Pmd: An optimal transportation-based user distance for
                  recommender systems},
  author       = {Meng, Yitong and Dai, Xinyan and Yan, Xiao and Cheng,
                  James and Liu, Weiwen and Guo, Jun and Liao, Benben and
                  Chen, Guangyong},
  booktitle    = {Advances in Information Retrieval: 42nd European
                  Conference on IR Research, ECIR 2020, Lisbon, Portugal,
                  April 14--17, 2020, Proceedings, Part II 42},
  pages        = {272--280},
  year         = {2020},
  organization = {Springer}
}

@inproceedings{das2007google,
  title     = {Google news personalization: scalable online collaborative
               filtering},
  author    = {Das, Abhinandan S and Datar, Mayur and Garg, Ashutosh and
               Rajaram, Shyam},
  booktitle = {Proceedings of the 16th international conference on World
               Wide Web},
  pages     = {271--280},
  year      = {2007}
}

@inproceedings{arya1993approximate,
  title        = {Approximate nearest neighbor queries in fixed
                  dimensions.},
  author       = {Arya, Sunil and Mount, David M},
  booktitle    = {SODA},
  volume       = {93},
  pages        = {271--280},
  year         = {1993},
  organization = {Citeseer}
}

@article{arya1998optimal,
  title     = {An optimal algorithm for approximate nearest neighbor
               searching fixed dimensions},
  author    = {Arya, Sunil and Mount, David M and Netanyahu, Nathan S and
               Silverman, Ruth and Wu, Angela Y},
  journal   = {Journal of the ACM (JACM)},
  volume    = {45},
  number    = {6},
  pages     = {891--923},
  year      = {1998},
  publisher = {ACM New York, NY, USA}
}

@inproceedings{zhu2019accelerating,
  title        = {Accelerating large-scale molecular similarity search
                  through exploiting high performance computing},
  author       = {Zhu, Chun Jiang and Zhu, Tan and Li, Haining and Bi, Jinbo
                  and Song, Minghu},
  booktitle    = {2019 IEEE International Conference on Bioinformatics and
                  Biomedicine (BIBM)},
  pages        = {330--333},
  year         = {2019},
  organization = {IEEE}
}

@article{dobson2023scaling,
  title   = {Scaling Graph-Based ANNS Algorithms to Billion-Size
             Datasets: A Comparative Analysis},
  author  = {Dobson, Magdalen and Shen, Zheqi and Blelloch, Guy E and
             Dhulipala, Laxman and Gu, Yan and Simhadri, Harsha Vardhan
             and Sun, Yihan},
  journal = {arXiv preprint arXiv:2305.04359},
  year    = {2023}
}

@article{peng2023efficient,
  title     = {Efficient approximate nearest neighbor search in
               multi-dimensional databases},
  author    = {Peng, Yun and Choi, Byron and Chan, Tsz Nam and Yang,
               Jianye and Xu, Jianliang},
  journal   = {Proceedings of the ACM on Management of Data},
  volume    = {1},
  number    = {1},
  pages     = {1--27},
  year      = {2023},
  publisher = {ACM New York, NY, USA}
}

@inproceedings{hermenier2009entropy,
  title     = {Entropy: a consolidation manager for clusters},
  author    = {Hermenier, Fabien and Lorca, Xavier and Menaud, Jean-Marc
               and Muller, Gilles and Lawall, Julia},
  booktitle = {Proceedings of the 2009 ACM SIGPLAN/SIGOPS international
               conference on Virtual execution environments},
  pages     = {41--50},
  year      = {2009}
}

@article{lin2013consolidated,
  title     = {Consolidated cluster systems for data centers in the cloud
               age: a survey and analysis},
  author    = {Lin, Jian and Zha, Li and Xu, Zhiwei},
  journal   = {Frontiers of Computer Science},
  volume    = {7},
  pages     = {1--19},
  year      = {2013},
  publisher = {Springer}
}

@inproceedings{lee2014resource,
  author    = {Lee, Byungjun and Oh, Kyung Hwan and Park, Hee Jung and
               Kim, Ung Mo and Youn, Hee Yong},
  booktitle = {2014 International Conference on Cyber-Enabled Distributed
               Computing and Knowledge Discovery},
  title     = {Resource Reallocation of Virtual Machine in Cloud
               Computing with MCDM Algorithm},
  year      = {2014},
  volume    = {},
  number    = {},
  pages     = {470-477}
}

@inproceedings{asai2023retrieval,
  title     = {Retrieval-based language models and applications},
  author    = {Asai, Akari and Min, Sewon and Zhong, Zexuan and Chen,
               Danqi},
  booktitle = {Proceedings of the 61st Annual Meeting of the Association
               for Computational Linguistics (Volume 6: Tutorial
               Abstracts)},
  pages     = {41--46},
  year      = {2023}
}

@article{lewis2020retrieval,
  title   = {Retrieval-augmented generation for knowledge-intensive nlp
             tasks},
  author  = {Lewis, Patrick and Perez, Ethan and Piktus, Aleksandra and
             Petroni, Fabio and Karpukhin, Vladimir and Goyal, Naman and
             K{\"u}ttler, Heinrich and Lewis, Mike and Yih, Wen-tau and
             Rockt{\"a}schel, Tim and others},
  journal = {Advances in neural information processing systems},
  volume  = {33},
  pages   = {9459--9474},
  year    = {2020}
}

@article{o1996log,
  title     = {The log-structured merge-tree (LSM-tree)},
  author    = {O’Neil, Patrick and Cheng, Edward and Gawlick, Dieter
               and O’Neil, Elizabeth},
  journal   = {Acta Informatica},
  volume    = {33},
  pages     = {351--385},
  year      = {1996},
  publisher = {Springer}
}

@article{liu2024retrievalattention,
  title   = {Retrievalattention: Accelerating long-context llm
             inference via vector retrieval},
  author  = {Liu, Di and Chen, Meng and Lu, Baotong and Jiang, Huiqiang
             and Han, Zhenhua and Zhang, Qianxi and Chen, Qi and Zhang,
             Chengruidong and Ding, Bailu and Zhang, Kai and others},
  journal = {arXiv preprint arXiv:2409.10516},
  year    = {2024}
}

@article{chen2025maximum,
	title={Maximum Inner Product is Query-Scaled Nearest Neighbor},
	author={Chen, Tingyang and Fu, Cong and Wang, Kun and Ke, Xiangyu and Gao, Yunjun and Zhou, Wenchao and Ni, Yabo and Zeng, Anxiang},
	journal={Proceedings of the VLDB Endowment},
	volume={18},
	number={6},
	pages={1770--1783},
	year={2025},
	publisher={VLDB Endowment}
}

@inproceedings{yang2024effectiveag,
	title={Effective and general distance computation for approximate nearest neighbor search},
	author={Yang, Mingyu and Li, Wentao and Jin, Jiabao and Zhong, Xiaoyao and Wang, Xiangyu and Shen, Zhitao and Jia, Wei and Wang, Wei},
	booktitle={2025 IEEE 41st International Conference on Data Engineering (ICDE)},
	pages={1098--1110},
	year={2025},
	organization={IEEE}
}

@article{zhong2025vsag,
	title={VSAG: An Optimized Search Framework for Graph-Based Approximate Nearest Neighbor Search},
	author={Zhong, Xiaoyao and Li, Haotian and Jin, Jiabao and Yang, Mingyu and Chu, Deming and Wang, Xiangyu and Shen, Zhitao and Jia, Wei and Gu, George and Xie, Yi and Lin, Xuemin and Shen, Heng Tao and Song, Jingkuan and Cheng, Peng},
	journal={Proceedings of the VLDB Endowment},
	volume={18},
	number={12},
	pages={5017--5030},
	year={2025},
	publisher={VLDB Endowment}
}

@article{pugh1990skip,
  title     = {Skip lists: a probabilistic alternative to balanced
               trees},
  author    = {Pugh, William},
  journal   = {Communications of the ACM},
  volume    = {33},
  number    = {6},
  pages     = {668--676},
  year      = {1990},
  publisher = {ACM New York, NY, USA}
}

@article{yu2025topology,
	title={A topology-aware localized update strategy for graph-based ann index},
	author={Yu, Song and Lin, Shengyuan and Gong, Shufeng and Xie, Yongqing and Liu, Ruicheng and Zhou, Yijie and Sun, Ji and Zhang, Yanfeng and Li, Guoliang and Yu, Ge},
	journal={Proceedings of the VLDB Endowment},
	volume={19},
	number={3},
	pages={495--508},
	year={2025},
	publisher={VLDB Endowment}
}

@article{cai2024navigating,
  title     = {Navigating labels and vectors: A unified approach to
               filtered approximate nearest neighbor search},
  author    = {Cai, Yuzheng and Shi, Jiayang and Chen, Yizhuo and Zheng,
               Weiguo},
  journal   = {Proceedings of the ACM on Management of Data},
  volume    = {2},
  number    = {6},
  pages     = {1--27},
  year      = {2024},
  publisher = {ACM New York, NY, USA}
}

@article{singh2021freshdiskann,
  title   = {Freshdiskann: A fast and accurate graph-based ann index
             for streaming similarity search},
  author  = {Singh, Aditi and Subramanya, Suhas Jayaram and
             Krishnaswamy, Ravishankar and Simhadri, Harsha Vardhan},
  journal = {arXiv preprint arXiv:2105.09613},
  year    = {2021}
}

@inproceedings{xu2023spfresh,
  title     = {Spfresh: Incremental in-place update for billion-scale
               vector search},
  author    = {Xu, Yuming and Liang, Hengyu and Li, Jin and Xu, Shuotao
               and Chen, Qi and Zhang, Qianxi and Li, Cheng and Yang,
               Ziyue and Yang, Fan and Yang, Yuqing and others},
  booktitle = {Proceedings of the 29th Symposium on Operating Systems
               Principles},
  pages     = {545--561},
  year      = {2023}
}

@article{xu2025place,
  title   = {In-Place Updates of a Graph Index for Streaming
             Approximate Nearest Neighbor Search},
  author  = {Xu, Haike and Manohar, Magdalen Dobson and Bernstein,
             Philip A and Chandramouli, Badrish and Wen, Richard and
             Simhadri, Harsha Vardhan},
  journal = {arXiv preprint arXiv:2502.13826},
  year    = {2025}
}

@article{zhong2025lsm,
  title   = {LSM-VEC: A Large-Scale Disk-Based System for Dynamic
             Vector Search},
  author  = {Zhong, Shurui and Mo, Dingheng and Luo, Siqiang},
  journal = {arXiv preprint arXiv:2505.17152},
  year    = {2025}
}

@article{chen2024singlestore,
  title     = {Singlestore-v: An integrated vector database system in
               singlestore},
  author    = {Chen, Cheng and Jin, Chenzhe and Zhang, Yunan and
               Podolsky, Sasha and Wu, Chun and Wang, Szu-Po and Hanson,
               Eric and Sun, Zhou and Walzer, Robert and Wang, Jianguo},
  journal   = {Proceedings of the VLDB Endowment},
  volume    = {17},
  number    = {12},
  pages     = {3772--3785},
  year      = {2024},
  publisher = {VLDB Endowment}
}

@inproceedings{niu2025blendhouse,
  title        = {BlendHouse: A Cloud-Native Vector Database System in
                  ByteHouse},
  author       = {Niu, Zhaojie and Tian, Xinhui and Peng, Xindong and Chen,
                  Xing},
  booktitle    = {2025 IEEE 41st International Conference on Data
                  Engineering (ICDE)},
  pages        = {4332--4345},
  year         = {2025},
  organization = {IEEE}
}

@inproceedings{xian2024vector,
  title     = {Vector search with OpenAI embeddings: Lucene is all you
               need},
  author    = {Xian, Jasper and Teofili, Tommaso and Pradeep, Ronak and
               Lin, Jimmy},
  booktitle = {Proceedings of the 17th ACM International Conference on
               Web Search and Data Mining},
  pages     = {1090--1093},
  year      = {2024}
}

@article{aguerrebere2023similarity,
	title={Similarity Search in the Blink of an Eye with Compressed Indices},
	author={Aguerrebere, Cecilia and Bhati, Ishwar Singh and Hildebrand, Mark and Tepper, Mariano and Willke, Theodore},
	journal={Proceedings of the VLDB Endowment},
	volume={16},
	number={11},
	pages={3433--3446},
	year={2023},
	publisher={VLDB Endowment}
}

@inproceedings{yang2025effective,
  title        = {Effective and general distance computation for approximate
                  nearest neighbor search},
  author       = {Yang, Mingyu and Li, Wentao and Jin, Jiabao and Zhong,
                  Xiaoyao and Wang, Xiangyu and Shen, Zhitao and Jia, Wei and
                  Wang, Wei},
  booktitle    = {2025 IEEE 41st International Conference on Data
                  Engineering (ICDE)},
  pages        = {1098--1110},
  year         = {2025},
  organization = {IEEE}
}

@article{li2025sindi,
  title   = {SINDI: an Efficient Index for Approximate Maximum Inner
             Product Search on Sparse Vectors},
  author  = {Li, Ruoxuan and Zhong, Xiaoyao and Jin, Jiabao and Cheng,
             Peng and Ni, Wangze and Chen, Lei and Shen, Zhitao and Jia,
             Wei and Wang, Xiangyu and Lin, Xuemin and others},
  journal = {2026 IEEE 42nd International Conference on Data
  Engineering (ICDE)},
  year    = {2026}
}

@article{zhong2025enhancegraph,
  title   = {EnhanceGraph: A Continuously Enhanced Graph-based Index
             for High-dimensional Approximate Nearest Neighbor Search},
  author  = {Zhong, Xiaoyao and Jin, Jiabao and Cheng, Peng and Yang,
             Mingyu and Chen, Lei and Li, Haoyang and Shen, Zhitao and
             Lin, Xuemin and Shen, Heng Tao and Song, Jingkuan},
  journal = {arXiv preprint arXiv:2506.13144},
  year    = {2025}
}

@article{yang2024fast,
  title   = {Fast High-dimensional Approximate Nearest Neighbor Search
             with Efficient Index Time and Space},
  author  = {Yang, Mingyu and Li, Wentao and Wang, Wei},
  journal = {arXiv preprint arXiv:2411.06158},
  year    = {2024}
}

@inproceedings{prokhorenkova2020graph,
  title        = {Graph-based nearest neighbor search: From practice to
                  theory},
  author       = {Prokhorenkova, Liudmila and Shekhovtsov, Aleksandr},
  booktitle    = {International Conference on Machine Learning},
  pages        = {7803--7813},
  year         = {2020},
  organization = {PMLR}
}

@article{yu2025approximate,
  title   = {Approximate Nearest Neighbor Search of Large Scale Vectors
             on Distributed Storage},
  author  = {Yu, Kun and Jin, Jiabao and Zhong, Xiaoyao and Cheng, Peng
             and Chen, Lei and Shen, Zhitao and Song, Jingkuan and Shen,
             Hengtao and Lin, Xuemin},
  journal = {arXiv preprint arXiv:2510.17326},
  year    = {2025}
}

@inproceedings{li2023more,
  title     = {More than capacity: Performance-oriented evolution of
               pangu in alibaba},
  author    = {Li, Qiang and Xiang, Qiao and Wang, Yuxin and Song, Haohao
               and Wen, Ridi and Yao, Wenhui and Dong, Yuanyuan and Zhao,
               Shuqi and Huang, Shuo and Zhu, Zhaosheng and others},
  booktitle = {21st USENIX Conference on File and Storage Technologies
               (FAST 23)},
  pages     = {331--346},
  year      = {2023}
}

@article{navarro2002searching,
  title     = {Searching in metric spaces by spatial approximation},
  author    = {Navarro, Gonzalo},
  journal   = {The VLDB Journal},
  volume    = {11},
  number    = {1},
  pages     = {28--46},
  year      = {2002},
  publisher = {Springer}
}

@article{azizi2025graph,
  title     = {Graph-based vector search: An experimental evaluation of
               the state-of-the-art},
  author    = {Azizi, Ilias and Echihabi, Karima and Palpanas, Themis},
  journal   = {Proceedings of the ACM on Management of Data},
  volume    = {3},
  number    = {1},
  pages     = {1--31},
  year      = {2025},
  publisher = {ACM New York, NY, USA}
}

@online{Alipay2025,
  title        = {Alipay},
  organization = {Ant Group},
  year         = {2025},
  url          = {https://www.antgroup.com}
}

\received{July 2025}
\received[revised]{October 2025}
\received[accepted]{November 2025}

\balance

\AtEndDocument{\label{TotPages}}

\end{document}